\documentclass[aps,prd,twocolumn,a4paper,superscriptaddress,10pt]{revtex4-1}
\usepackage{graphicx}
\usepackage{multirow}
\usepackage{latexsym}
\usepackage{hyperref}
\usepackage[british]{babel}
\usepackage{amsmath}
\usepackage{wrapfig}
\usepackage{amsmath}
\usepackage{xcolor}

\begin{document}

\title{Multitension strings in high-resolution U(1)$\times$U(1) simulations}
\author{J. R. C. C. C. Correia}
\email{Jose.Correia@astro.up.pt}
\affiliation{Centro de Astrof\'{\i}sica da Universidade do Porto, Rua das Estrelas, 4150-762 Porto, Portugal}
\affiliation{Instituto de Astrof\'{\i}sica e Ci\^encias do Espa\c co, Universidade do Porto, Rua das Estrelas, 4150-762 Porto, Portugal}
\affiliation{Faculdade de Ci\^encias, Universidade do Porto, Rua do Campo Alegre 687, 4169-007 Porto, Portugal}
\author{C. J. A. P. Martins}
\email{Carlos.Martins@astro.up.pt}
\affiliation{Centro de Astrof\'{\i}sica da Universidade do Porto, Rua das Estrelas, 4150-762 Porto, Portugal}
\affiliation{Instituto de Astrof\'{\i}sica e Ci\^encias do Espa\c co, Universidade do Porto, Rua das Estrelas, 4150-762 Porto, Portugal}

\date{\today}

\begin{abstract}
Topological defects are a fossil relic of early Universe phase transitions, with cosmic strings being the best motivated example. While in most cases one studies Nambu-Goto or Abelian-Higgs strings, one also expects that cosmologically realistic strings should have additional degrees of freedom in their worldsheets, one specific example being superstrings from Type IIB superstring theory. Here we continue the scientific exploitation of our recently developed multi-GPU field theory cosmic strings code to study the evolution of U(1)$\times$U(1) multitension networks, which are a numerically convenient proxy: these contain two lowest-tension strings networks able to interact and form bound states, providing a convenient first approximation to the behaviour expected from cosmic superstrings. We start with a discussion of our code validation, including a comparison of the the evolution of these networks under three different assumptions: physical evolution (using the true equations of motion), the constant comoving width assumption (frequently used in the literature) and also the numerically convenient core growth case. We rely on the largest field theory simulations of this model so far, specifically $4096^3$, $\Delta x = 0.5$ boxes. We present robust evidence of scaling for the lightest strings, measured through a complete and self-consistent set of correlation length and velocity diagnostics. We also find a linearly growing average length of the bound state segments, consistent with a scaling behaviour. (In previously reported lower-resolution simulations, such behaviour had only been identified with carefully engineered initial conditions, rich in those segments.) Finally, while we see no evidence of a large population of bound states forming at early stages of the network evolution, we do present tentative evidence for an asymptotic constant value of the fraction of bound states, with this value being different in the radiation and the matter eras. Our work demonstrates that our GPU-accelerated field theory code can by successfully extended beyond the simple Abelian-Higgs approximation, and enables future detailed studies of realistic string networks and of their observational signatures.
\end{abstract}
\maketitle
\allowdisplaybreaks

\section{\label{intr}Introduction}

Topological defects are a possible fingerprint of early Universe phase transitions \cite{Kibble:1976sj}. Depending on the nature of the symmetry breaking pattern, different defects with different co-dimensions can form, the basic types being domain walls, cosmic strings and monopoles. The line-like cosmic strings are the most benign, in the sense that they are not expected to overclose the Universe and are ubiquitous in Grand Unified Theories. They are also expected to contribute to cosmological backgrounds, most notably the Cosmic Microwave Background (where they can contribute mostly in B-mode polarization) and the Stochastic Gravitational Wave Background.

Given that strings can contribute to such backgrounds and that their tension is directly related to the energy scale at which the phase transition occurred, searches for defects can either provide a smoking gun for theories of new physics, or constrain their model parameters. This makes them a primary target for both current and future observational probes, such as LISA \cite{LISA}, the SKAO \cite{Weltman:2018zrl}, or next-generation cosmic microwave background facilities \cite{CORE}. However, the heavy computational limitations of string simulations strongly restrict our current ability to provide reliable and accurate predictions for their observational fingerprints.

This is compounded by analogous limitations to the exploration of cosmic strings with additional degrees of freedom. Indeed, while in most cases one simulates Nambu-Goto or Abelian-Higgs strings, one also expects that realistic strings should have additional degrees of freedom in their worldsheets, including charges and currents \cite{Witten,Peter0,Peter1,Sarangi:2002yt,Peter2,Peter3}. Perhaps the best such example are superstrings from Type IIB superstring theory \cite{Witten}, stretched to horizon scales which can form a plethora of bound states; for a review see \cite{Sarangi:2002yt}. These superstrings are expected products of string inflationary scenarios. While they do not correspond to topologically stable solutions of the equations of motions of some field theory, they can still obey a topological stability condition and, for all practical purposes, form an interconnected network of strings, stretched to horizon scales---but including some significant twists to the usual cosmic string phenomenology.

In this work we continue the scientific exploitation of our recently developed multi-GPU field theory cosmic strings code \cite{Correia:2018gew,Correia:2020yqg} capable of simulating the largest ever Abelian-Higgs lattices. We tackle the evolution of multitension networks, envisaged as a numerically convenient first approximation to the behaviour expected from cosmic superstrings. Specifically, we have extended our code to the $U(1) \times U(1)$ model of Saffin \cite{Saffin:2005cs}. With our choices of the model parameters (to be detailed below) there are two lowest-tension strings networks, with equal tensions, that will form and will be able to interact and---presumably---also to form bound states.

We must clearly note that what follows is obviously field theory based, and thus cannot capture all the aspects of cosmic superstring phenomenology: the tension spectra will not be the one expected therein, no supersymmetry is included in the model, and the effects of extra dimensions are not included, resulting in less exotic intercommuting probabilities. However, our relatively simple model can already allow us to quantitatively investigate several aspects that are necessary stepping stones towards more realistic simulations which we intend to carry out in the future. Specifically, in this work we address in detail the impact of higher resolution and dynamic range on the network's scaling properties, and especially on the formation and properties of the bound states. In passing, and in the context of our code validation procedure, we also explore the impact of the Press-Ryden-Spergel (PRS) approximation \cite{PRS} on the formation and abundance of bound states in the network.

Our work builds upon, and significantly extends, two earlier works which have also simulated the Saffin model \cite{Urrestilla:2007yw,Lizarraga:2016hpd}. Firstly, Urrestilla \& Vilenkin \cite{Urrestilla:2007yw} relied on small $512^3, \Delta x=1.0, \Delta t=0.2$ simulations. While they clearly show that the network is not freezing (as one might naively have expected), the small dynamic range only allowed very tentative evidence of scaling, though interestingly this seemed to apply to all components, including the bound states (which they suggest are only a small fraction of the network's energy). These limitations are compounded by a relatively limited set of diagnostics: there are no velocity measurements, and even the correlation length slopes are reported without statistical uncertainties. Some of these limitations were mitigated by Lizarraga \& Urrestilla \cite{Lizarraga:2016hpd}, who used $1024^3, \Delta x=0.5, \Delta t=0.1$ simulations. Here some tentative velocity measurements are provided, and there is some exploration of the possible role of initial conditions, specifically for the case of the bound states: in addition to the natural case where bound states form dynamically, they also simulate the opposite and 'artificial' case where a network is numerically engineered in order to start with a large number of bound states. Again they report scaling of all components, but it is not clear whether the natural and artificial networks are approaching the same scaling solution, although the fraction of bound states at late times is small in both types of simulations. Importantly, all these simulation rely on the PRS algorithm.

Our goal here is to shed light on these issues, relying on an extensive set of high-resolution $4096^3, \Delta x=0.5, \Delta t=0.1$ simulations, as well as on a set of multiple numerical diagnostics for the correlation lengths and velocities of all the string components in the simulations. (These diagnostics include some redundancy, to allow consistency tests of our results.) As part of the code validation, we assess the possible impact of the PRS algorithm, including on the formation and stability of the bound states. We therefore provide a comparison of simulations using the PRS algorithm with those using physical evolution (i.e. the true equations of motion); in passing we also briefly explore the case of core growth evolution (also often used in the literature, for numerical convenience).

The work reported herein relied on about 125 000 node hours (corresponding to 8.5 million core hours) at Piz Daint, one of Europe's largest supercomputers, with the vast majority of this time corresponding to the larger $4096^3$ simulations (a number of smaller simulations were also done, as reported in what follows). Since one early purpose of the present work is code validation and verification, we have focused on the equal tensions case. The logical next step will be to simulate unequal tensions, which the code can of course do, but this is left for subsequent work, together with detailed comparisons with some analytic results available in the literature \cite{CKS,Pourtsidou:2010gu,Zipping,PhysRevD.99.063516}.

What follows is divided into five additional sections. We start with an introduction which discusses the field theory model being considered, followed by a section highlighting the simulation itself, including the discretization and methods for characterization of the network. We then proceed to present comparisons to previous works in the literature, providing a validation of our extended code. Finally we present our results, which constitute the highest-resolution $(p\,q)$ string simulations to date and, in a nutshell, confirm the scaling of these networks. In passing we also comment on the results of the exploration of the aforementioned numerical issues. A short final section summarizes the work conducted and discusses possible new avenues of investigation.

\section{\label{prac}Field theory cosmic strings and bound states}

Attempts to simulate bound state $(p,q)$ strings will inevitably fall into one of two categories. The first would be extensions of Nambu-Goto cosmic string simulations, but such simulations with more than one string type do not exist so far. The second are field theory simulations, of which the simplest model to implement corresponds to the dual local $U(1)$ model of Saffin \cite{Saffin:2005cs}. Consider the following Lagrangian
\begin{align}
    \mathcal{L} &= |D_{\mu}\phi|^2 -\frac{1}{4} F^{\mu\nu}F_{\mu\nu} \nonumber \\
    &+|\mathcal{D}_{\mu}\psi|^2 -\frac{1}{4} G^{\mu\nu}G_{\mu\nu} -V(|\phi|,|\psi|) 
\end{align}
with two complex scalar fields, $\phi$ and $\psi$, and two $U(1)$ gauge fields, $A_{\mu}$ and $B_{\mu}$, with corresponding gauge field strengths $G_{\mu \nu}$ and $F_{\mu \nu}$. The covariant derivatives, gauge field strengths and potential are given by
\begin{align}
    D_{\mu} = \partial_{\mu} -ie_p A_{\mu} && \mathcal{D}_{\mu} = \partial_{\mu} -ie_q B_{\mu}
\end{align}
\begin{align}
    F_{\mu\nu} = \partial_{\mu}A_{\nu} - \partial_{\nu}A_{\mu} && G_{\mu\nu} = \partial_{\mu}B_{\nu} - \partial_{\nu}B_{\mu}
\end{align}
\begin{align}
    V(|\phi|,|\psi|) &= \frac{\lambda_p}{4}(|\phi|^2 -\sigma_p^2)^2 +\frac{\lambda_q}{4}(|\psi|^2 - \sigma_q^2)^2 \nonumber \\
                     &-\kappa (|\phi|^2 - \sigma_p^2)(|\psi|^2 - \sigma_q^2) 
\end{align}
where $\lambda_{p,q}$ are scalar couplings, $e_{p,q}$ are gauge couplings and $\kappa$ is the coupling between the two scalar fields. If these parameters are such that $0 < \kappa < \frac{1}{2}\sqrt{\lambda_p \lambda_q}$ then the vacuum manifold will be non-trivial in the two sectors, supporting the existence of two types of strings strings and, due to the non-zero value of $\kappa$, also of bound state strings.

Before proceeding, we should highlight a caveat: this model cannot capture all the proper physics of cosmic superstrings. For instance, it clearly lacks supersymmetry \cite{Jeannerot1,Jeannerot2}, and the intercommuting probabilities and the tension spectra will not be the expected ones for superstrings \cite{Jones1,Jones2}. Still, a field theory simulation of this model can provide a plethora of useful information, including on the abundances of the bound states and the properties of scaling of each network component.

Previous studies of this model have been reported \cite{Urrestilla:2007yw,Lizarraga:2016hpd}, but some potentially key aspects have been neglected so far. The main one is to study these networks under physical evolution conditions (as opposed to using the PRS algorithm), which is particularly relevant for a proper assessment of bound state formation. The limited dynamic range of earlier works (at least when compared to what is achievable in contemporary state-of-the-art facilities) also warrants improvements. As previously mentioned, our dynamic range in the present work is much larger ($4096^3$ at $dx=0.5$) and we will also carry out physical evolution simulations. Nevertheless, we will attempt to make choices which enable us to compare our results to both previous works (and note where differences arise), as this is an important component of our code validation process. That said, both works also differ between them at times; our numerical choices, including most definitions of length and velocity estimates, will be closer to the definitions of the most recent of the two previous works \cite{Lizarraga:2016hpd}.

We begin with the aforementioned $U(1)\times U(1)$, originally locally invariant Lagrangian density. From the variation of the action under the assumption of a Friedmann-Lema\^{\i}tre-Roberson-Walker metric and the temporal gauge ($A_0 = 0$), one obtains the equations of motion
\begin{equation}
\ddot{\phi} + 2\frac{\dot{a}}{a}\dot{\phi} = D^jD_j\phi -a^2 \phi \bigg[ \frac{\lambda_p}{2} (|\phi|^2 - \sigma_p^2) - \kappa (|\psi|^2 - \sigma_q^2) \bigg] 
\end{equation}
\begin{equation}
\ddot{\psi} + 2\frac{\dot{a}}{a}\dot{\psi} = D^jD_j\psi -a^2 \psi \bigg[ \frac{\lambda_q}{2} (|\psi|^2 - \sigma_q^2) - \kappa (|\phi|^2 - \sigma_p^2) \bigg] 
\end{equation}
\begin{equation}
\dot{F}_{0j} = \partial_j F_{ij} -2a^2 e_p^2 Im[\phi^* D_j \phi]
\end{equation}
\begin{equation}
\dot{G}_{0j} = \partial_j G_{ij} -2a^2 e_q^2 Im[\psi^* D_j \psi]
\end{equation}
along with a two copies of Gauss's law, for each sector,
\begin{equation}
\partial_i F_{0i} = 2 a^2 e_p^2 Im[\phi^* \dot{\phi}]\phi
\end{equation}
\begin{equation}
\partial_i G_{0i} = 2 a^2 e_q^2 Im[\psi^* \dot{\psi}]\psi
\end{equation}
which will be tested in the validation section. As one may expect, all of the previous equations are symmetric under exchanges of $\phi \leftrightarrow \psi$, $A_\mu \leftrightarrow B_\mu$, $F_{ij} \leftrightarrow G_{ij}$ and under the corresponding exchanges of couplings $\lambda_p \leftrightarrow \lambda_q$ and $e_p \leftrightarrow e_q$. We will assume criticality in the two sectors, which corresponds to the condition
\begin{equation}
\frac{\lambda_{p,q}}{2e_{p,q}^2} = 1\,,
\end{equation}
and for simplicity we will further assume both symmetry breaking scales to be unity $\sigma_{p,q} = 1$.

\section{\label{overview}Simulation Overview}

In these simulations, much like in the Abelian-Higgs and domain walls cases, the comoving defect width would nominally behave as $a^{-2}$ for both the scalar and gauge widths. The canonical workaround for this issue is to adopt the PRS comoving width controlling prescription \cite{PRS}, where all coupling constants are made to vary as
\begin{align}
\lambda_{p,q} = \lambda_{p,q0} a^{-2(1-\beta)} && e = e_{p,q} a^{-(1-\beta)} &&  \kappa = \kappa_0 a^{-2(1-\beta)}
\end{align}
and where, following \cite{Lizarraga:2016hpd} $\kappa$ is made to vary in the same way as $\lambda_{p,q}$. The parameter $\beta$ can be set to $0$ (constant comoving width, i.e. PRS), $1$ (recovering the original equations of motion), or to other values; in particular the choice $\beta<0$ implies core growth, which despite being clearly unphysical can be numerically useful \cite{Correia:2020gkj}. One of our purposes in the present work is to use our code validation to provide a comparison of the impact of the PRS approximations, as compared to the true physical evolution.

This generic prescription implies that the equations of motion can now be re-written as
\begin{equation}
\ddot{\phi} + 2\frac{\dot{a}}{a}\dot{\phi} = D^jD_j\phi -a^{2\beta}  \phi \bigg[ \frac{\lambda_p}{2} (|\phi|^2 - \sigma_p^2) -\kappa(|\psi|^2 - \sigma_q^2)\bigg]
\end{equation}
\begin{equation}
\ddot{\psi} + 2\frac{\dot{a}}{a}\dot{\psi} = D^jD_j\psi -a^{2\beta}  \psi \bigg[ \frac{\lambda_q}{2} (|\psi|^2 - \sigma_q^2) -\kappa(|\phi|^2 - \sigma_p^2)\bigg]
\end{equation}
\begin{equation}
\dot{F}_{0j} + 2(1-\beta)\frac{\dot{a}}{a}F_{0j} = \partial_j F_{ij} -2a^{2\beta} e_p^2 Im[\phi^* D_j \phi]
\end{equation}
\begin{equation}
\dot{G}_{0j} + 2(1-\beta)\frac{\dot{a}}{a}G_{0j} = \partial_j G_{ij} -2a^{2\beta} e_q^2 Im[\psi^* D_j \psi]\,.
\end{equation}

This is very similar to writing two discretized evolution equations for the two Abelian-Higgs strings, but with an additional term due to the existence of the coupling in the potential. The evolution equations for updating the conjugate momenta of the scalar fields in both string sectors are executed first in each timestep via
\begin{equation}
\begin{split}
(1+\delta)\Pi^{x,\eta+\frac{1}{2}} &= (1-\delta)\Pi^{x,\eta-\frac{1}{2}}+\Delta\eta  \bigg[ D_j^-D_j^+\phi^{x,\eta} \\ &-a_\eta^{2\beta}\phi^{x,\eta} [ \frac{\lambda_{p0}}{2} (|\phi^{x,\eta}|^2-\sigma_p^2) \\& -\kappa(|\psi^{x, \eta}|^2 - \sigma_q^2) ] \bigg]
\end{split}
\end{equation}
\begin{equation}
\begin{split}
(1+\delta)\Psi^{x,\eta+\frac{1}{2}} &= (1-\delta)\Psi^{x,\eta-\frac{1}{2}}+\Delta\eta  \bigg[D_j^-D_j^+\psi^{x,\eta} \\ &-a_\eta^{2\beta}\psi^{x,\eta} [\frac{\lambda_{q0}}{2} (|\psi^{x,\eta}|^2-\sigma_q^2) \\& -\kappa(|\phi^{x, \eta}|^2 - \sigma_p^2) ]  \bigg]
\end{split}
\end{equation}
where $\delta$ is defined as
\begin{equation}
\delta=\frac{1}{2} \alpha \frac{dlna}{dln\eta}\frac{\Delta \eta}{\eta} = \frac{1}{2} \alpha \frac{m \Delta \eta}{(1-m)\eta}\,.
\end{equation} 
and sets the strength of Hubble damping on the scalar fields. Throughout this work the constant $\alpha$, which is the factor multiplying the term in $\dot{\phi}$ in the equations of motion, has the value $\alpha=2$. To obtain the last equality we have assumed power-law expanding universes, specifically with a scale factor $a\propto t^{m}$; in this work we will report on simulations in the radiation era ($m=1/2$) and the matter era ($m=2/3$).

Next we have the evolution equations for the gauge conjugate momenta $E$ and $H$, 
\begin{equation}
\begin{split}
(1+\omega)E^{x,\eta+\frac{1}{2}}_i &=  (1-\omega)E^{x,\eta-\frac{1}{2}}_i +\Delta\eta [-\partial_i^- F_{ij} \\
&+ 2e_{p0}^{2}a^{2\beta}_\eta Im[\phi^* D_i^+ \phi]^{x,\eta} ]
\end{split} 
\end{equation}
\begin{equation}
\begin{split}
(1+\omega)H^{x,\eta+\frac{1}{2}}_i &=  (1-\omega)H^{x,\eta-\frac{1}{2}}_i +\Delta\eta [-\partial_i^- G_{ij} \\
&+ 2e_{q0}^{2}a^{2\beta}_\eta Im[\psi^* D_i^+ \psi]^{x,\eta} ]
\end{split} 
\end{equation}
where the parameter $\omega$, defined as
\begin{equation}
\omega=\delta(1-\beta)\,,
\end{equation}
introduces an unphysical damping of the gauge field for any value of $\beta \neq 1$. In the case of $\beta=1$, the damping vanishes as is expected in the physically correct limit.
To finish updating all the fields in a given timestep, one uses following prescription
\begin{equation}
\phi^{x,\eta+1} = \phi^{x,\eta} + \Delta \eta \Pi^{x,\eta+\frac{1}{2}}
\end{equation}
\begin{equation}
\psi^{x,\eta+1} = \psi^{x,\eta} + \Delta \eta \Psi^{x,\eta+\frac{1}{2}}
\end{equation}
\begin{equation}
A^{x,\eta+1}_i = A^{x,\eta}_i + \Delta \eta E^{x,\eta+\frac{1}{2}}_i
\end{equation}
\begin{equation}
B^{x,\eta+1}_i = B^{x,\eta}_i + \Delta \eta H^{x,\eta+\frac{1}{2}}_i\,,
\end{equation}
again for all non-conjugate fields in all sectors.

One important caveat in defect simulations is that the dynamical range that can be simulated and used to characterize certain behaviors is limited. Therefore, if one assumes that a scaling solution is an attractor and should eventually be reached, it is desirable that the numerical simulation approaches this solution as fast as possible, There is no unique way to proceed in terms of generating initial conditions, but we can partially rely on previous analogous works \cite{Urrestilla:2007yw,Lizarraga:2016hpd} while applying some modifications of our own. Specifically, we generate simple initial conditions, with both complex scalar fields having magnitudes set by the vacuum expectation value and with random phases. We will also apply a diffusive phase to smooth out the large gradients present in such initial conditions, followed by a damping period to form string networks more quickly. The specific details of how these operations are applied (and when, in terms of conformal time) are different from the previous authors, the differences being mainly motivated by our much larger box sizes.

We start with the diffusive cooling equations, taken from the previous cooling functions used in Abelian-Higgs for the first sector
\begin{equation}
  \dot{\phi} = D^jD_j\phi -\frac{\lambda_{p0}}{2} (|\phi|^2 - \sigma_p)\phi 
  \end{equation}
  \begin{equation}
  \dot{F}_{0j} = \partial_j F_{ij} -2 e_{p0}^2 Im[\phi^* D_j \phi]\,
\end{equation}
and interchangeably for the second sector by, again, performing appropriate substitutions ($\phi \rightarrow \sigma$, $A_\mu \rightarrow B_\mu$, $\lambda_p \rightarrow \lambda_q$ and $e_p \rightarrow e_q$),
\begin{equation}
  \dot{\psi} = D^jD_j\psi -\frac{\lambda_{q0}}{2} (|\psi|^2 - \sigma_q)\psi 
  \end{equation}
  \begin{equation}
  \dot{G}_{0j} = \partial_j G_{ij} -2 e_{q0}^2 Im[\psi^* D_j \psi]\,.
\end{equation}
Note that while this method of cooling is equal to what was previously implemented for the Abelian-Higgs case \cite{Correia:2020gkj} (if one ignores the presence of the coupling term in the potential) it is different from what was applied in \cite{Urrestilla:2007yw,Lizarraga:2016hpd} which merely average each scalar field over nearest neighbors 30 times. In practice, both serve the purpose of smoothing over the initial conditions gradients and we saw no indication throughout the validation procedure that using either method produces a discernible impact on network evolution. This cooling is applied from an initial conformal time of $\eta=-10.0$, until $\eta=1.0$, in timesteps of size $\Delta \eta = 1/30$.

Next comes a damping period, which facilitates the approach to scaling. Although previous work on $U(1)\times U(1)$-string network simulations have used the above cosmological discretized equations, for this period they introduced a fixed Hubble damping factor $\gamma = \frac{\dot{a}}{a}$ of either $0.2$ \cite{Urrestilla:2007yw} or $0.4$ \cite{Lizarraga:2016hpd}. In our case the highly damped cosmological evolution will be set by varying the expansion rate. Generically, we consider power laws in physical time, again parametrized as $a\propto t^m$. For instance if we initially set a high expansion rate $m=0.9$ and apply damping for a short, early period of time (from $\eta=1.0$ until $\eta=5.0$), the damping factor will be sufficiently large to quickly relax the fields into a network. The fact that this period is short enables most of the dynamic range of the network simulation to still be available for evolution in a matter or radiation dominated epoch, further enhancing our dynamic range gains. We will set the duration of this phase such that $\gamma$ never falls below $0.4$. Although \cite{Urrestilla:2007yw} suggested that a sufficiently high damping would aid in the formation of bound states (at an initial stage), we see no evidence of a large population of bound states forming at early stages of the network evolution.

\subsection{\label{avnet}Network simulation diagnostics}

Besides a first-pass validation with the Gauss law and from an inspection of field isosurfaces, we will also define suitable length estimators for each network type, such that we can assess scaling (via the evolution of the mean string separations) and compute bound state abundances. These diagnostics follow from our recent work on Abelian-Higgs simulations \cite{Correia:2020yqg,Correia:2020gkj,Correia:2019bdl}. Specifically, we rely on length estimators based on windings, and therefore the mean string separations are given by
\begin{align}
  \xi_{p} = \sqrt{\frac{ \mathcal{V} }{L_{p}}} && \xi_{q} = \sqrt{\frac{ \mathcal{V} }{L_{q}}} && \xi_{pq} = \sqrt{\frac{ \mathcal{V} }{L_{pq}}}
\end{align}
where $L_p$ and $L_q$ are given by the sum of non-zero plaquettes computed in the respective sector (so either using $\phi$, $A_\mu$ or $\psi$, $B_\mu$) with the length of string in bound states $L_{pq}$ subtracted. On the other hand, $L_{pq}$ corresponds to cells where both types of winding overlap. We leave the details of this computation to the next section. For now, we note that with the choice of parameters made for the validation procedure (corresponding to strings of equal tension), we do not observe any bound state with a winding of two or higher (in any sector), meaning that $pq$-strings are, at least for the intents and purposes of this section, (1,1) bound states.
 
In principle this is sufficient to search for and identify a scaling behavior for each relevant string species, although we can additionally define length estimators for the full network (all string types included together). This was done in \cite{Lizarraga:2016hpd} via a combined winding estimator,
\begin{equation}
\xi_{W} = \sqrt{\frac{ \mathcal{V} }{ L_{p} + L_{q} + L_{pq} } }\,.
\end{equation}
This can also be done via the canonical Lagrangian length estimator used throughout the literature in Abelian-Higgs simulations,
\begin{equation}
\xi_{\mathcal{L}} = \sqrt{ \frac{-\mu V}{ \sum \mathcal{L}_x } }\,,
\end{equation}
where $V$ is the box volume, $\mu$ is the tension of either basic string type and $\sum \mathcal{L}_x$ is a  lattice-wide sum of the Langrangian, computed at every site. The tension for the (1,1) state is different, however, and due to the fact that the most abundant string types have exactly the same tension this estimator still gives a good indication of the scaling behavior of the network. We do note that if we wish to test the specific case where the basic constituent strings have unequal tensions, this estimator might not be the most reasonable choice. 

Let us also define some additional quantities complementing the length estimator. Similar to \cite{Lizarraga:2016hpd}, one can compute two fractions which allow us to quantify the relative abundance of bound states. First is the total fraction of bound states $f_{total}$, defined as
\begin{equation}
f_{total} = \frac{L_{pq}}{L_{pq} + L_p + L_q}\,,
\end{equation}
which is expected not to exceed more than a few percent when scaling is reached for both the matter and radiation epochs. We can also compute this abundance relative to the length of one of the constituent strings---for instance relative to $L_p$. In this case, and again following the definition and notation of \cite{Lizarraga:2016hpd}, we have
\begin{equation}
f_{p} = \frac{L_{pq}}{L_p}\,.
\end{equation}
We remark that although our present goal is merely to study the abundance of bound states in these simulations and how their dynamics may be affected by certain numerical choices, we will use the velocity estimators of \cite{Lizarraga:2016hpd} as an additional validation source.

Measuring string network velocities is also important, both for the purpose of testing whether or not the networks are scaling and for the purpose of calibrating quantitative analytic models of defect evolution \cite{Book}. Here we use the scalar field velocity estimator, for which the mean string velocity is given by
\begin{equation}
\label{eq:defvPhiSS}
<v^2>_{\phi} = \frac{2R}{1+R}\,,
\end{equation}
with $R$ given by
\begin{equation}
\label{eq:defRSS}
R = \frac{\sum_x ( |\Pi|^2 + |\Psi|^2 ) \mathcal{W}}{\sum_{x,i} ( |D^+_{x,i} \phi|^2 + |D^+_{x,i} \psi|^2 ) \mathcal{W} }\,,
\end{equation}
with $\mathcal{W}$ being a weight function, meant to merely localize the estimators around strings. We will refer to this as a the field-based velocity estimator. The weight function can be used to specify that we wish to compute the mean velocity of all strings, merely by choosing it to be equal to the Lagrangian, or the mean velocity of only bound-state strings, by choosing the weight function to be given by the interaction potential. We note that this choice of $R$ extends its previous (and usual) definition \cite{Lizarraga:2016hpd,Correia:2019bdl}, since it takes into account both scalar fields (or both string sectors).

\subsection{\label{pqseg}Identifying (p,q) segments}

The computation of the total length of $pq$-strings in the simulation box may seem to be reduced to the relatively simple task of detecting plaquettes of exactly double winding (one winding of each type). However, this is not sufficient: there can be some accidental displacements of windings, such that the two strings still overlap, but the windings are situated one lattice site apart. On the other hand, one can also have accidental crossings at any given timestep, of say only one plaquette, which in the next timestep no longer overlap, and in such cases we can reasonably state that no bound-state string formed. In order to solve these issues, the authors of \cite{Lizarraga:2016hpd} have allowed p- and q-strings to be considered as overlapping if they are within a transverse distance of 4 lattice units, while in the earlier case of \cite{Urrestilla:2007yw} a maximum inter-segment distance (rather than the transverse distance) of 5 was adopted. In addition, both works also considered a minimum threshold on the length of segments, requiring that proper bound states have a minimum of either $L_{pq}=3$ and $L_{pq} = 20$, respectively. 

We now describe our two methods for the computation of the length estimator, which slightly differ from those above (and improve them, in a sense to be described below), but give rise to similar conclusions. A first method is based on using cells pierced by strings, without specifying the plaquettes themselves. This is less memory intensive, although it gives no information about the orientation of different strings in either the same cell or in neighbouring cells. This first method is made via a custom CUDA kernel and it should be understood as no more than a fast, approximate (and less robust) computation of the length itself. A complementary second method requires the usage of the \textit{in situ} capabilities of our simulation \cite{Correia:2020yqg,Correia:2021mgf}. This implies that the second method is more time-consuming, as a result of being more I/O-intensive, although the method itself is more robust, for reasons that we shall explain next.

In our first method we merely store in two separate arrays (one for each sector) if a cell is pierced by a winding (with the content of each cell set by the total magnitude of the windings that pierce it). We then verify if cells pierced in both arrays are either non-zero at the same location, or displaced by one cell in any direction. We then sum the number of cells pierced by both types of strings and use this as an indication of the number of cells that make up the length.

This method is less robust overall for two reasons. The first reason is that number of cells of an overlap region should give rise to a collection of segments of length $\Delta x \times (N_{cells} - 1)$. Since this kernel merely detects overlaps site by site, but doesn't attempt to cluster cells into regions (this in fact would have to be done a posteriori, via some clustering algorithm), we directly use the number of cells overall. We may therefore expect an off-by-one systematic error. This is solved by the robust method, as collections of cells are separated by regions. The second reason is that since no clustering is performed in the fast method, it is not possible to apply a lower threshold on the number of cells which a segment should contain to be considered a proper bound state. This again is easily solved in the robust method.

We could argue that these biases might not be significant, or at least have a reasonable chance of yielding adequate results: $L_{pq}$ should be dominated by the larger segments (and not by small few-cell segments) and having an off-by-one error in large segments will not incur a significant difference towards the total string length. Nonetheless, the robust method also has an advantage as we shall see next: we can choose the overlap to occur at greater distances than just one cell. This is due to the flexibility of the filter that identifies overlaps between two data sources in the \textit{Paraview} software which we use. However more robust and flexible the second method may be, the first one already yields reasonable results; specifically, they are already in agreement with literature results.

\begin{figure*}
  \centering
\includegraphics[width=1.0\columnwidth]{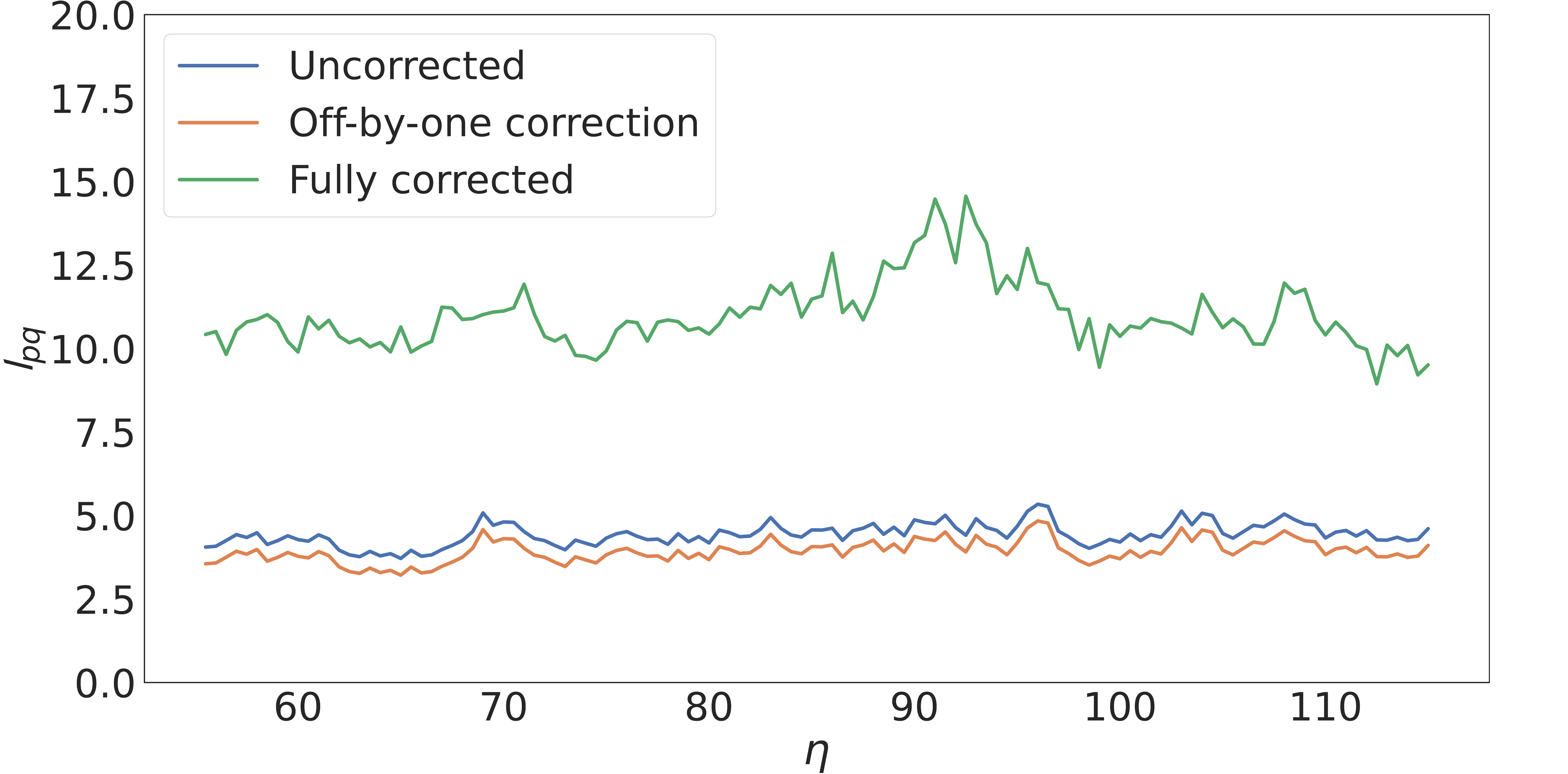}
\includegraphics[width=1.0\columnwidth]{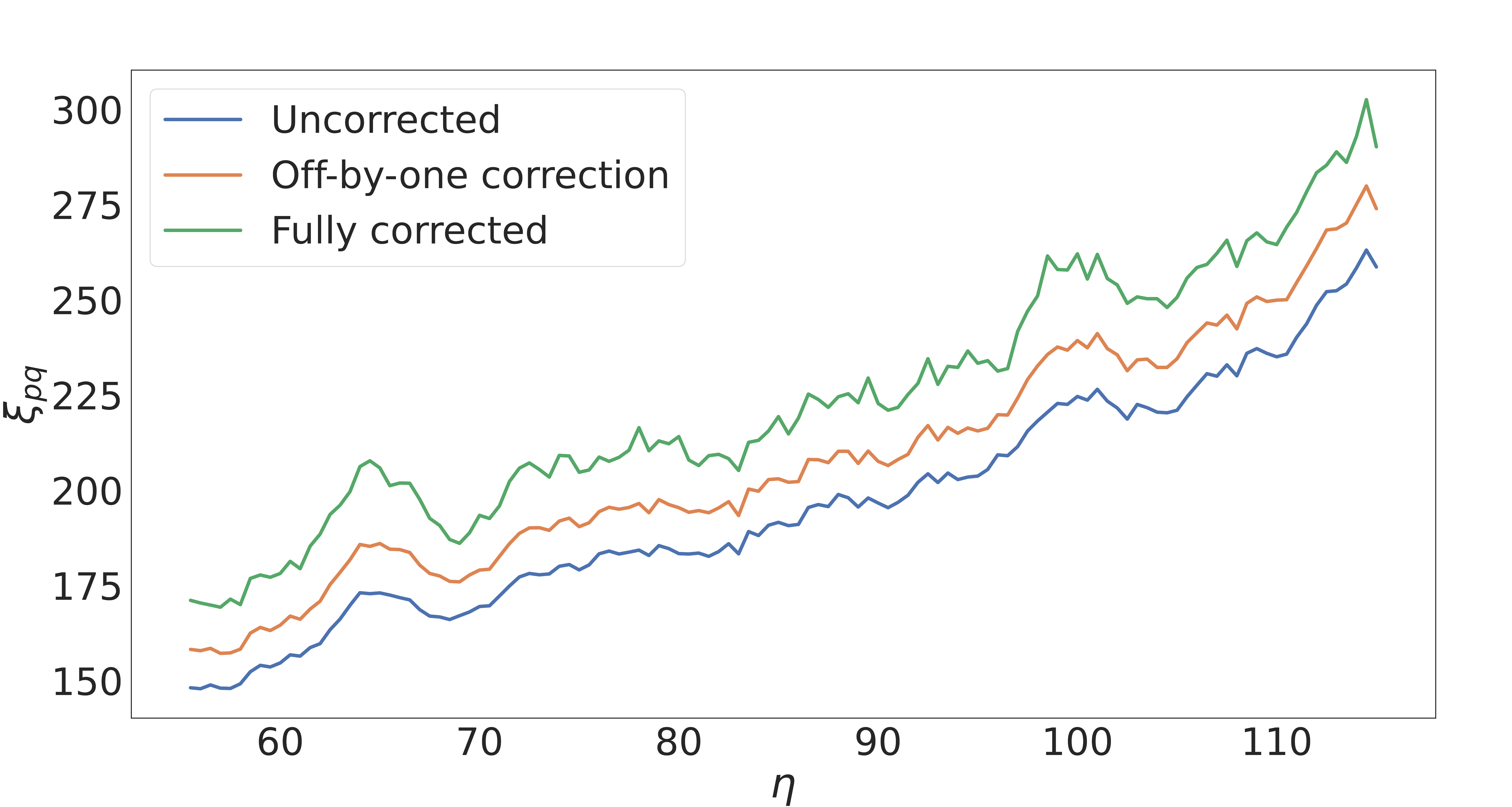}
\includegraphics[width=1.0\columnwidth]{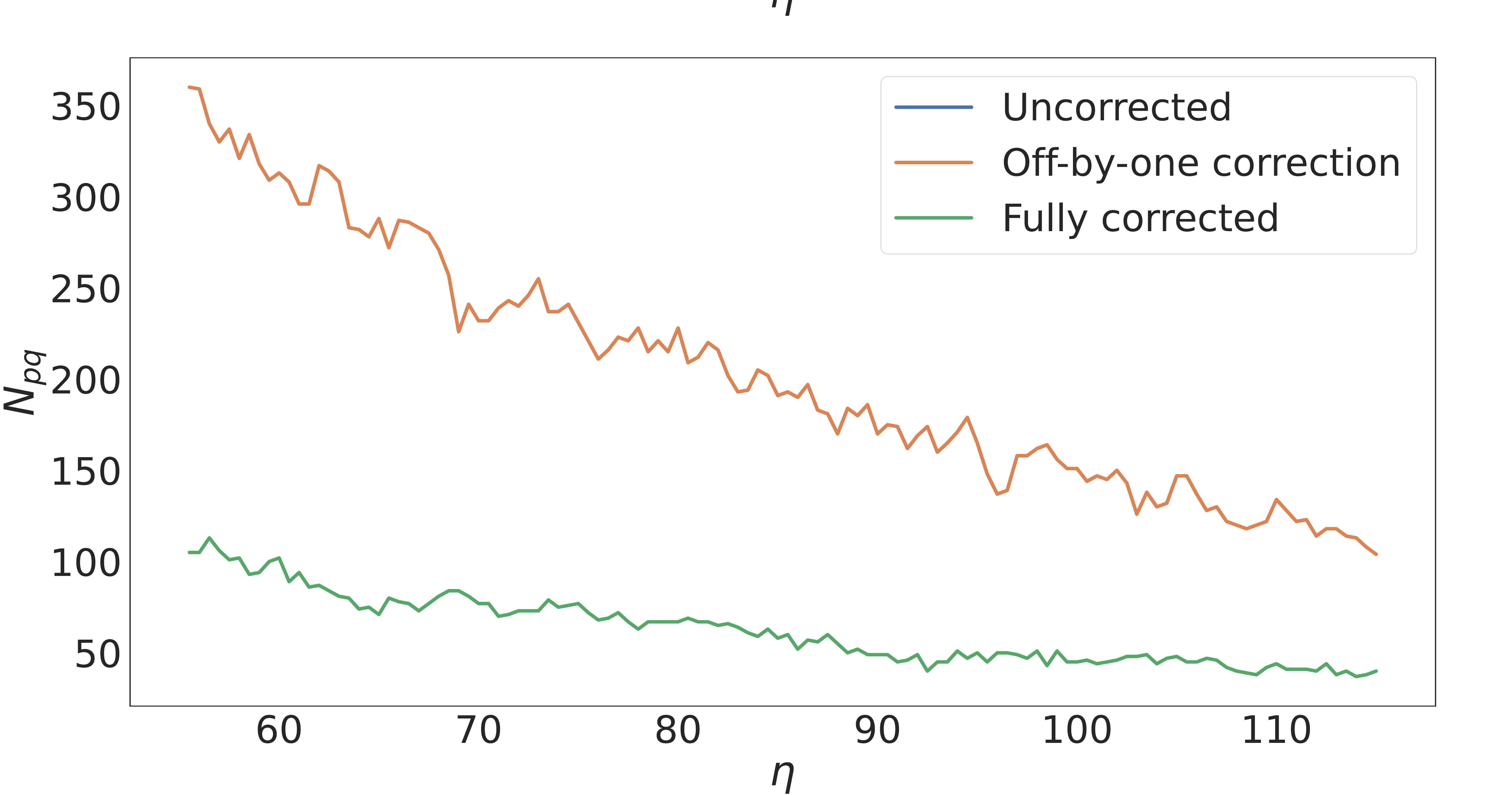}
\includegraphics[width=1.0\columnwidth]{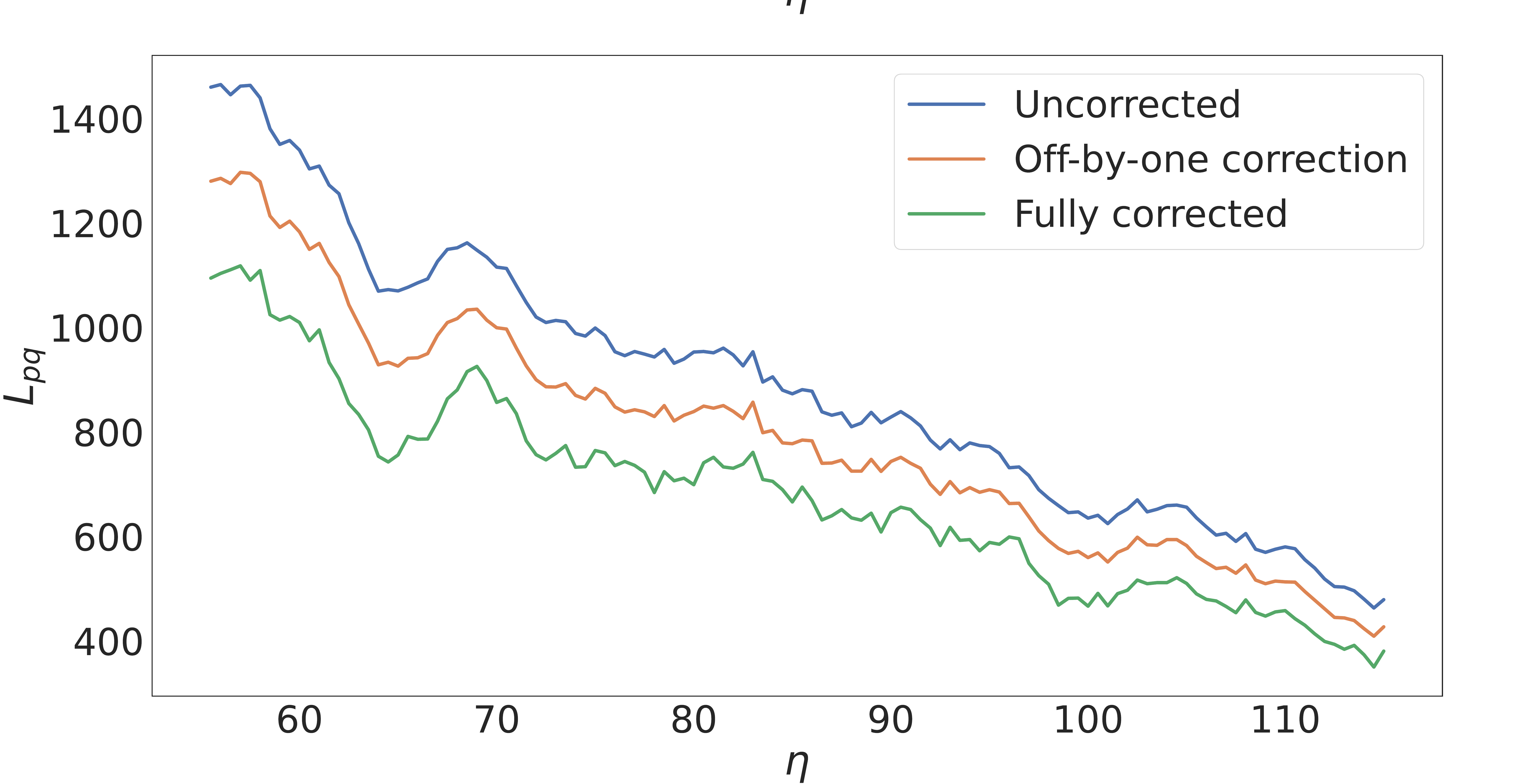}
\caption{The impact of corrections from the robust method on the mean segment length $l_{pq}$ on the upper left panel, the evolution of mean $pq$-string separation $\xi_{pq}$ on the upper right panel, the number of $pq$ segments $N_{pq}$ on the lower left panel and on the total segment length $L_{pq}$ for the lower right panel, for a small ($512^3$, $\Delta x=0.5$), matter epoch simulation. These panels assume an overlap of one cell in all directions and a minimum threshold on the segment length of $l_{pq}=3.5$.\label{figure01}}
\end{figure*}

\begin{figure*}
\includegraphics[width=1.0\columnwidth]{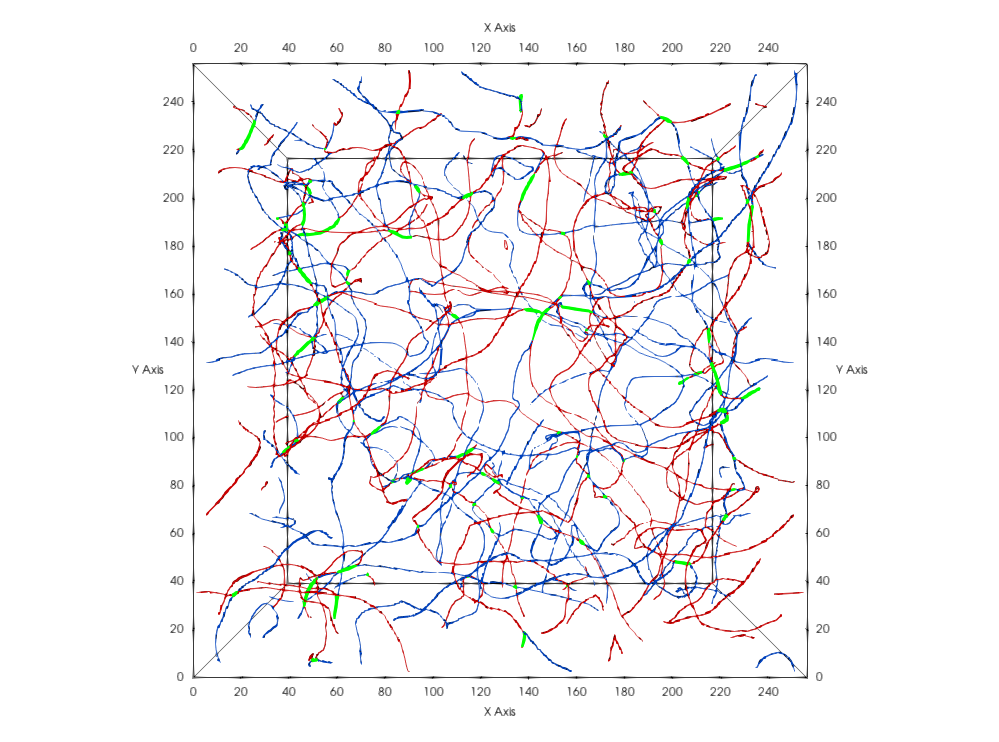}
\includegraphics[width=1.0\columnwidth]{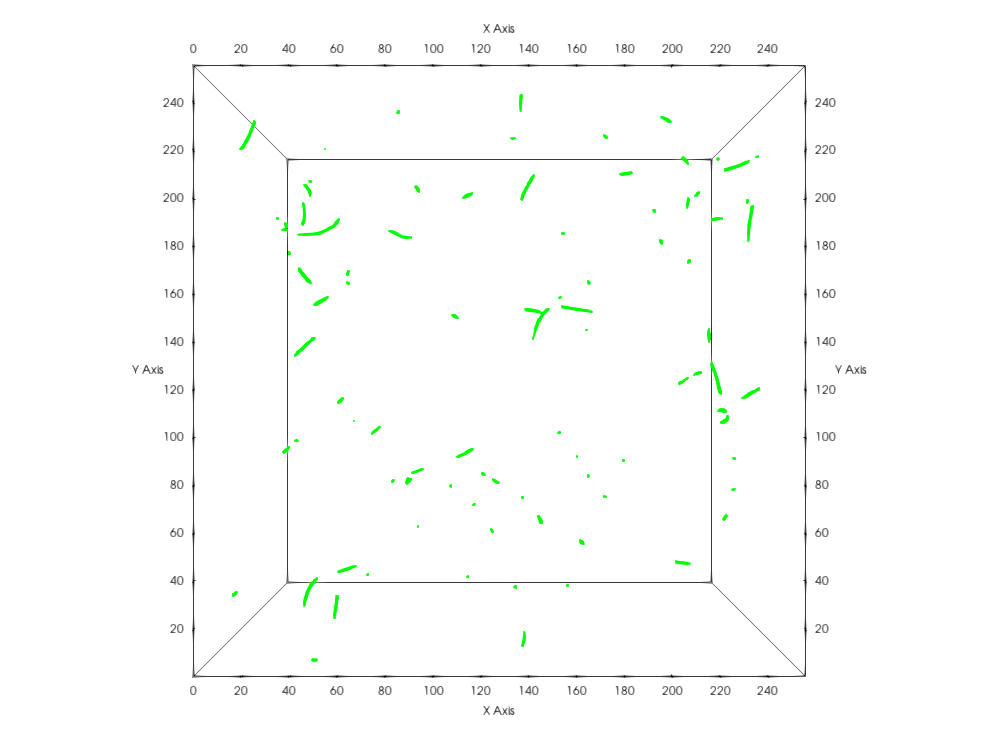}
\includegraphics[width=1.0\columnwidth]{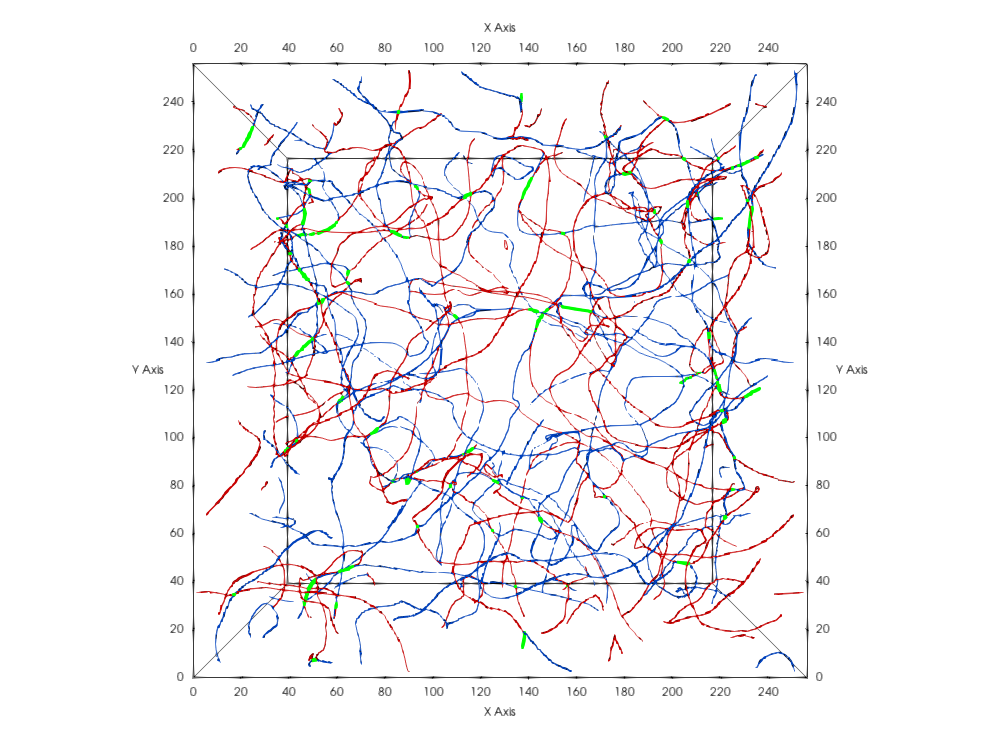}
\includegraphics[width=1.0\columnwidth]{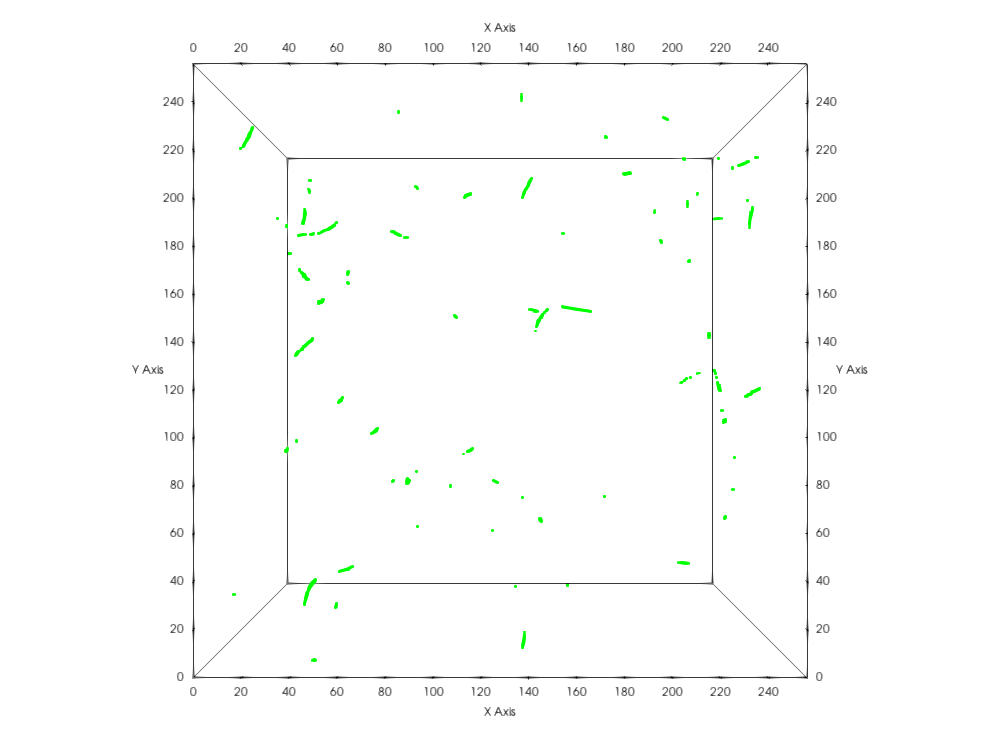}
\includegraphics[width=1.0\columnwidth]{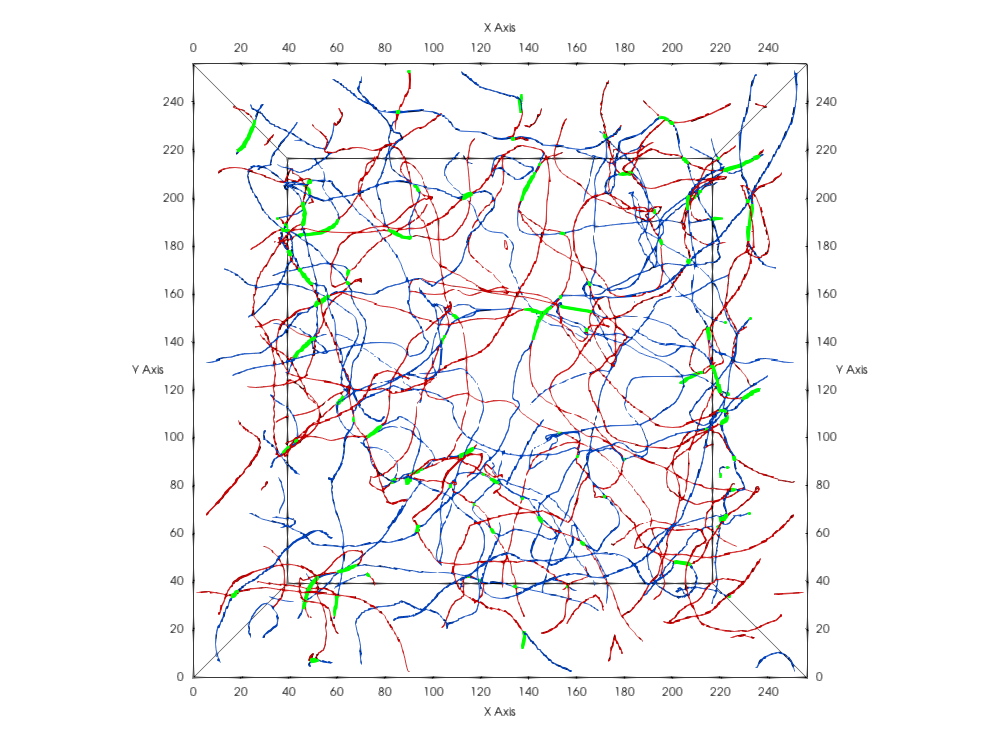}
\includegraphics[width=1.0\columnwidth]{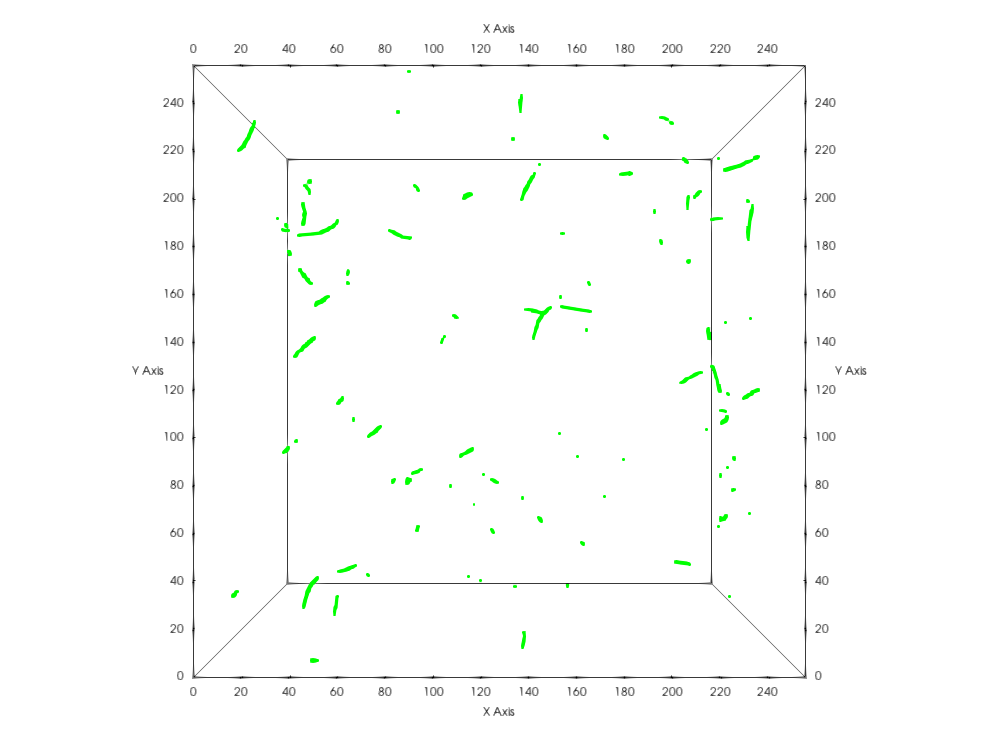}
\caption{Visualizations of a snapshot of a $512^3$, $\Delta x=0.5$ simulation at conformal time $\eta=101.0$ in the matter epoch. Cells pierced by p-strings are color-coded in blue, those pierced by q-strings in red and $pq$-strings in green. In the left-hand-side panels we display all three string types; in the right-hand-side panels we show only $pq$-strings. The identification of bound states is done either via the interaction potential isosurface $\kappa(|\phi|^2-\sigma_p^2)(|\psi|^2-\sigma_q^2)=0.855$ (on the top) or via winding cell overlaps (requiring exact overlaps the middle panel, and one cell of tolerance in the lower panel, both obtained via the fast method). We are not excluding regions of cells using a minimum threshold on the length of the bound state. \label{figure02}}
\end{figure*}

Before we move to the characterization of the robust method, there is an additional quantity which has been used to characterize pq-string field theory simulations, which is the average physical length of $pq$-segments. It can be defined as
\begin{equation}
  l_{pq} = \frac{L_{pq} }{N_{pq}}
\end{equation}
where $N_{pq}$ is the number of $pq$-segments and $L_{pq}$ is now defined as $L_{pq} = (N_{cells}-1) \times \Delta x$. We additionally consider only regions with $N_{cells}$ larger than $41$ (equivalently, $L_{pq} > 20.0$) for large high resolution lattices ($4096^3$). Here we note another disadvantage of the fast method: since it does not classify windings into their respective segments, it also does not compute $N_{pq}$, meaning that in order to study such a quantity we must use the robust method.

The robust method heavily relies on our recently developed \textit{in situ} visualization tools \cite{Correia:2020yqg,Correia:2021mgf}. The catalyst adaptor was changed to output cells pierced by windings of the first sector, windings of the second sector, the interaction potential $V_{int}$, and cells identified as pierced by $pq$-segments via the fast method. We further note that the value of each cell will depend on the absolute number of windings that pierce a cell through plaquettes: one $p$-string at criticality with $\lambda_p = 2e_p^2 = 2$ will pierce two faces of a cell, resulting in having an output value of $2$. This is important as strings with higher windings, say a (2,0) string, would appear as a collection of cells valued at $4$ for p-winding cells, but with no corresponding cells for q-windings. Note that this also leads us to confirm that no (2,0) or (0,2) or even other higher winding bound states were found. In total, this means we end up with four (optional) output files per timestep: one per winding type (p, q or pq) and one for the interaction potential isosurfaces. All outputs are then passed to a custom pipeline, adapted from a previous centerlines script, for which a detailed discussion can be found in \cite{Correia:2020yqg,Correia:2021mgf}, which visualizes not only an isosurface of the interaction potential but also both types of cells, and finds overlapping cells ($pq$-strings). A connectivity filter is then applied to separate each individual string segment. From this we can perform (and output) the computations of $L_{pq}$ and $N_{pq}$, by also correcting for the possible systematic error sources described above.

To illustrate how these corrections affect both the mean string separation $\xi_{pq}$ and the mean segment length $l_{pq}$ we can take a small resolution lattice ($512^3$, $\Delta x=0.5$), evolved in the matter era and apply corrections (minimum $l_{pq}=3.5$, off-by-one, assuming always one-cell overlap) in order to assess their impact. The comparison can be found in figure \ref{figure01} and we can note two important details. One is the fact that the mean string separation $\xi_{pq}$ shows rather similar evolution in all cases: indeed, the successive corrections mostly shift the mean string separation to larger values. The second aspect to notice is that, qualitatively, we can see the larger impact of the threshold correction on the mean segment length $l_{pq}$.

We further remark that adjusting the size of the overlap region is crucial for a better bound state identification, and that this is immediately obvious from the comparison of the top (interaction potential) and middle (exact cell overlap) and bottom panels (one-cell over of threshold) of figure \ref{figure02}. Although it can be argued that both identification mechanisms are reasonable, it is also clear that in some cases not allowing some margin for the overlap inevitably fails to identify all cells pierced by bound states. In passing we also note that, looking at the right-hand side panels of this figure, it is not at all clear that the set of $pq$-string segments can be reasonably assumed to form a Brownian network. This will be important for subsequent analytic modeling of the dynamics of these segments.

\section{\label{val}Code Validation and preliminary results}

We now proceed with a comparison of the present simulations with the two relevant references in the literature \cite{Urrestilla:2007yw,Lizarraga:2016hpd}. We will use the same lattice spacing as \cite{Lizarraga:2016hpd}, $\Delta x = 0.5$, which is double the lattice spacing of \cite{Urrestilla:2007yw}. The modified cooling and damping as applied here imply that we do not require a large lattice for this validation, and therefore the box size is kept at $1024^3$. This also has the advantage of providing the same dynamic range presented in the literature. In the following section we will present our $4096^3$ production runs.

As explained previously, the simulation parameters lead to equal tensions for the constituent strings and criticality for each sector: $\lambda_{p0} = \lambda_{q0} = 2$, $e_{p0} = e_{q0} = 1$ and $\lambda_{p0} = \lambda_{q0} = 2e_{p0}^2 = 2e_{q0}^2$. In the validation we will also force all constants to scale such that the comoving width of strings is kept constant throughout (implying $\beta=0$, on in other words, the PRS algorithm). Moreover, both symmetry breaking scales are equal and set to unity, $\sigma_p = \sigma_q = 1$. These parameter choices mean that the basic strings have tensions $\mu_p = \mu_q = 2 \pi \sigma_p = 2\pi \sigma_q$, unlike the cosmic superstring case where $\mu_p = \mu_F$ and $\mu_q = \mu_F /g_s$, with $\mu_F$ being the fundamental string tension. The coupling constant $\kappa_0$ will be kept at $0.9$; according to \cite{Lizarraga:2016hpd}, this leads to no significant change in the amount of $pq$-string abundances with respect to $\kappa=0.95$. In the validation we considered all the introduced estimators for the mean string separation and the mean velocity squared. Before we do so, however, there are two simple but necessary checks to be performed: firstly the verification of Gauss' law for each $U(1)_L$ sector, and secondly a visual confirmation of the existence of more than two string networks. 

\begin{figure}
  \centering
\includegraphics[width=1.0\columnwidth]{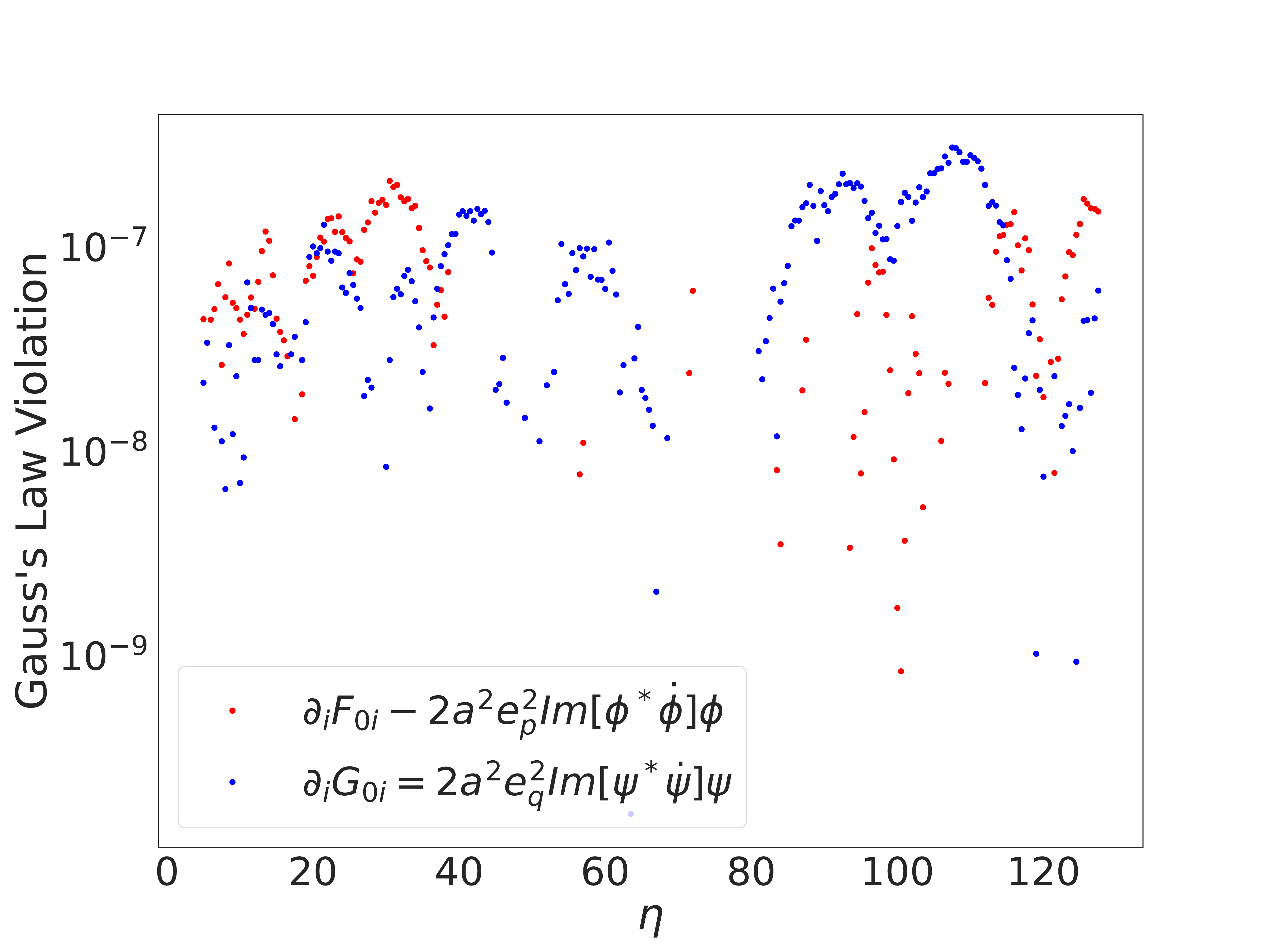}
\caption{Numerical measurement of the violation of Gauss' law, throughout the evolution of a $512^3$, $\Delta x=0.5$ matter epoch simulation. Red points refer to the $p$-string sector, and blue points to the $q$-string sector.\label{figure03}}
\end{figure}

The verification of Gauss's law for each sector can be seen in figure \ref{figure03} for a matter era run with $\beta=0$, starting at conformal time $\eta=1.0$, throughout the damping phase and subsequent cosmological evolution. As can be observed both versions of Gauss's law are preserved, at worst to $10^{-7}$. Note that this is true both in the damping and the cosmological phase: there is no a priori reason for this not to be true for the damping phase.

\begin{figure*}
\includegraphics[width=1.0\columnwidth]{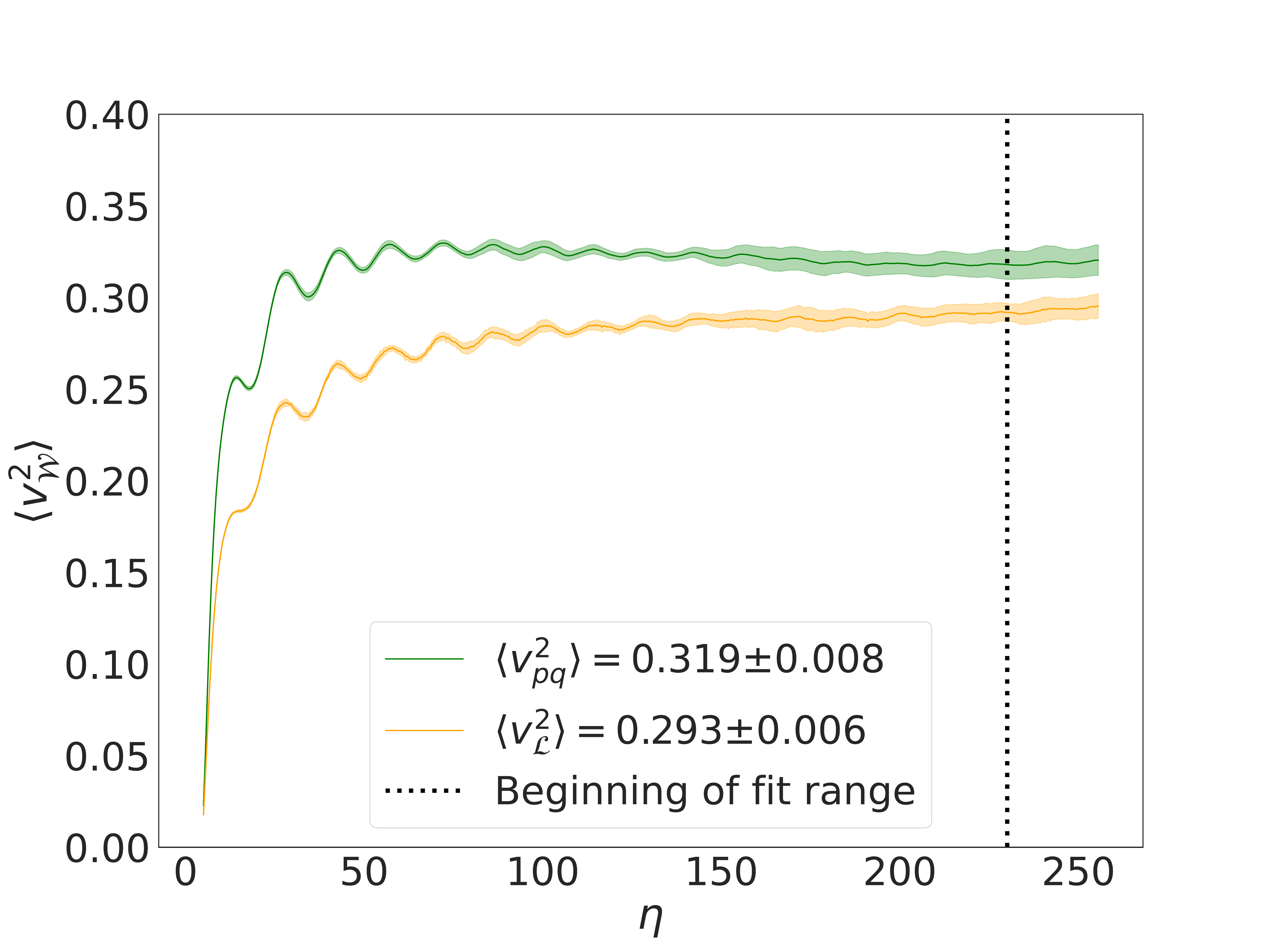}
\includegraphics[width=1.0\columnwidth]{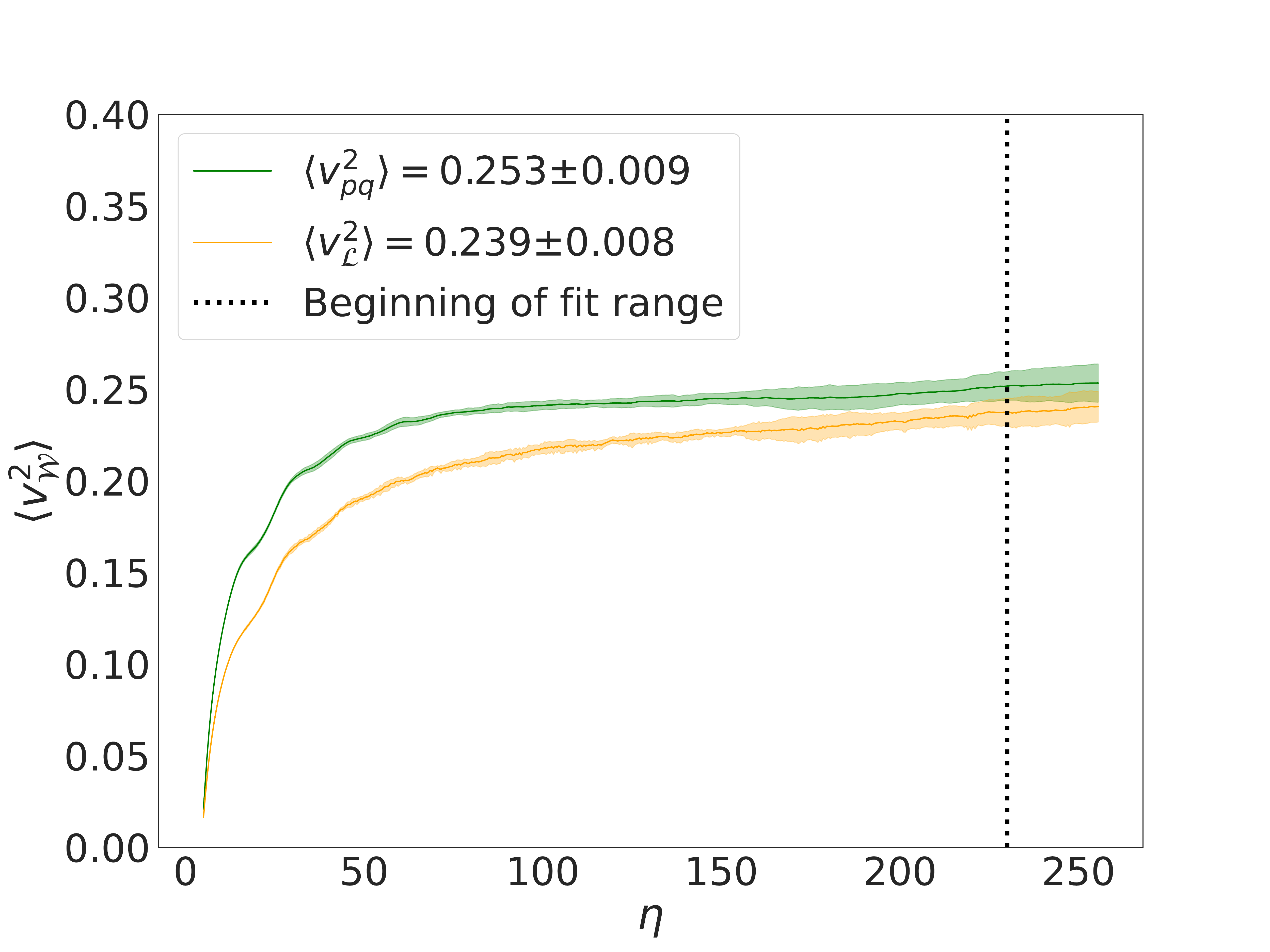}
\caption{The evolution of the mean squared velocity $\langle v^2_{\mathcal{W}} \rangle$ for either the full network (by specifying the full Lagrangian as a weight function, in orange) or the pq-segments (by using the interaction potential as the weight, in green). Left and right panels correspond to the radiation and matter epochs, respectively. \label{figure04} }
\end{figure*}

\begin{figure*}
\includegraphics[width=1.0\columnwidth]{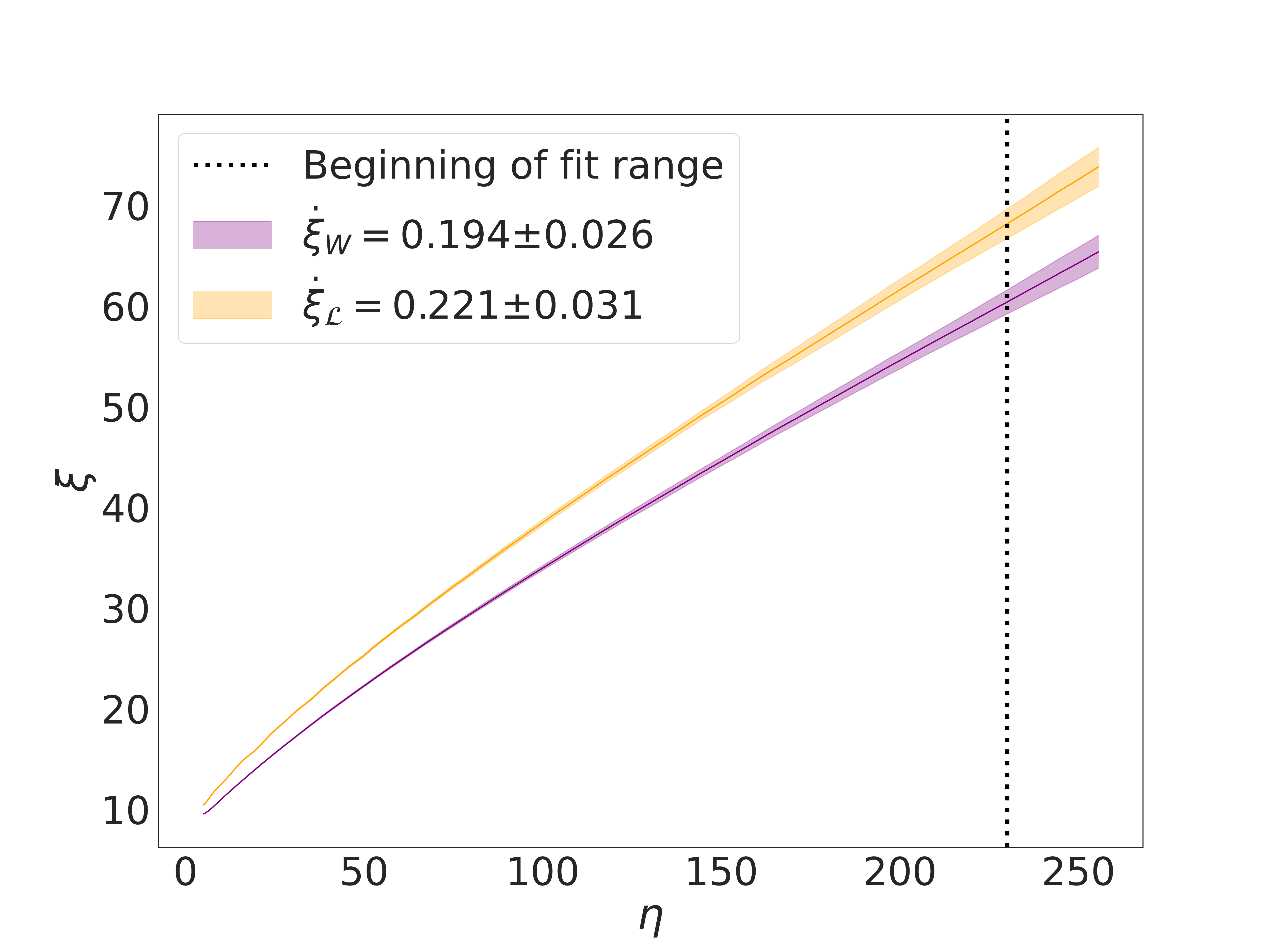}
\includegraphics[width=1.0\columnwidth]{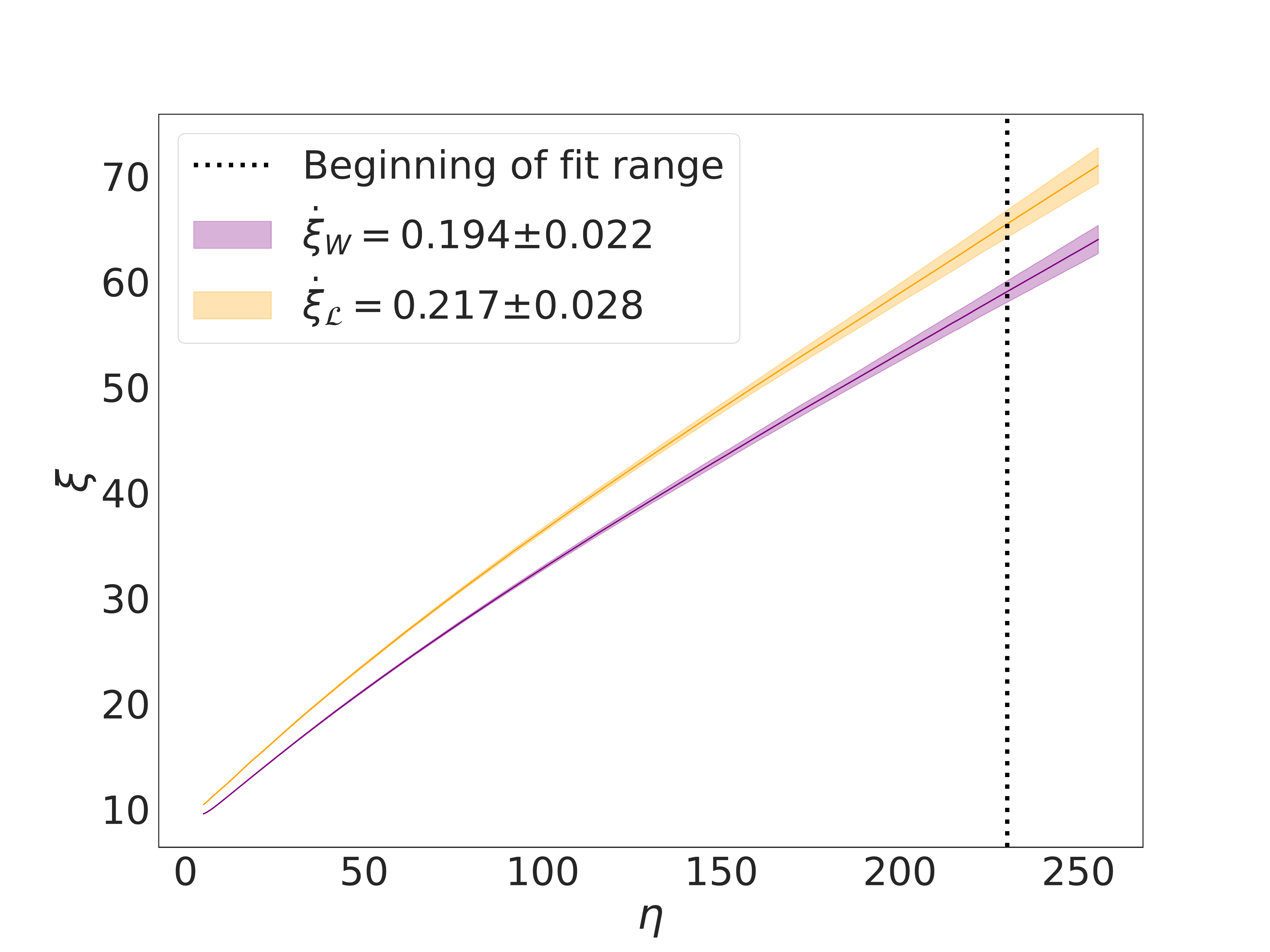}
\caption{The the evolution of the mean string separation for the full network, either using the Lagrangian length estimator (in orange) or the Winding length estimator (in purple) for both the radiation and matter epochs (left and right-hand-side panels, respectively). In the case of the winding estimator, the length of $pq$-segments is computed using the fast method. \label{figure05}}
\end{figure*}

The evolution of both velocity estimators is illustrated in figure \ref{figure04}, and that of the mean string separations in figure \ref{figure05}, with the corresponding asymptotic values computed for the conformal time range $\eta \in [230.0-256.0]$. Equivalently, the same information of these asymptotic values can be found in table \ref{table1} and table \ref{table2}, respectively, along with the asymptotic values of our earlier simulations of the simpler Abelian-Higgs networks (only for the full network, with the same velocity estimator) and for the directly comparable figures of \cite{Lizarraga:2016hpd}. We note that \cite{Urrestilla:2007yw} does not report any velocity estimates. 

Starting with $\langle v^2_{\mathcal{L}} \rangle$, we note that the velocities in our simulations are indeed compatible, within their $1-\sigma$ uncertainties, although lower than, the ones presented in \cite{Lizarraga:2016hpd}. For $\langle v^2_{pq} \rangle$, there is an important difference to note: in \cite{Lizarraga:2016hpd} the velocities are given in a range, since in the conformal time period where the network reaches scaling (based on full network estimators), the velocity of $pq$-strings decreases. Here we observe a different behavior: the velocities slightly increase throughout this range. Possibly, this can be attributed to the different preparation of initial conditions, specifically due to the different damping applied in our case and theirs. If we compute the values from the last $20.0$ conformal time units, however, we can compare to the lower bound of the ranges provided in \cite{Lizarraga:2016hpd}. Assuming uncertainties comparable to ours for this lower range, we can see that velocities are compatible in radiation epoch while underestimated in matter era, by about $10\%$. This can again be attributed to the different damping periods, given that it is not entirely clear from the figure whether or not the velocity has completely stabilized and thus requires more dynamic range to reach its asymptotic value.

\begin{table*}
  \begin{center}
    \begin{tabular}{|c|c|c|c|c|c|c|}
      \hline
Size, $\Delta x$ & m & $\langle v^2 \rangle_{pq}$ & $\langle v^2 \rangle_\mathcal{L}$ & Reference\\
        \hline
  $1024^3,\Delta x = 0.5$ & 1/2 & -               & $0.272\pm0.002$ & Abelian-Higgs \cite{Correia:2020yqg} \\
  $1024^3,\Delta x = 0.5$ & 1/2 & $0.319\pm0.008$ & $0.293\pm0.006$ & This section\\
  $1024^3,\Delta x = 0.5$ & 1/2 & $\sim 0.33$     & $0.306\pm0.004$ & \cite{Lizarraga:2016hpd}\\
  $512^3, \Delta x = 1.0$ & 1/2 & -             & -               & \cite{Urrestilla:2007yw}\\
      \hline
  $1024^3,\Delta x = 0.5$ & 2/3 & - &               $0.228\pm0.004$ & Abelian-Higgs \cite{Correia:2020yqg}\\
  $1024^3,\Delta x = 0.5$ & 2/3 & $0.247\pm0.006$ & $0.253\pm0.009$ & This section\\
  $1024^3,\Delta x = 0.5$ & 2/3 & $\sim 0.27$     & $0.264\pm0.006$ & \cite{Lizarraga:2016hpd}\\
  $512^3, \Delta x = 1.0$ & 2/3 & -               & -               & \cite{Urrestilla:2007yw}\\
      \hline
    \end{tabular}
\caption{The asymptotic values of the mean velocity squared $\langle v^2_\mathcal{W} \rangle$ for both the full network (weighted by the Lagrangian) and $pq$-segments (weighted by the interaction potential) for the simulations from this section, our previous Abelian-Higgs simulations \cite{Correia:2020yqg}, and $pq$-strings simulations from \cite{Lizarraga:2016hpd}.}
    \label{table1}
    \end{center}
\end{table*}

\begin{table*}
  \begin{center}
    \begin{tabular}{|c|c|c|c|c|c|c|c|}
      \hline
Size, $\Delta x$ & m & $\dot{\xi}_W$ & $\dot{\xi}_p$ & $\dot{\xi}_{pq}$  & Reference\\
        \hline
  $1024^3, \Delta x = 0.5$ & 1/2 & $0.280\pm0.026$ & $=\dot{\xi}_W$ & -  & Abelian-Higgs \cite{Correia:2020yqg}\\
  $1024^3, \Delta x = 0.5$ & 1/2 & $0.194\pm0.026$ & $0.270\pm0.050$ & $2.488\pm0.612$  & This section, fast method\\
  $1024^3, \Delta x = 0.5$ & 1/2 & -               & -               & -                & \cite{Lizarraga:2016hpd}\\
  $512^3, \Delta x = 1.0$  & 1/2 & $0.15$          & $0.22$          & -                & \cite{Urrestilla:2007yw}\\
      \hline
  $1024^3, \Delta x = 0.5$ & 1/2 & $0.279\pm0.016$ & $=\dot{\xi}_W$ & -  & Abelian-Higgs \cite{Correia:2020yqg}\\
  $1024^3, \Delta x = 0.5$ & 2/3 & $0.194\pm0.022$ & $0.277\pm0.042$ & $1.634\pm0.721$  & This section, fast method\\
  $1024^3, \Delta x = 0.5$ & 2/3 & -               & -               & -                & \cite{Lizarraga:2016hpd}\\
  $512^3, \Delta x = 1.0$  & 2/3 & $0.15$          & $0.21$          & -                & \cite{Urrestilla:2007yw}\\
      \hline
    \end{tabular}
\caption{The asymptotic rate of change of the mean string separation $\xi$ for both the full network using windings $\xi_W$, the rate of change for $p$-strings only, and the rate of change of $\xi_{pq}$ for $pq$-strings. For comparison we provide the results obtained from our previous Abelian-Higgs simulations \cite{Correia:2020yqg} (where the full network estimator is equivalent to only having a single string type, say $p$-strings) and those of the directly comparable work in the literature \cite{Urrestilla:2007yw} (for which no uncertainties were reported).}
    \label{table2}
    \end{center}
\end{table*}

\begin{figure*}
\includegraphics[width=1.0\columnwidth]{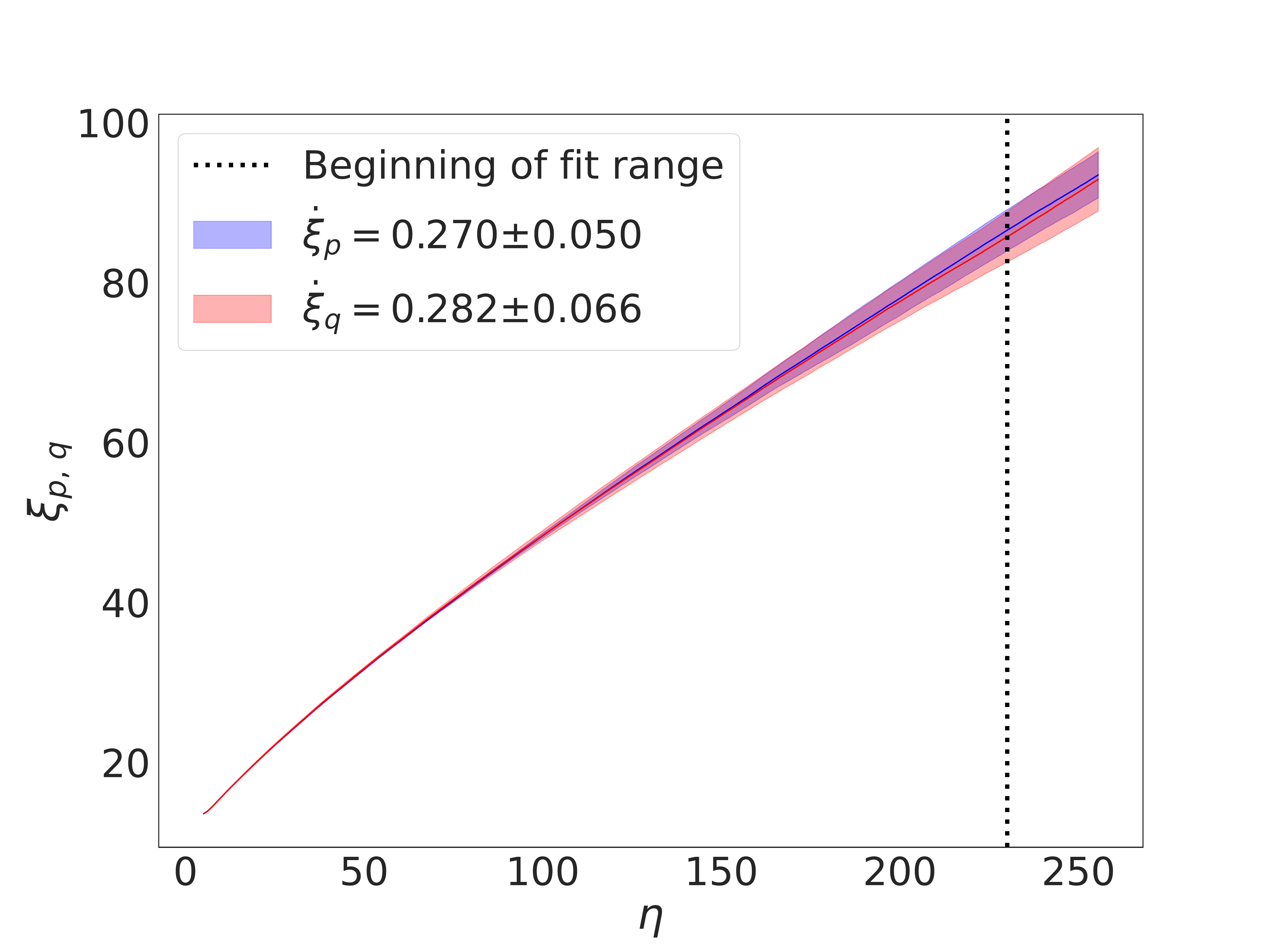}
\includegraphics[width=1.0\columnwidth]{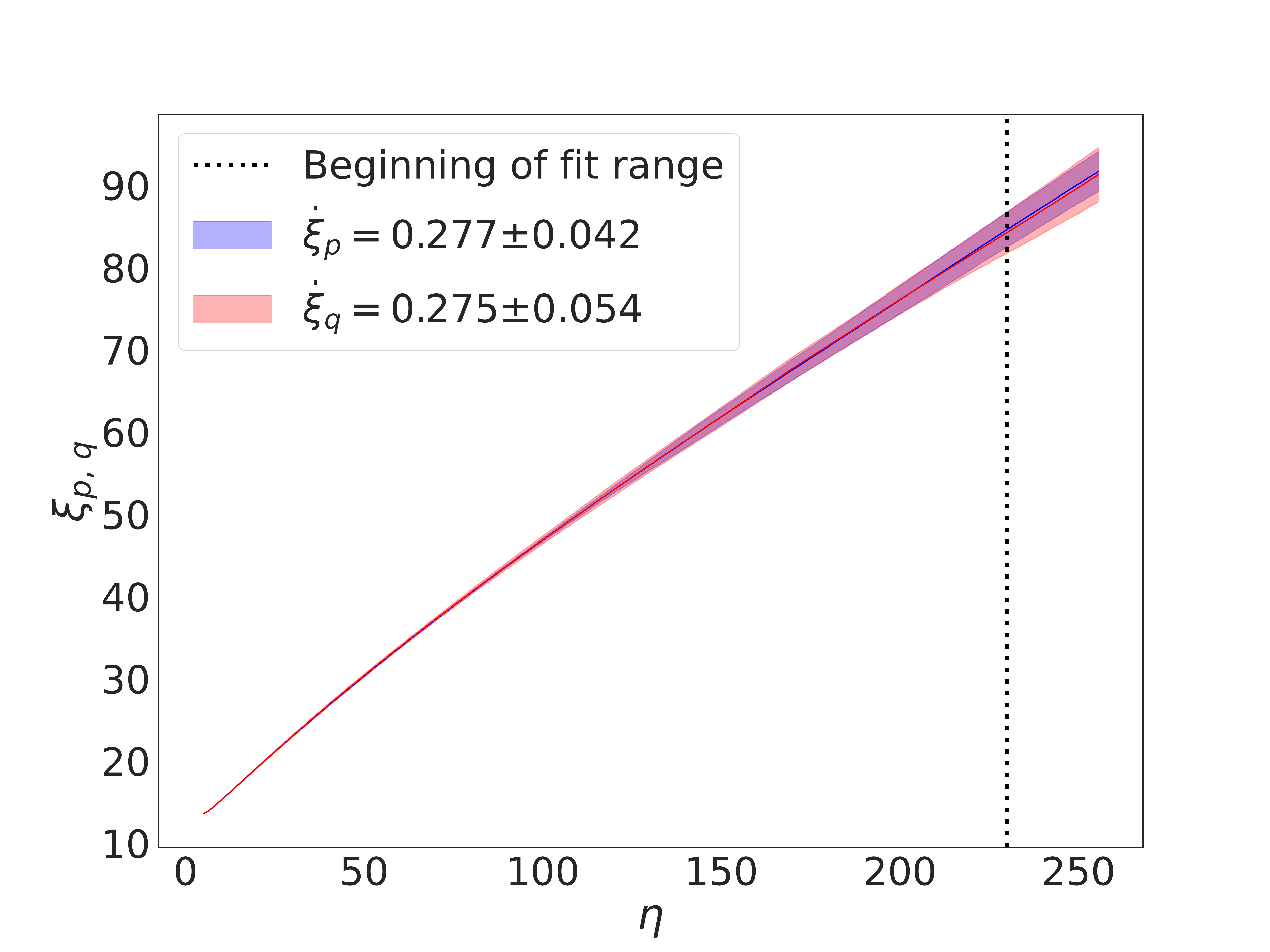}
\includegraphics[width=1.0\columnwidth]{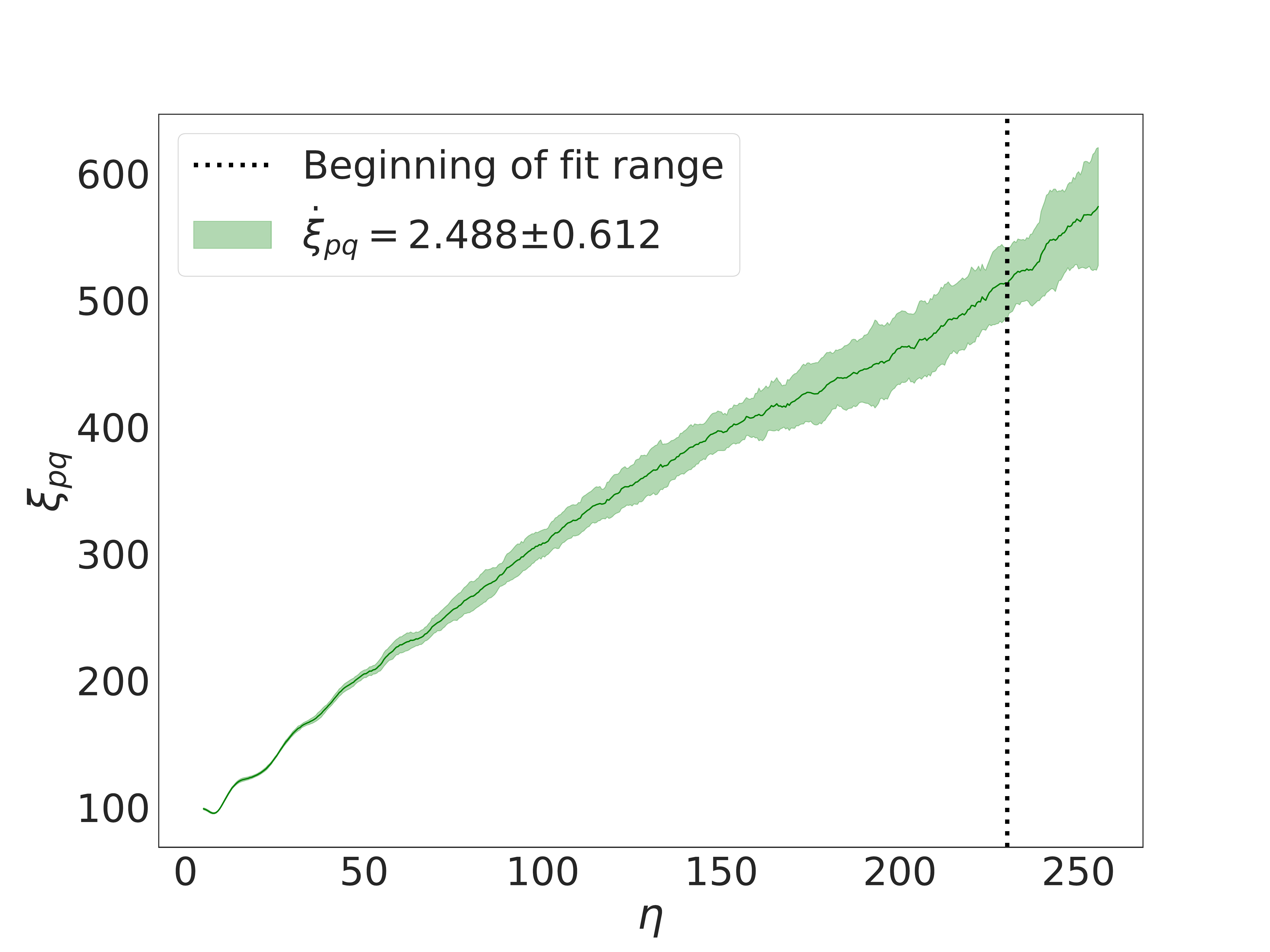}
\includegraphics[width=1.0\columnwidth]{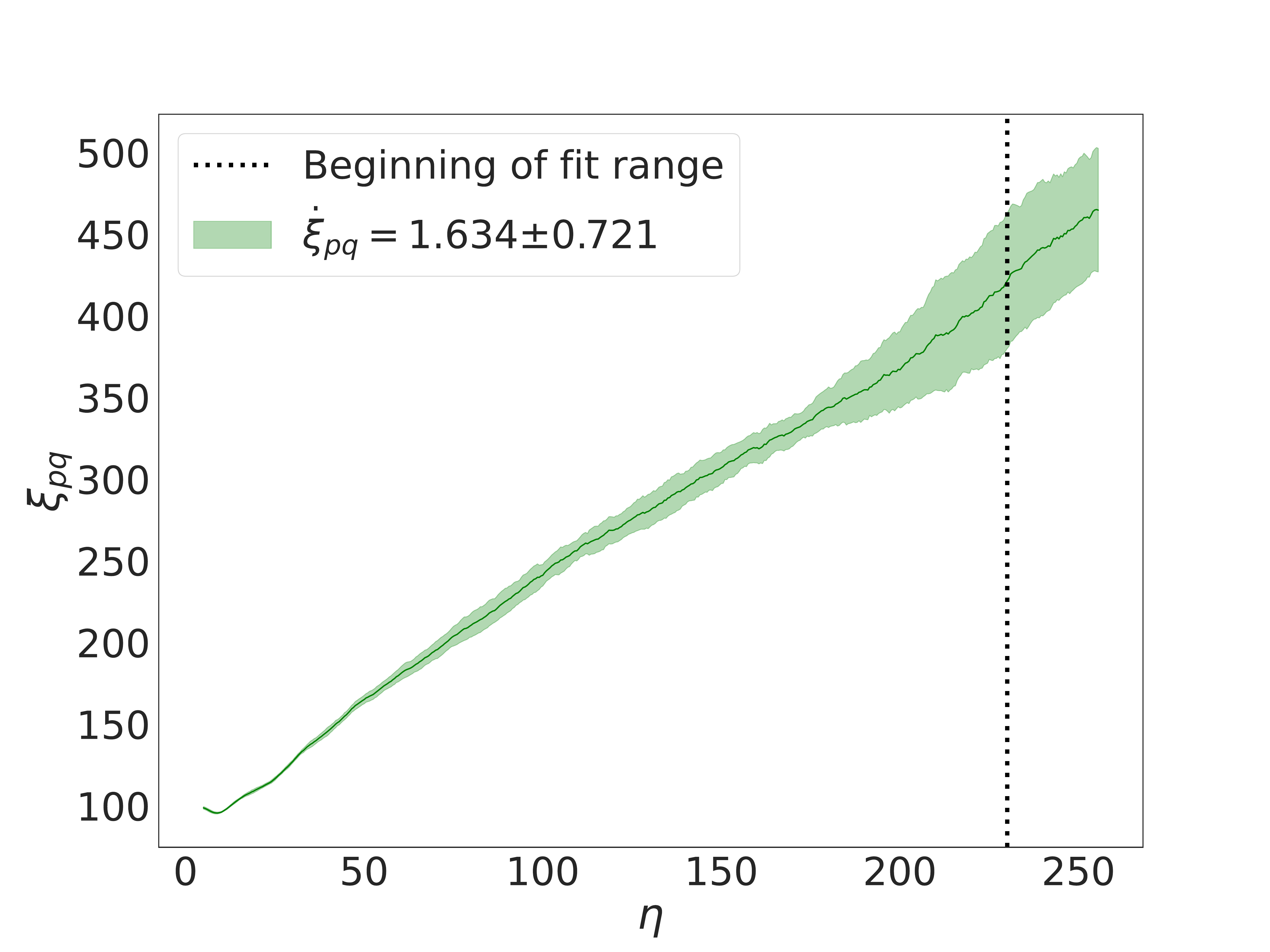}
\includegraphics[width=1.0\columnwidth]{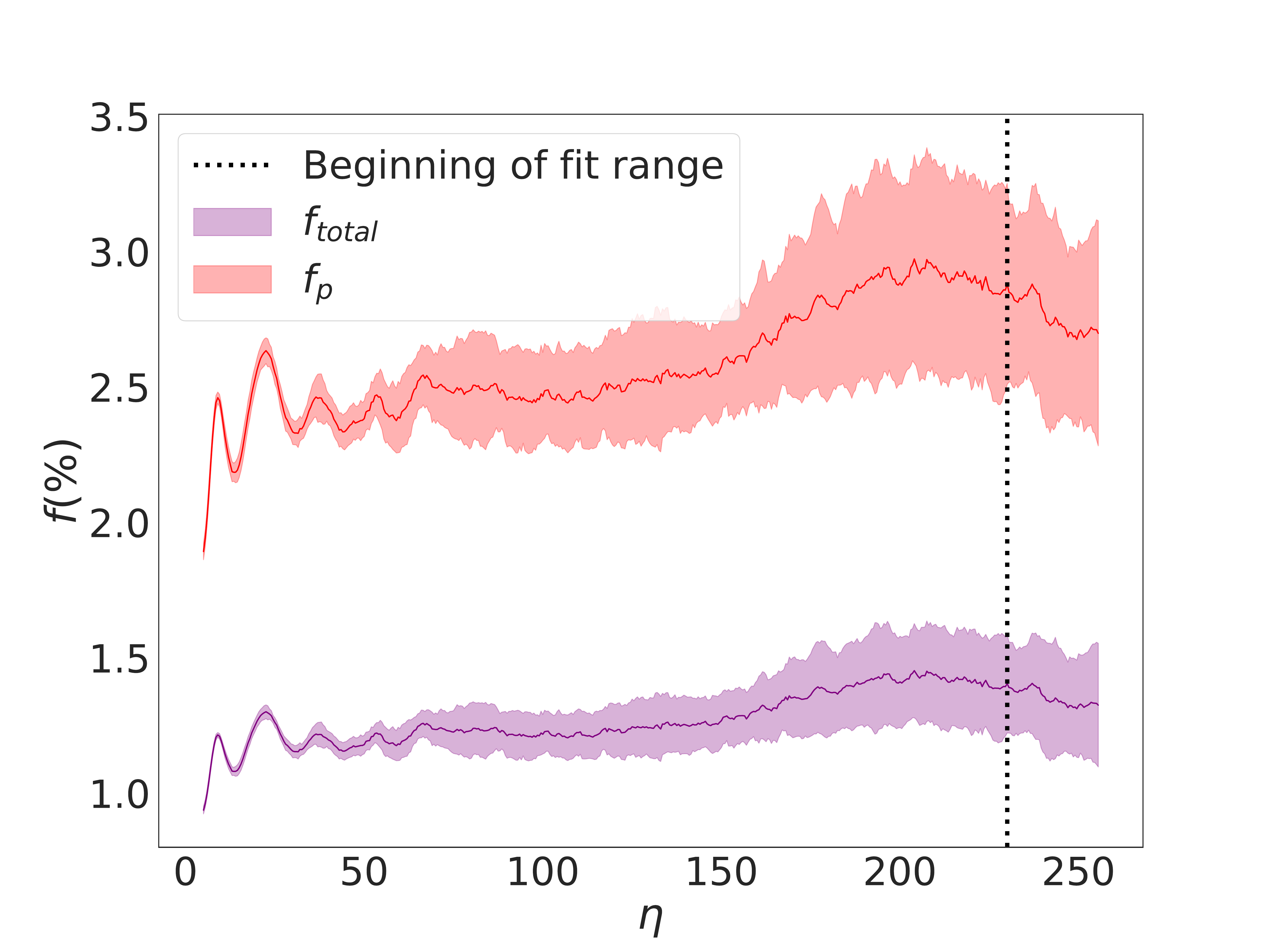}
\includegraphics[width=1.0\columnwidth]{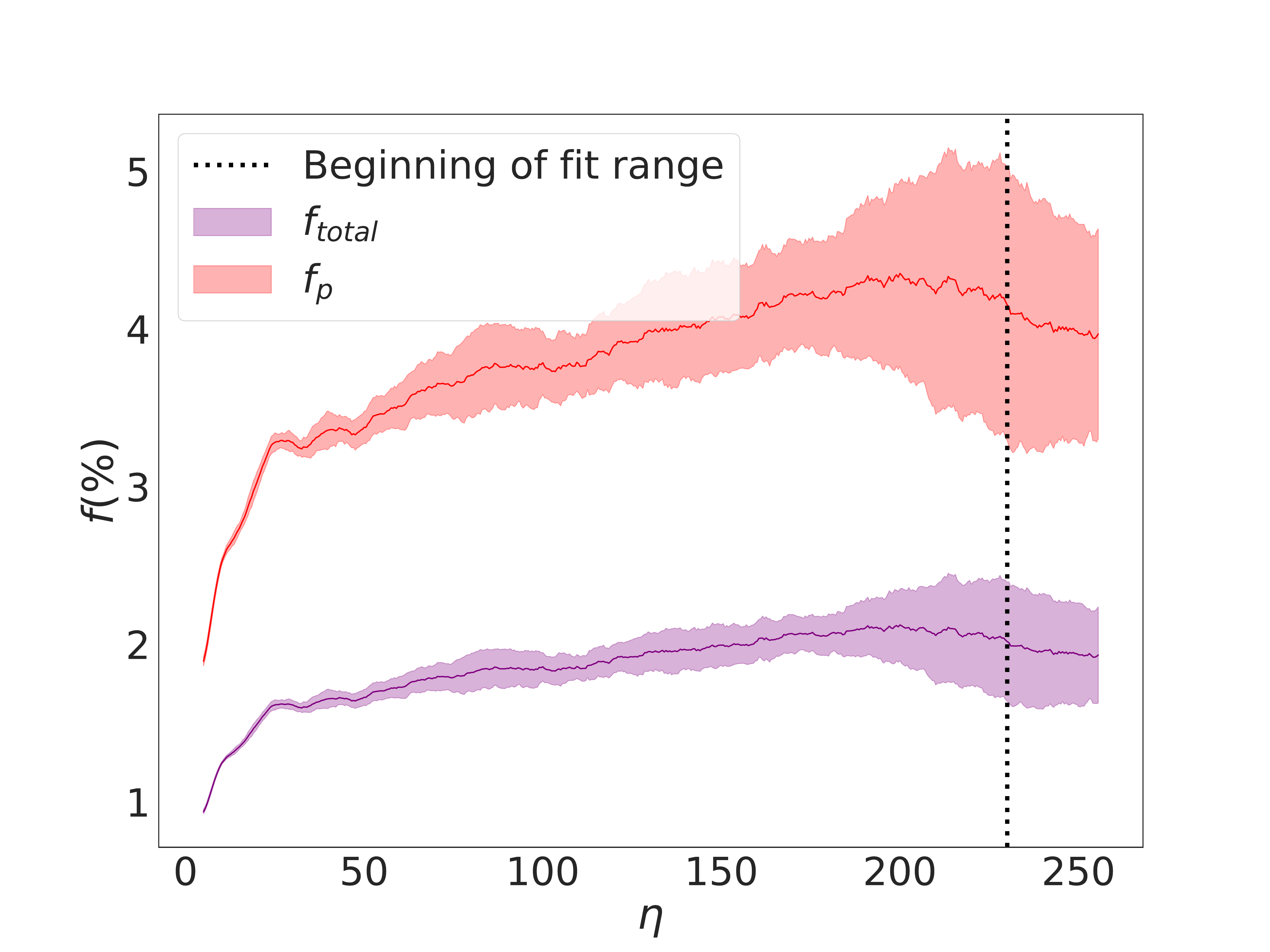}
\caption{The evolution of relevant quantities derived from the total length of pq-strings present in the box, with $L_{pq}$, derived from the fast computation method. The mean string separations $\xi_{p}$ and $\xi_{q}$ are shown in the two upper plots (in red and blue, respectively), $\xi_{pq}$ in the middle panels (in green), and the relative abundances of pq-strings in the lower panels $f_{total}$ and $f_p$ (in purple and red, respectively). As in previous figures, the left-hand-side corresponds to radiation epoch, the right-hand-side to matter epoch.\label{figure06}}
\end{figure*}

All mean string separation estimators (that is, $\xi_p$, $\xi_q$ and $\xi_{pq}$) and both fractions ($f_{total}$ and $f_{p}$) can be seen in figure \ref{figure06} in both the radiation and matter eras (left and right hand side panels, respectively). In all of these we have used the fast method to compute $L_{pq}$. All string separations achieve scaling, although the behavior of $\xi_{pq}$ is somewhat less smooth than that of $\xi_{p}$. Even though we could be tempted to attribute this effect to the two previously discussed systematic limitations, it might instead be related to the the low abundance of bound states. In fact, even in \cite{Lizarraga:2016hpd} the jagged behavior of $\xi_{pq}$ can be observed. We further note that the fraction $f_{total}$ is always within $1-2\%$, and it seems to be slightly larger in matter era (cf. the bottom panels of figure \ref{figure06}), although the difference might not be statistically significant. The fraction of bound states relative to the fraction of $p$-strings is also in agreement with the literature values, as in matter era it seems to be of about $4\%$. 

We can now discuss the asymptotic values of the various measures of the mean string separation. Note that this still assumes the fast computation method, and we will need to revisit this comparison with the robust method in the next section. Table \ref{table2} summarizes the comparison. For completeness we have added the values obtained in our earlier Abelian-Higgs simulations with the same box size, where the full network estimator and the $p$-string values are the same. For the full network and $p$-strings we compare these results to \cite{Urrestilla:2007yw}. Note that we do not have any literature values to compare $\dot{\xi}_{pq}$ with; in the aforementioned reference the typical distance between Y-junctions is computed in a very different way, using the number of $pq$-segments. The values of $\dot{\xi}_W$ and $\dot{\xi}_p$ are both somewhat larger than those presented by \cite{Urrestilla:2007yw}, but given the likely uncertainties (which \cite{Urrestilla:2007yw} does not explicitly report) the values are compatible. Also note that \cite{Urrestilla:2007yw} used a box size of $512^3$ with a lattice spacing $\Delta x = 1.0$, which could affect the reliability of their mean string separation estimates and of the evolution of the fields themselves. We do not think that the difference in damping and the difference in bound state identification could explain this tension as our fractions $f_{total}$ and $f_p$ are in agreement with \cite{Urrestilla:2007yw}. Moreover, given the small values of $f_{total}$ and $f_p$, the identification of bound states cannot impact $\xi_p$, $\xi_q$ and $\xi_W$ significantly (but note that this might not be the case with $\xi_{pq}$).

\section{\label{Results}Scaling properties and bound states}

We will now make use of the full capabilities of our GPU-accelerated code to explore the scaling properties of these defect networks, and in particular to assess the impact of various assumptions about the comoving string width on the network properties, including the abundance of bound states. Our analysis is based on simulations with lattice size of up to $4096^3$ keeping the same lattice spacing $\Delta x =0.5$ and conformal timestep size $\Delta \eta = 0.1$. These are by far the largest simulations of this model performed to date. A short animation of one of these simulations can be seen in \cite{Movie}.

Despite this enhanced dynamic range, the true comoving string width, $r_s$ would vary too quickly both in the radiation and the matter epochs (since $r_s \propto a^{-2\beta}$ and $\beta=1$) for us to be able to resolve the string network either at late or early times. Therefore, we will use the core growth prescription of \cite{Bevis:2006mj} to allow the string width to grow in the initial stages of the simulation by setting $\beta<0$. The way we set these values is to choose a maximum string width, normalize the scale factor such that the radius is unity at the end, and from these infer the maximum radius at some transition time (we can choose to fix either one or the other). The transition time is the timestep when the value of $\beta$ is switched from core growth to physical evolution. The value of $\beta$ for the growth phase is chosen such that the simulation begins with unit radius and reaches maximum radius at the transition time.

For the same choice of maximum radius, and again the same normalization at the end of the simulation, the radius would change more quickly in the matter era than in the radiation era for physical evolution, and as such the transition time would tend to be later. However, we would like to have the same dynamic range in both core growth and physical evolution, in both the matter and radiation eras, in order to ensure a comparison between the various settings that is as physically fair as is numerically possible. As such we fix the transition time, and instead vary the maximum radius. Additionally, before the growth phase, there is a period of diffusive and damped evolution. This is similar to what was applied in the validation section, with one small difference, which stems from the over-damping noted therein in the matter epoch. This difference consists in using a lower damping expansion rate, of $m_{\rm damp}=0.75$ for the damping period previous to the matter era evolution, while we use $m_{\rm damp}=0.95$ for the radiation era. During the damping period, and for simplicity, we make the choice of having constant comoving width ($r_s=1$). A summary of all parameter choices can be seen in table \ref{table3}. 

\begin{table}
  \begin{center}
    \begin{tabular}{|c|c|c|c|c|c|c|c|}
      \hline
m & $m_{\rm damp}$ & $\beta_{\rm growth}$ & $\beta$ & $\eta_{\rm tr}$ & $r_{s,max}$\\
        \hline
   1/2 & $0.95$ & $-0.25984202$   & $1$ & $343.0$  & $3.0$ \\
   1/2 & $0.95$ & $0.0$           & $0$ & $-$      & $1.0$ \\
   \hline
   2/3 & $0.75$ & $-0.25984981$   & $1$ & $343.0$  & $9.0$ \\
   2/3 & $0.75$ & $0.0$           & $0$ & $-$      & $1.0$ \\
      \hline
    \end{tabular}
\caption{A summary of the choices made in the early evolution of our $4096^3$ radiation and matter era simulations ($m=1/2$ and $m=2/3$, respectively): the power law exponent for the evolution of the scale factor during the initial damping phase ($m_{\rm damp}$), the values of $\beta$ in the core growth phase and the subsequent evolution, the conformal time at which one changes to physical evolution, and the maximum string radius for physical simulations ($\beta=1$) and constant comoving width ones ($\beta=0$).}
    \label{table3}
    \end{center}
\end{table}

For comparison purposes, we also evolve constant comoving width (PRS) simulations with the same periods of diffusive and damped evolution (that is, the same damping expansion rate, $m_{\rm damp}$). Moreover, the same set of 10 random seeds for the initial conditions is used for all the sets of simulations (matter and radiation, physical and constant width simulations); the statistical uncertainties which we report come from averaging these sets of 10 simulations.  As such we will have three different cases for us to compare $\beta=0$, $\beta=1$ and $\beta<0$. As lowering $\beta$ from $1$ breaks the time-invariance of the discretized action, one can also think of it as parameter that controls violation of energy conservation. Given the role of kinematics in junction dynamics \cite{PhysRevD.99.063516}, one could expect that changing $\beta$ impacts the formation/destruction of $pq$-segments. In this sense, the core growth period, which one may see as a necessary evil for large field theory simulations, ends up giving us an additional source of comparison.

\subsection{\label{fastRes} Computation of $L_{pq}$ with the fast method}

\begin{figure*}
\includegraphics[width=1.0\columnwidth]{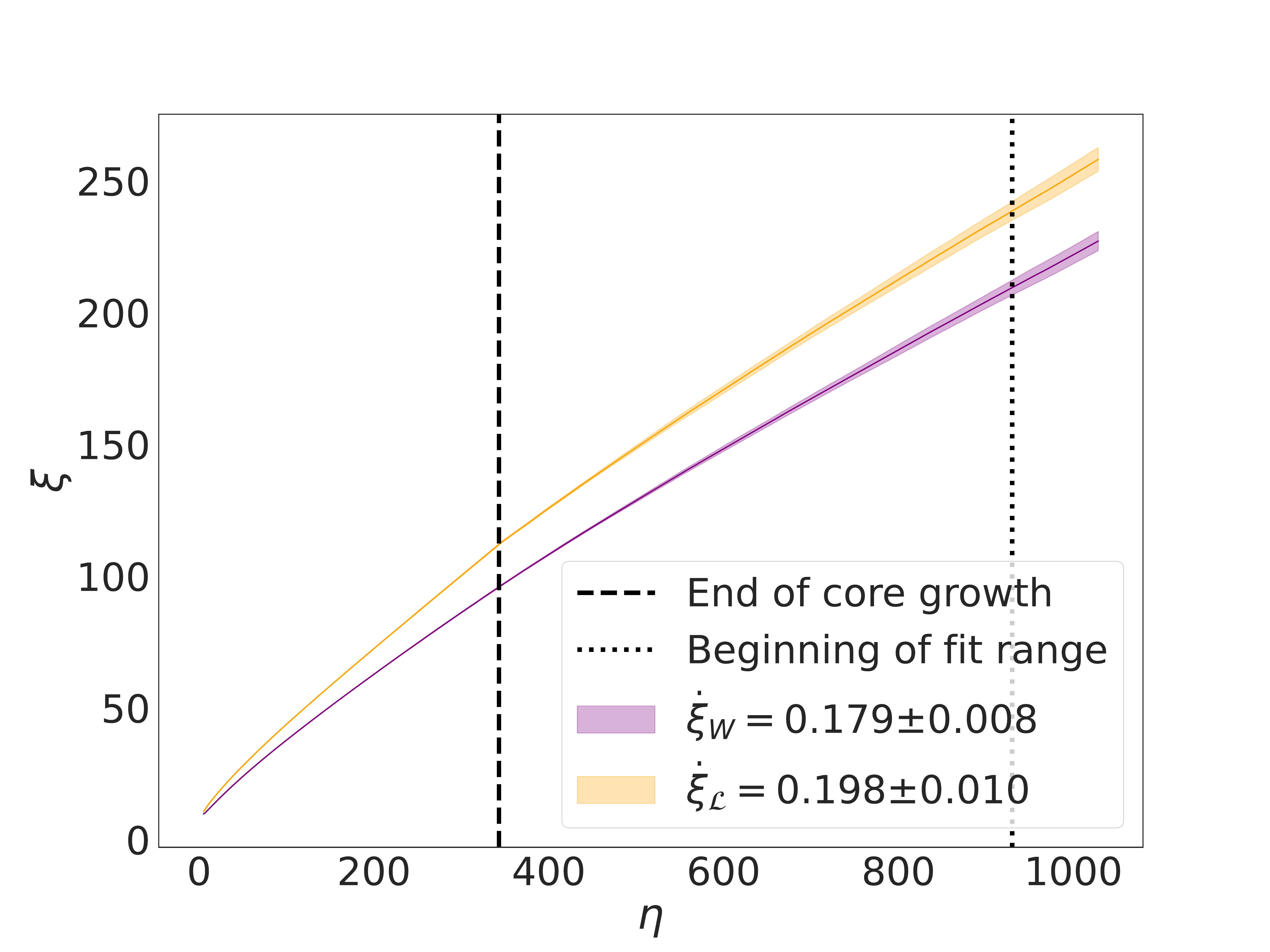}
\includegraphics[width=1.0\columnwidth]{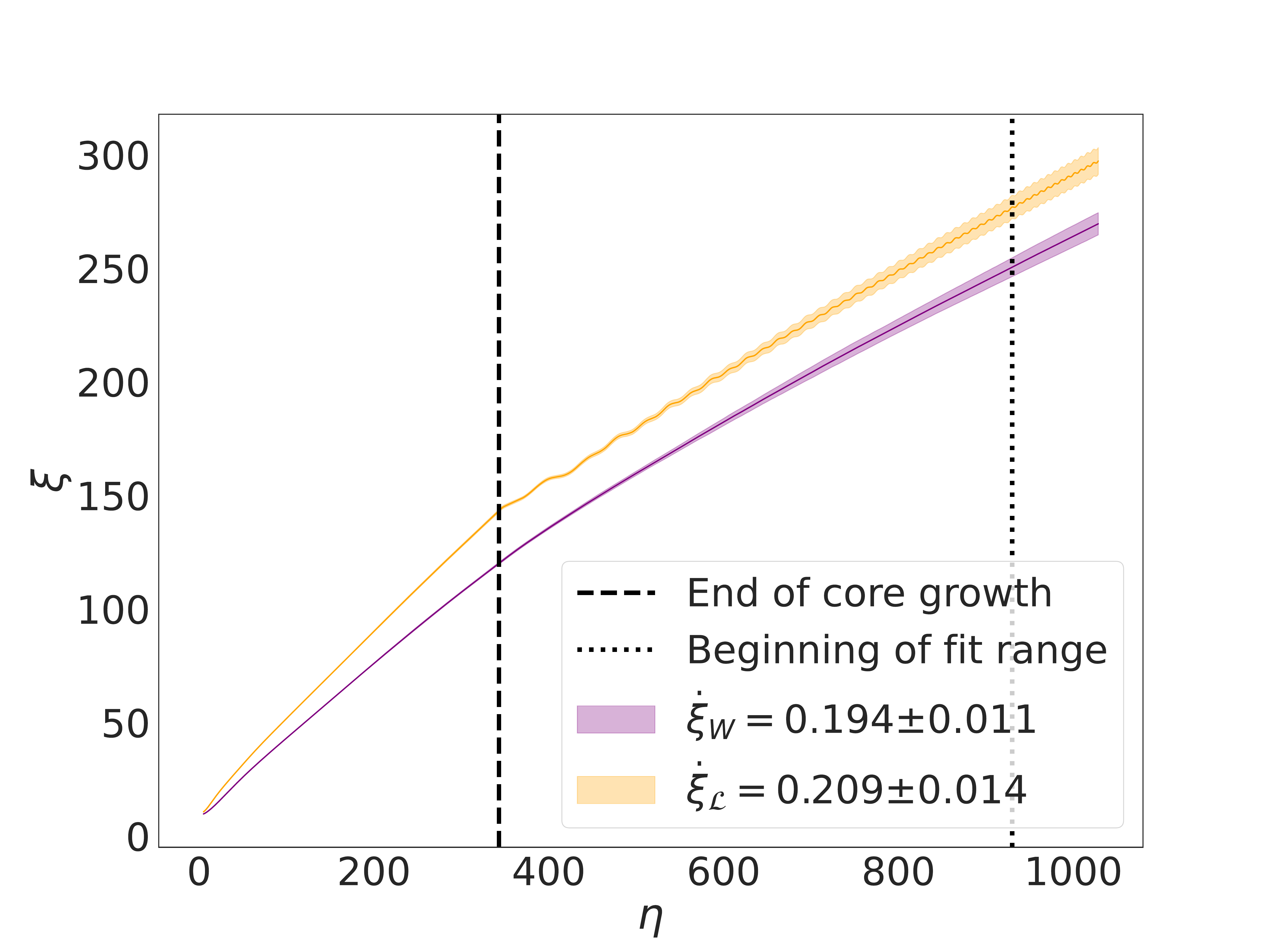}
\includegraphics[width=1.0\columnwidth]{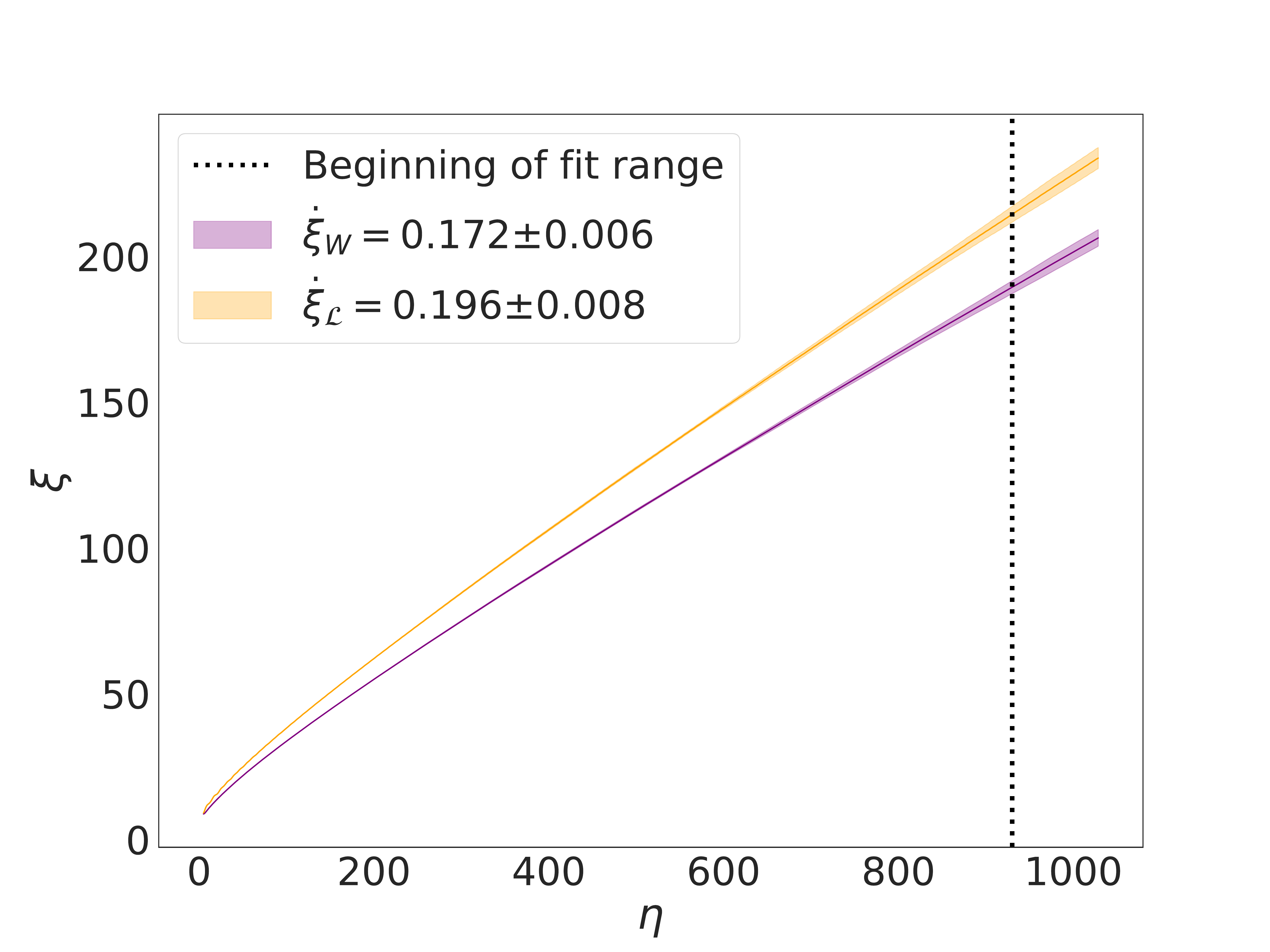}
\includegraphics[width=1.0\columnwidth]{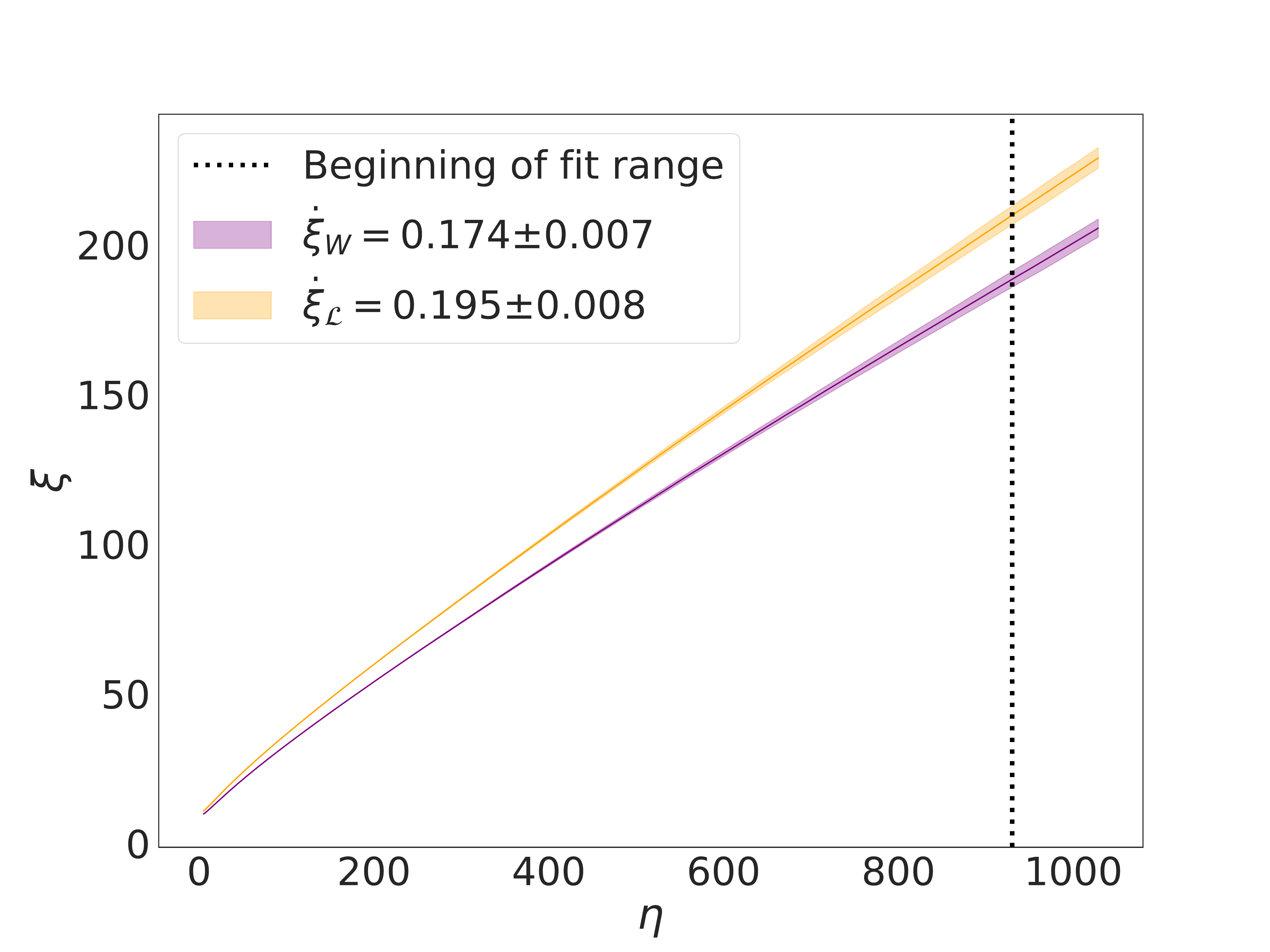}
\includegraphics[width=1.0\columnwidth]{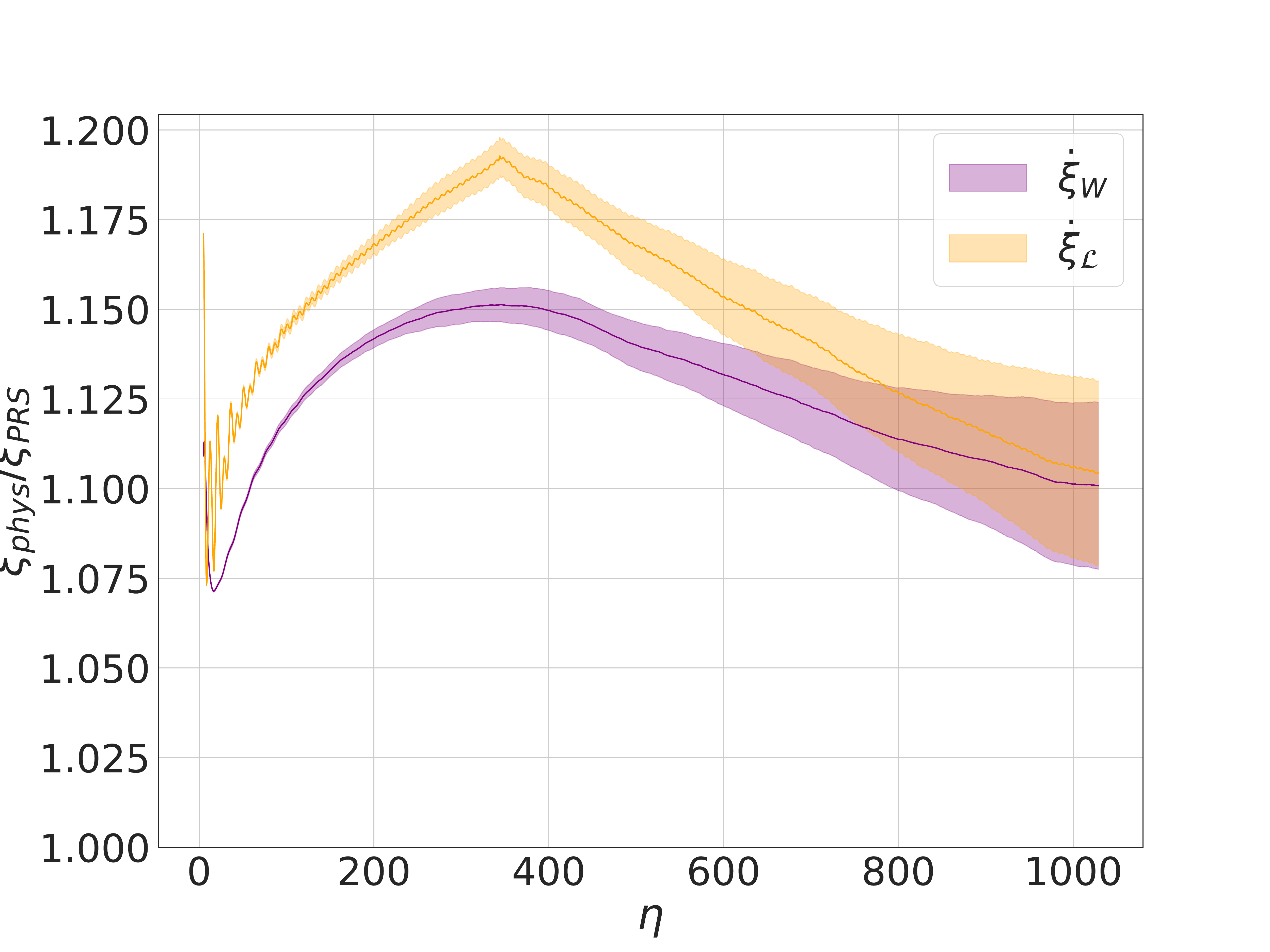}
\includegraphics[width=1.0\columnwidth]{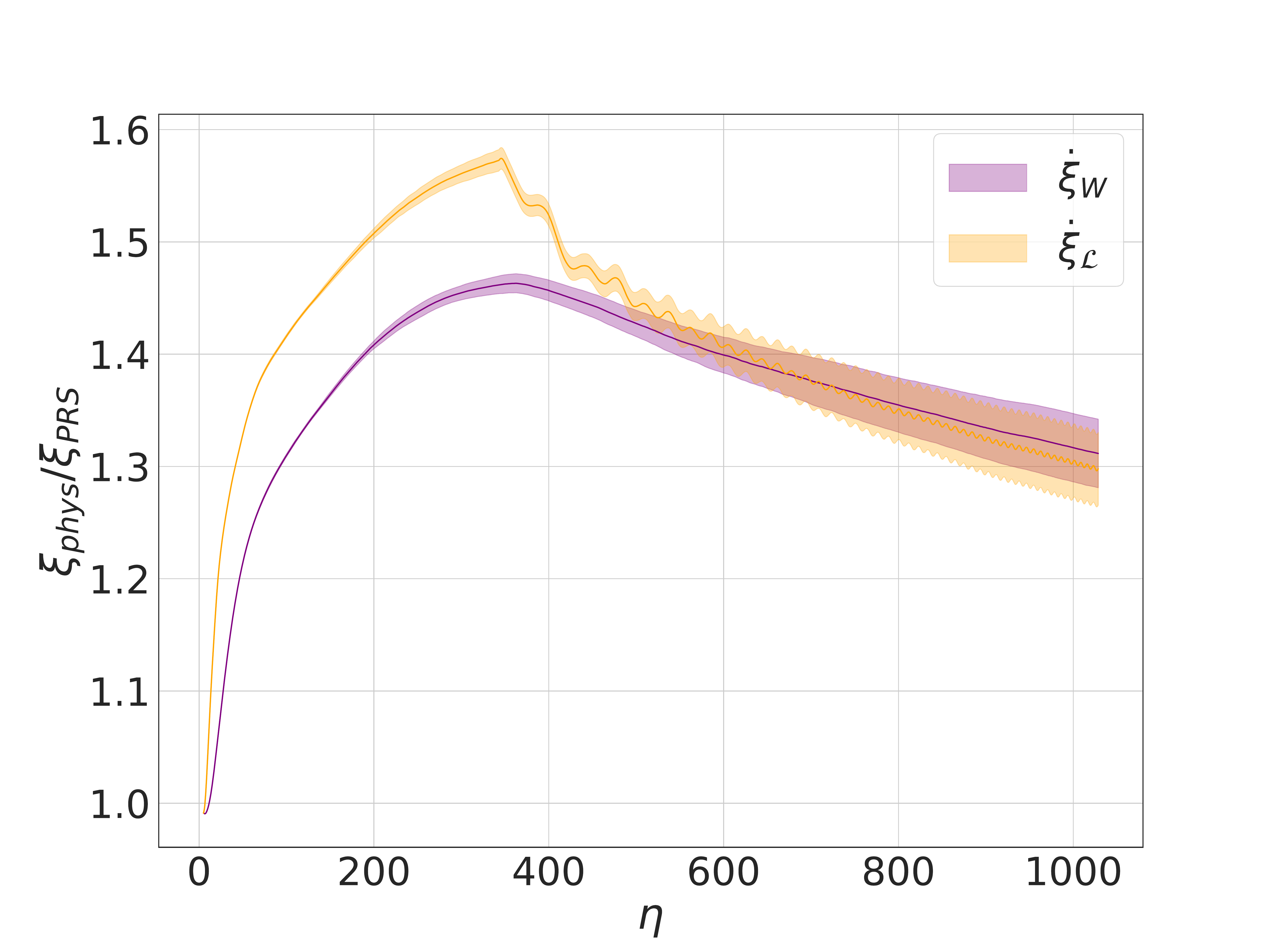}
\caption{The evolution of the mean string separation for the entire network (all string species included), using either the Lagrangian length estimator (in orange) or the winding length estimator (in purple). Left panels correspond to the radiation epoch, and right panels to the matter epoch. Top panels use core growth and subsequent physical evolution, while middle panels correspond to constant comoving width simulations. The bottom panels show the ratio of the values obtained in the physical and constant comoving width values. $L_{pq}$ is computed via the fast method throughout.\label{figure07}}
\end{figure*}

We begin with the full network mean string separation, shown in the panels of figure \ref{figure07} for both physical (including the early core growth phase) and constant comoving width (PRS) simulations, shown in the top and middle panels respectively, for both radiation and matter epoch. Overall there is reasonable agreement between physical and comoving width simulations in the asymptotic values of $\dot{\xi}$, which in this section is always computed for the conformal time range $\eta \in [950,1024]$. It is worthy of note that this agreement is best at the end of the simulations; in other words, although the two sets of simulations are converging towards approximately the same scaling solution (within statistical uncertainties), the way in which they approach it is different--- and more so in the radiation era than in the matter era. This is an important point to bear in mind when using simulations for calibrating analytic models of string evolution. We will return to this point in the conclusions.

Table \ref{table4} displays a comparison of this and other diagnostics. Again we see a $\sim 10\%$ difference between the two length estimators $\xi_{\mathcal{L}}$ and $\xi_W$. This agreement is not entirely unexpected, and neither is the existence of a different slope for the evolution in the core growth regime. This is particularly obvious if we compute the ratio (plotted in the bottom panels) between the quantities in the upper and middle panels: in the core growth phase the difference between it and the constant comoving width (PRS) case keeps increasing, while this tendency is inverted as soon as we shift to physical evolution.

\begin{table*}
  \small
  \begin{center}
    \begin{tabular}{|c|c|c|c|c|c|c|c|c|}
      \hline
$\beta$ & m & Size, $\Delta x$ & $\dot{\xi}_W$ & $\dot{\xi}_p$ & $\dot{\xi}_{pq}$  & Reference\\
        \hline
0 & 1/2 & $4096^3, \Delta x = 0.5$  & $0.172\pm0.006$ & $0.242\pm0.010$ & $1.501\pm0.375$  & This section, fast method\\
0 & 1/2 & $1024^3, \Delta x = 0.5$  & $0.194\pm0.026$ & $0.270\pm0.050$ & $2.488\pm0.612$  & Previous section, fast method\\
0 & 1/2 & $512^3, \Delta x = 1.0$   & $0.15$          & $0.22$          & -                & \cite{Urrestilla:2007yw}\\
1 & 1/2 & $4096^3, \Delta x = 0.5$  & $0.179\pm0.008$ & $0.267\pm0.015$ & $2.012\pm0.488$  & This section, fast method\\
    \hline
0 & 2/3 & $4096^3, \Delta x = 0.5$  & $0.174\pm0.007$ & $0.248\pm0.007$ & $1.460\pm0.240$  & This section, fast method\\
0 & 2/3 & $1024^3, \Delta x = 0.5$  & $0.194\pm0.022$ & $0.277\pm0.042$ & $1.634\pm0.721$  & Previous section, fast method\\
0 & 2/3 & $512^3, \Delta x = 1.0$   & $0.15$          & $0.21$          & -                & \cite{Urrestilla:2007yw}\\
1 & 2/3 & $4096^3, \Delta x = 0.5$  & $0.194\pm0.011$ & $0.286\pm0.021$ & $1.573\pm0.292$  & This section, fast method\\
      \hline
    \end{tabular}
\caption{The asymptotic rate of change of the mean string separation $\xi$, for the full network using windings $\xi_W$, the rate of change for $p$-strings only, and the rate of change of $\xi_{pq}$ for $pq$-strings. For comparison we provide both the literature values (which have no uncertainties reported) and the Abelian-Higgs values (where the full network estimator is equivalent to only having a single string type).}
    \label{table4}
    \end{center}
\end{table*}

\begin{figure*}[p]
\includegraphics[width=1.0\columnwidth]{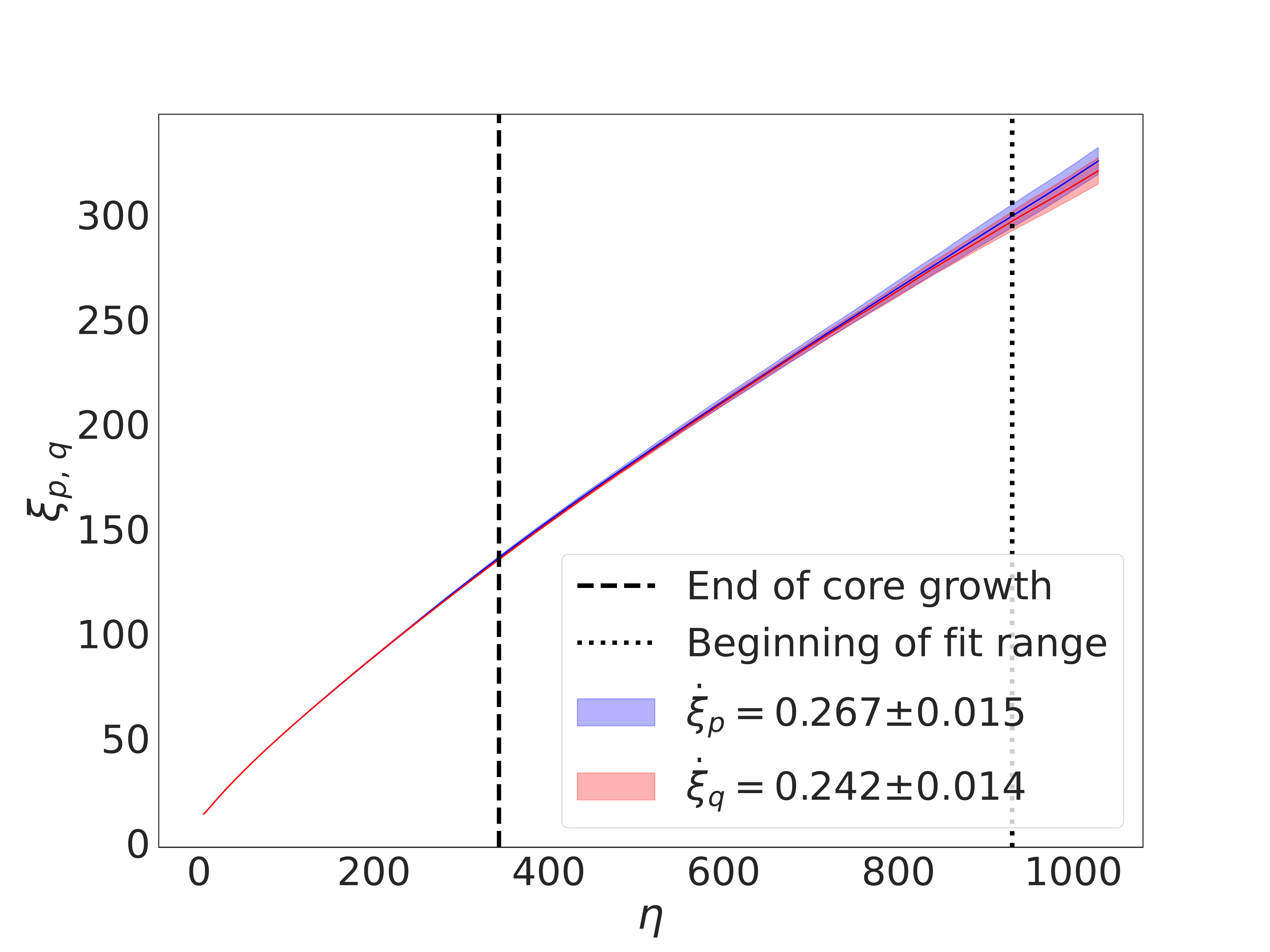}
\includegraphics[width=1.0\columnwidth]{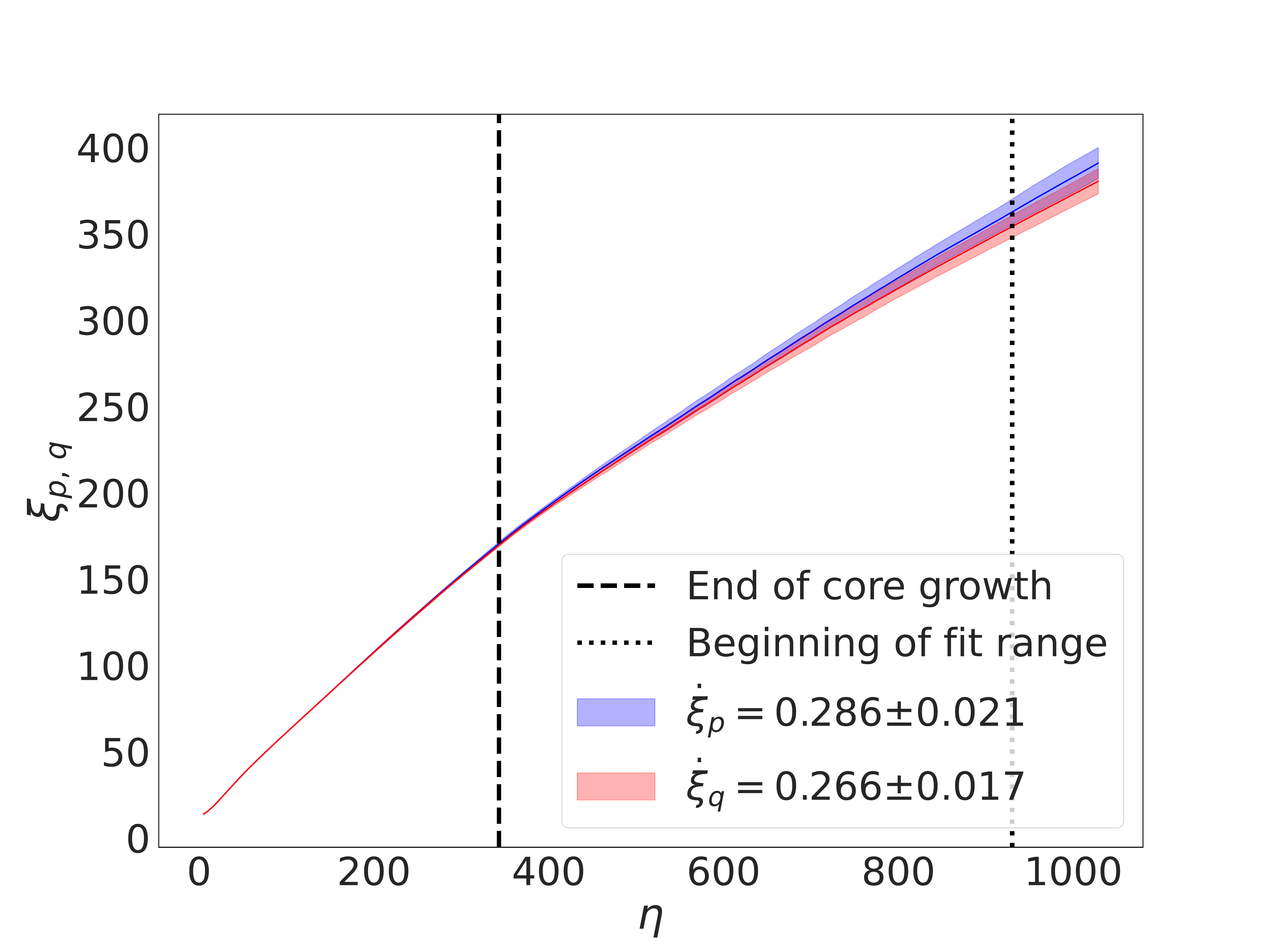}
\includegraphics[width=1.0\columnwidth]{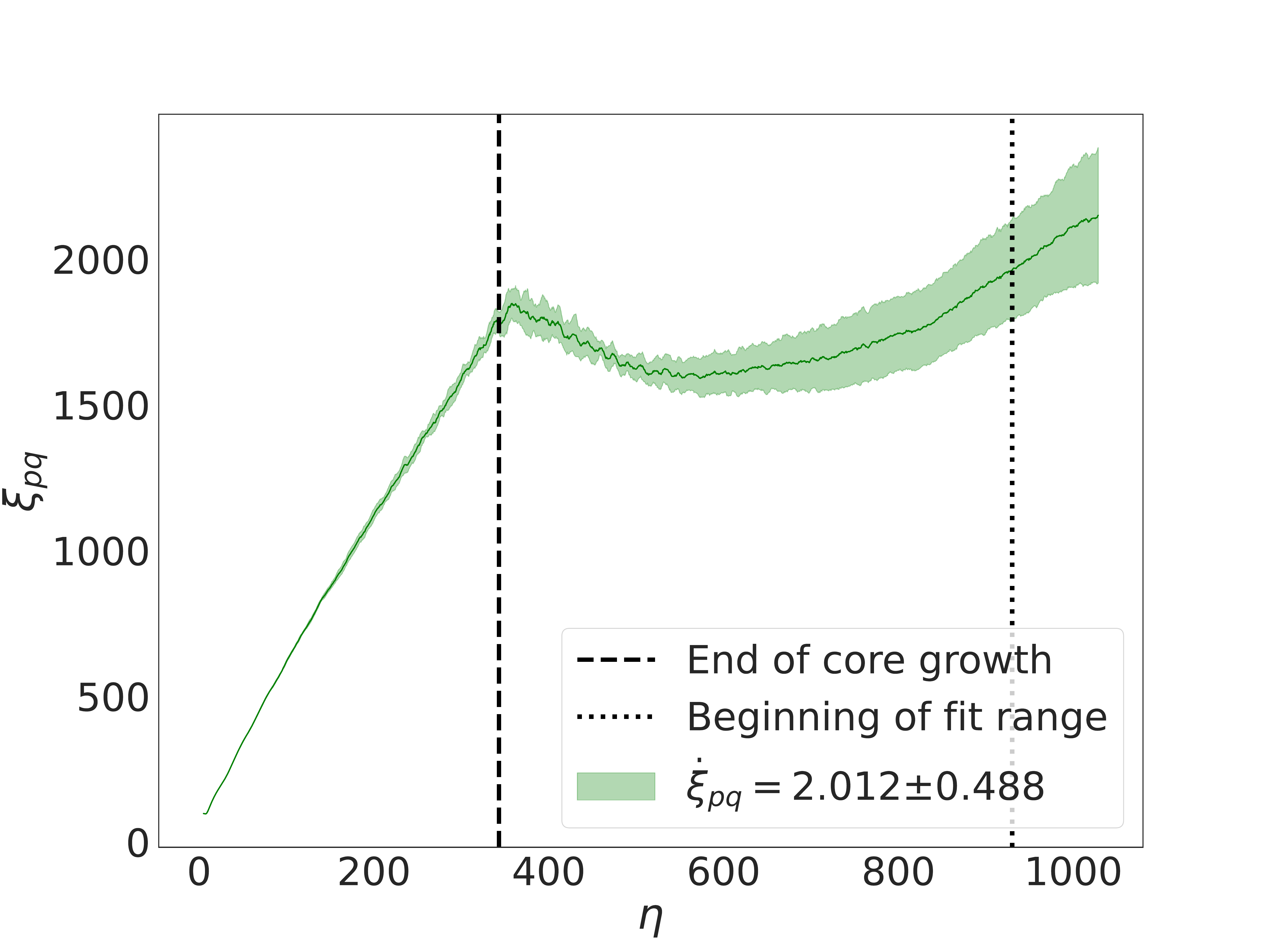}
\includegraphics[width=1.0\columnwidth]{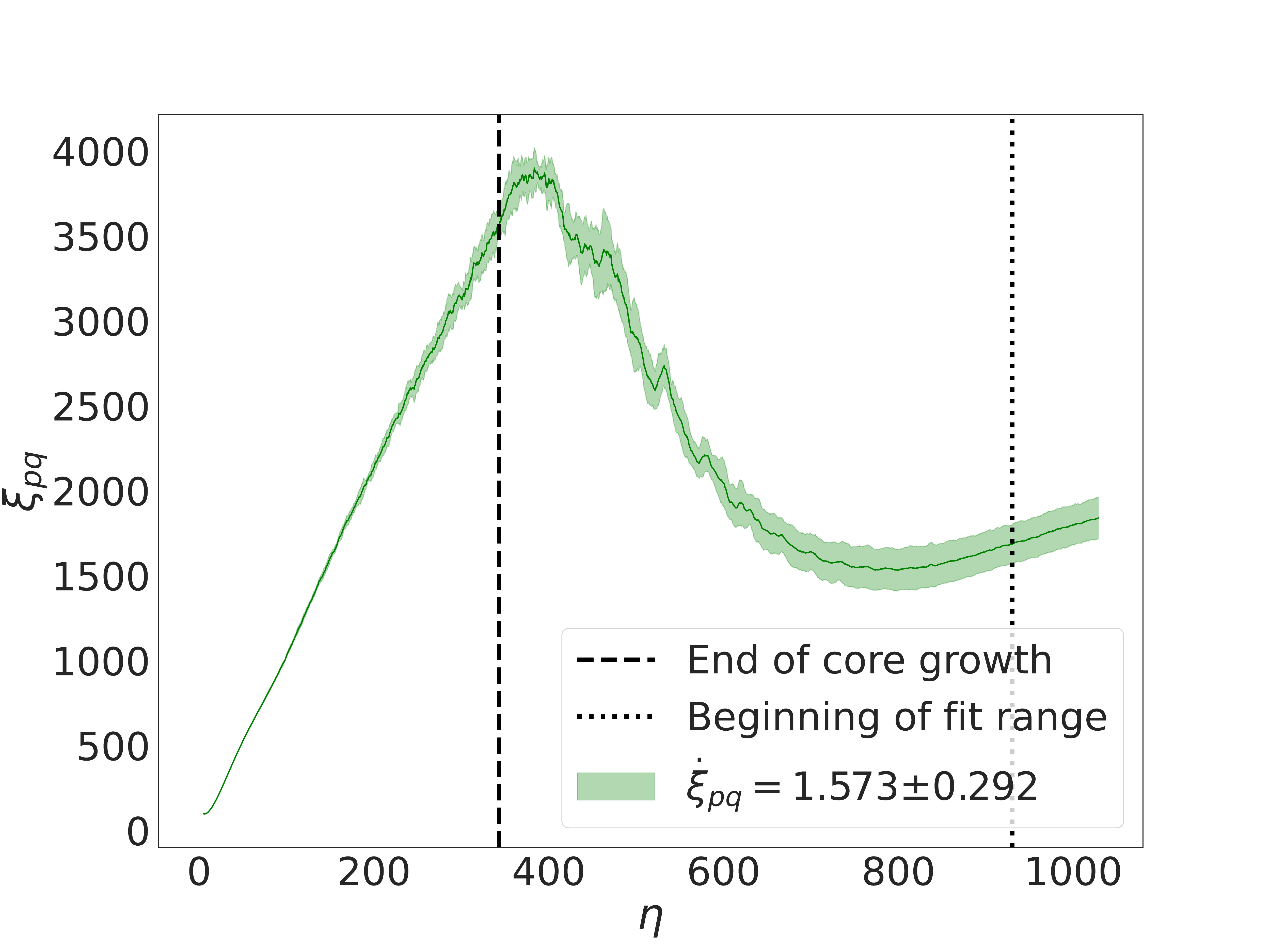}
\includegraphics[width=1.0\columnwidth]{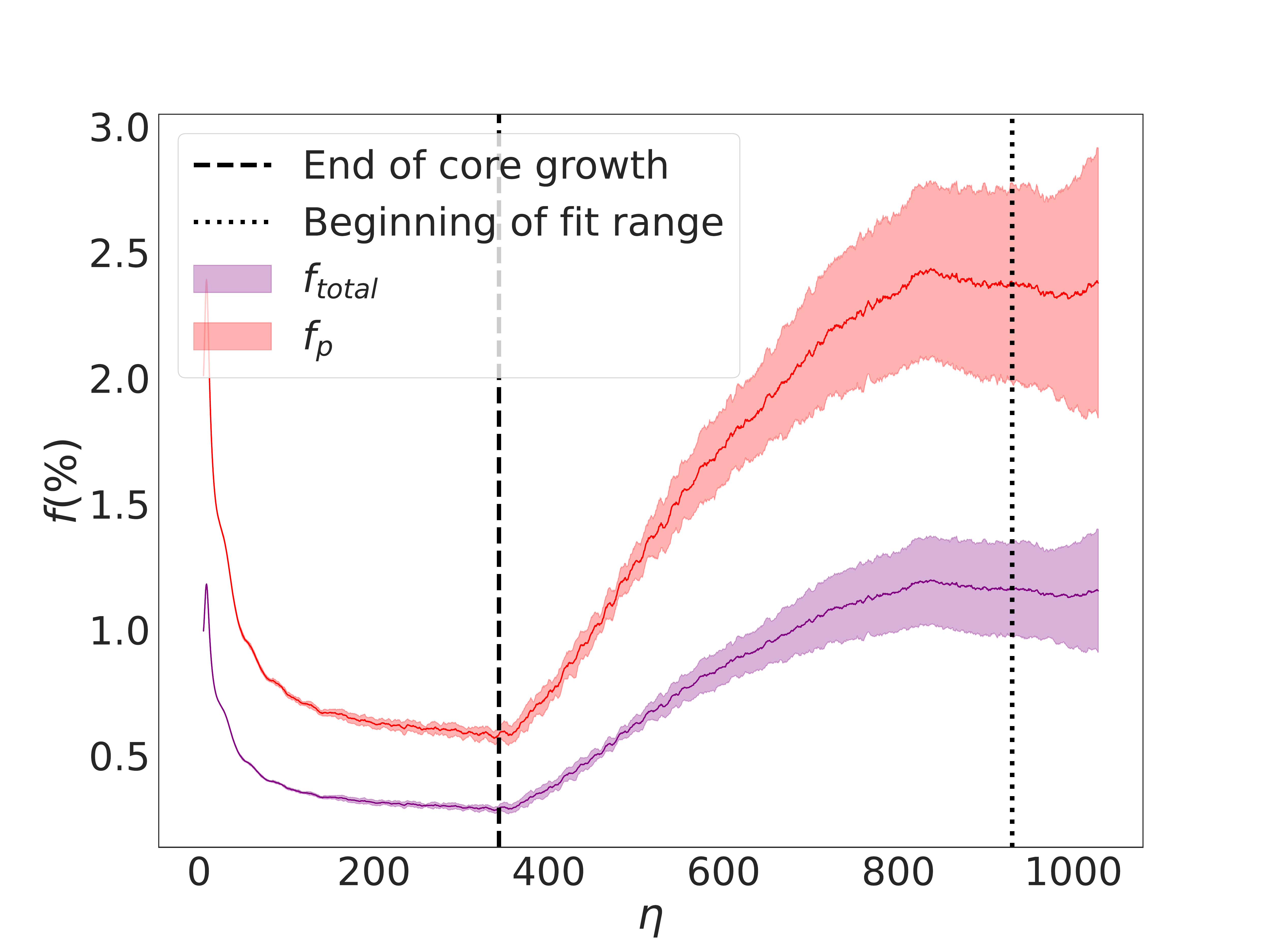}
\includegraphics[width=1.0\columnwidth]{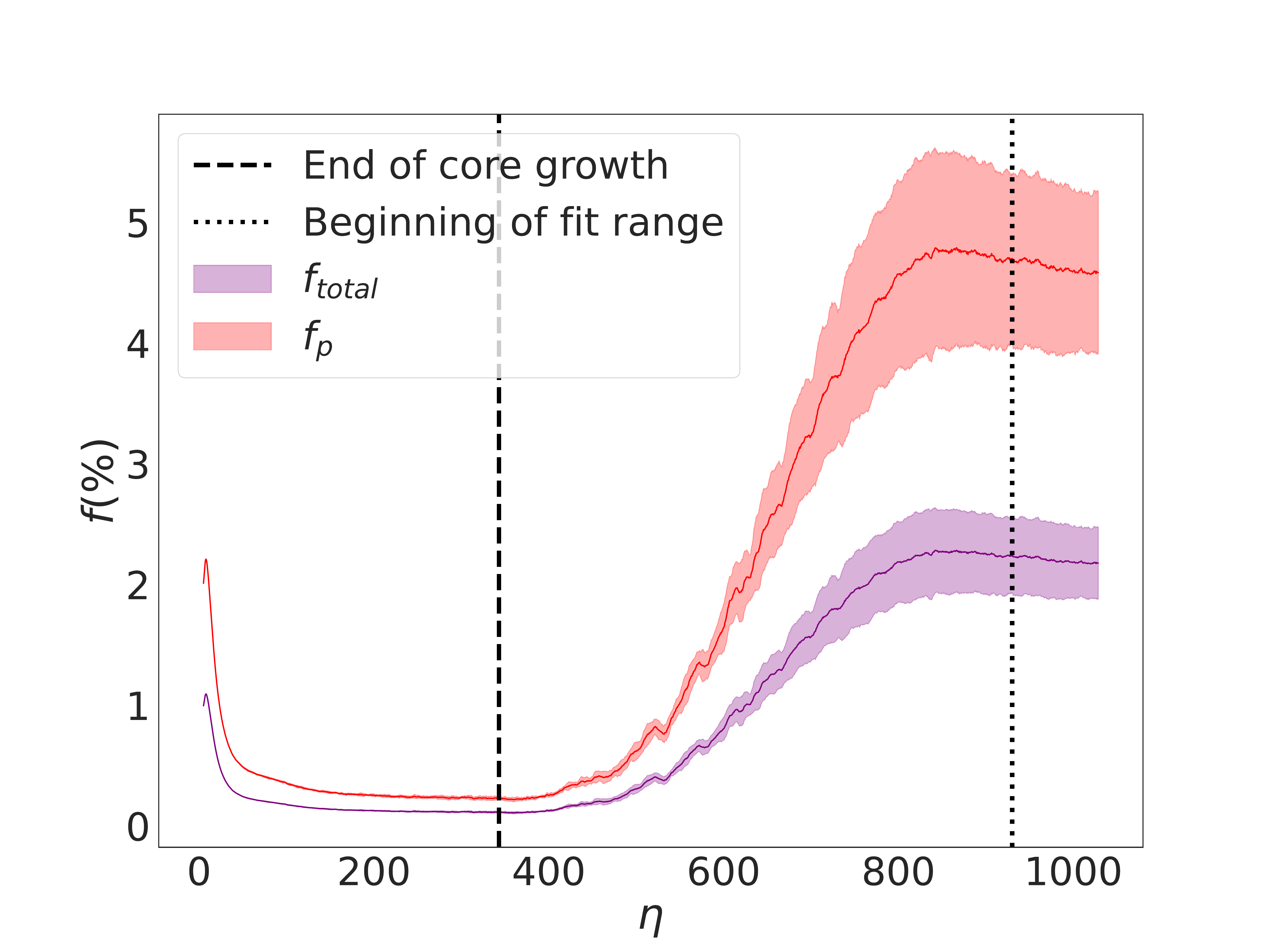}
\caption{The evolution of several quantities derived from the total length of pq-strings present in the box, $L_{pq}$, obtained from the fast computation method, for simulations with core growth up to conformal time $\eta=343.0$ and subsequent physical evolution. The mean string separations $\xi_{p}$ and $\xi_{q}$ are shown in the top panels (red and blue, respectively); $\xi_{pq}$ in the middle panels (in green), and the relative abundances of pq-strings in the lower panels ($f_{total}$ and $f_p$, in purple and red, respectively). Left and right side panels correspond to the radiation and matter epochs respectively.\label{figure08}}
\end{figure*}

\begin{figure*}[p]
\includegraphics[width=1.0\columnwidth]{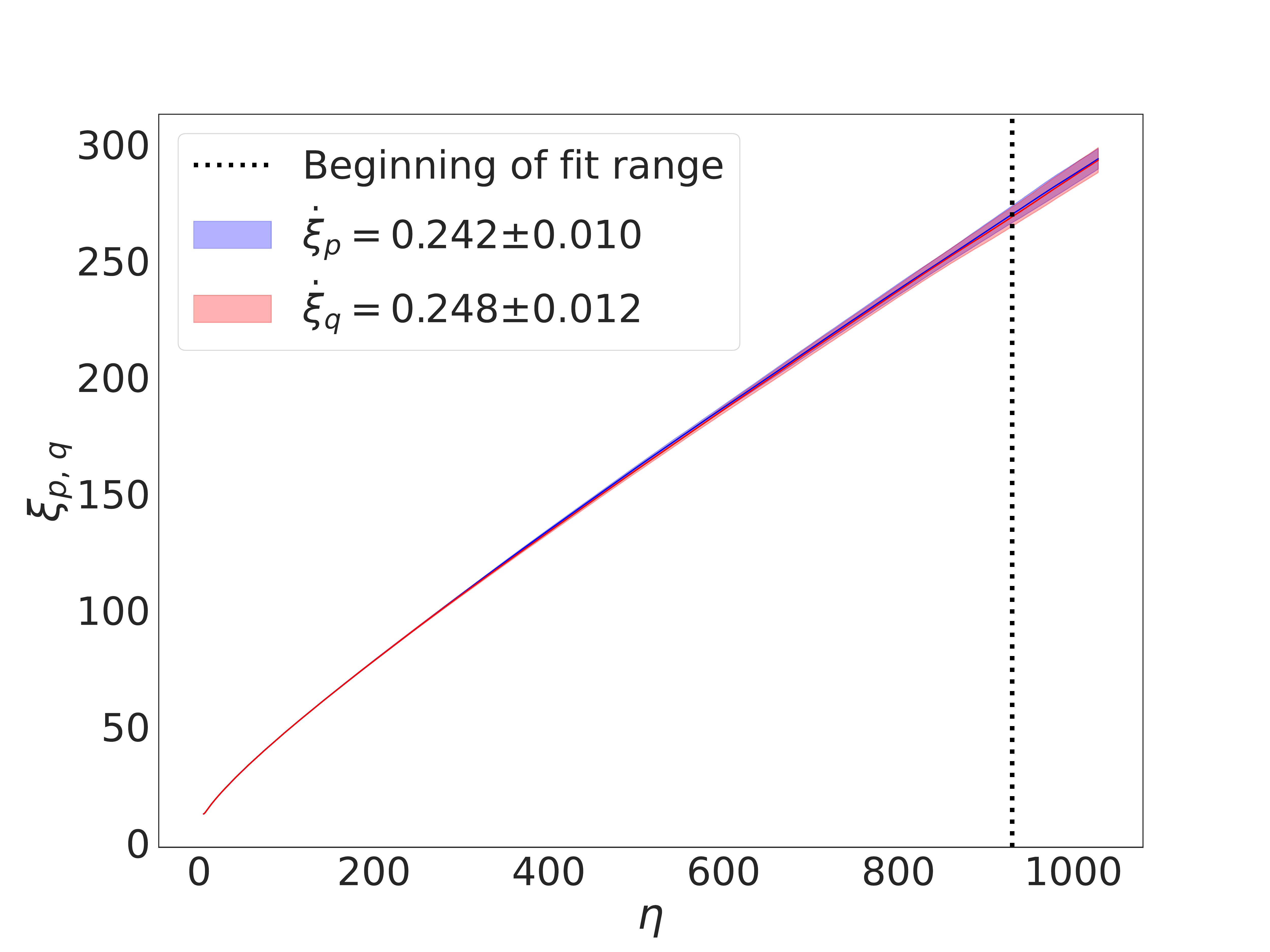}
\includegraphics[width=1.0\columnwidth]{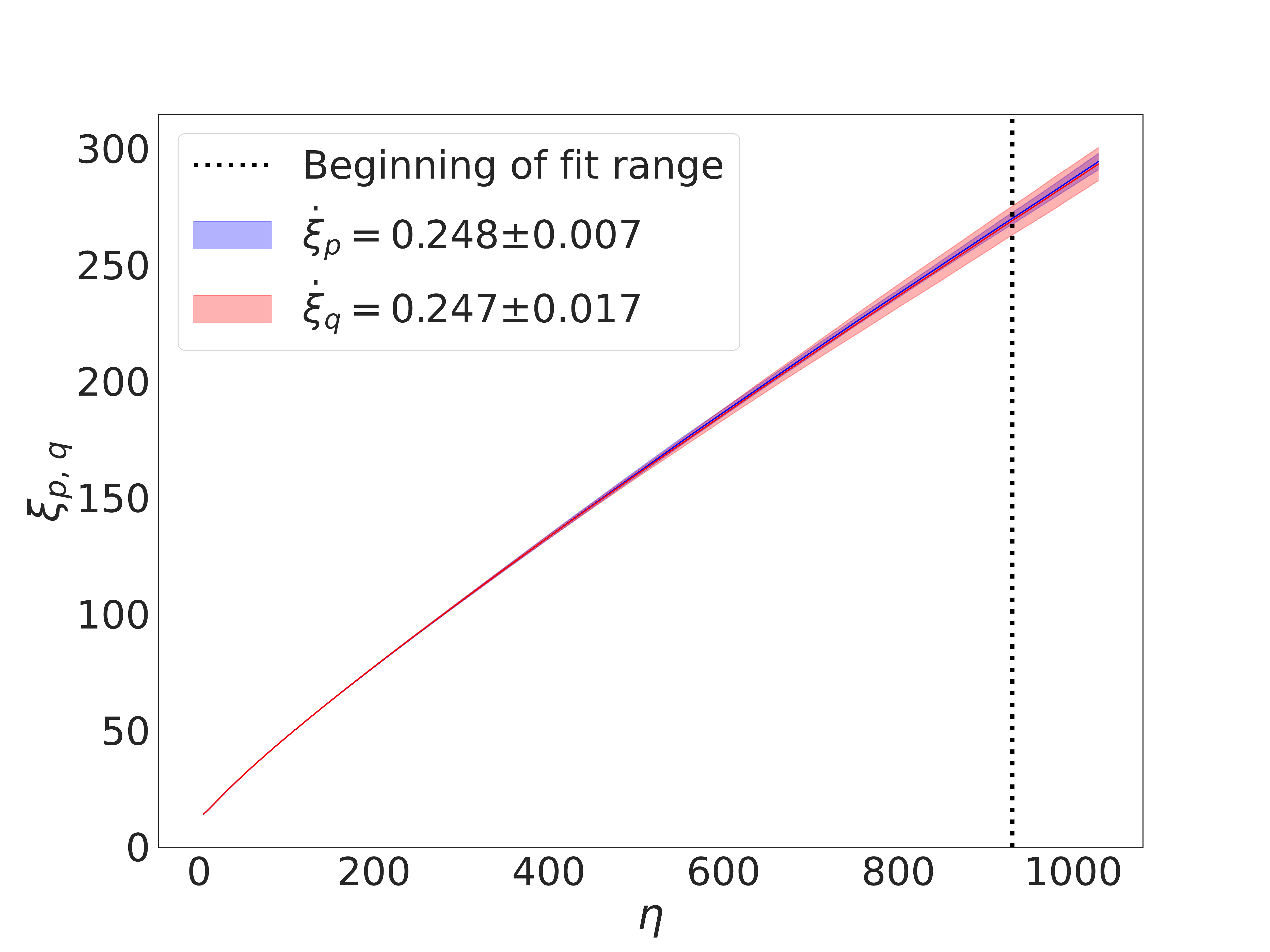}
\includegraphics[width=1.0\columnwidth]{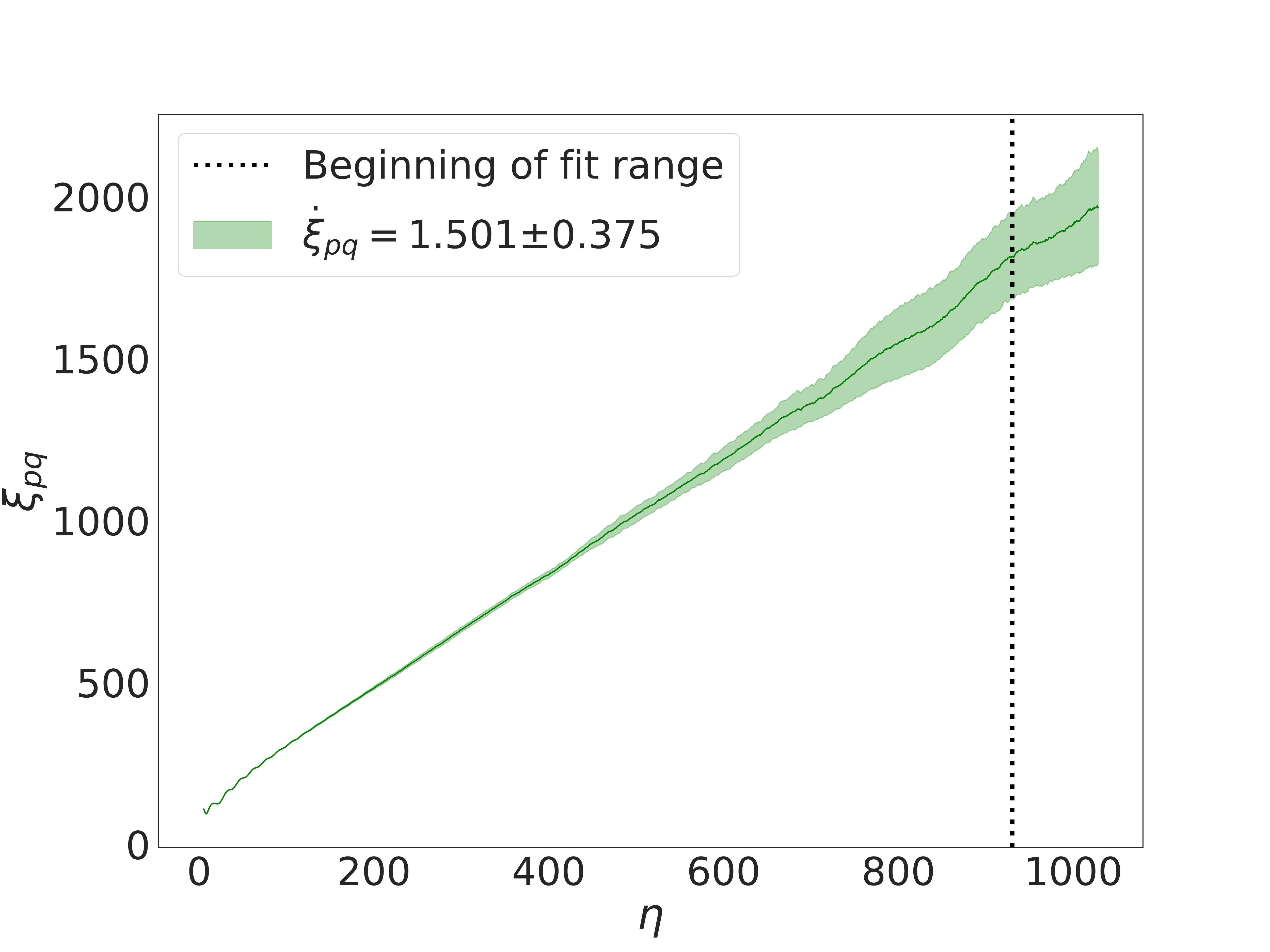}
\includegraphics[width=1.0\columnwidth]{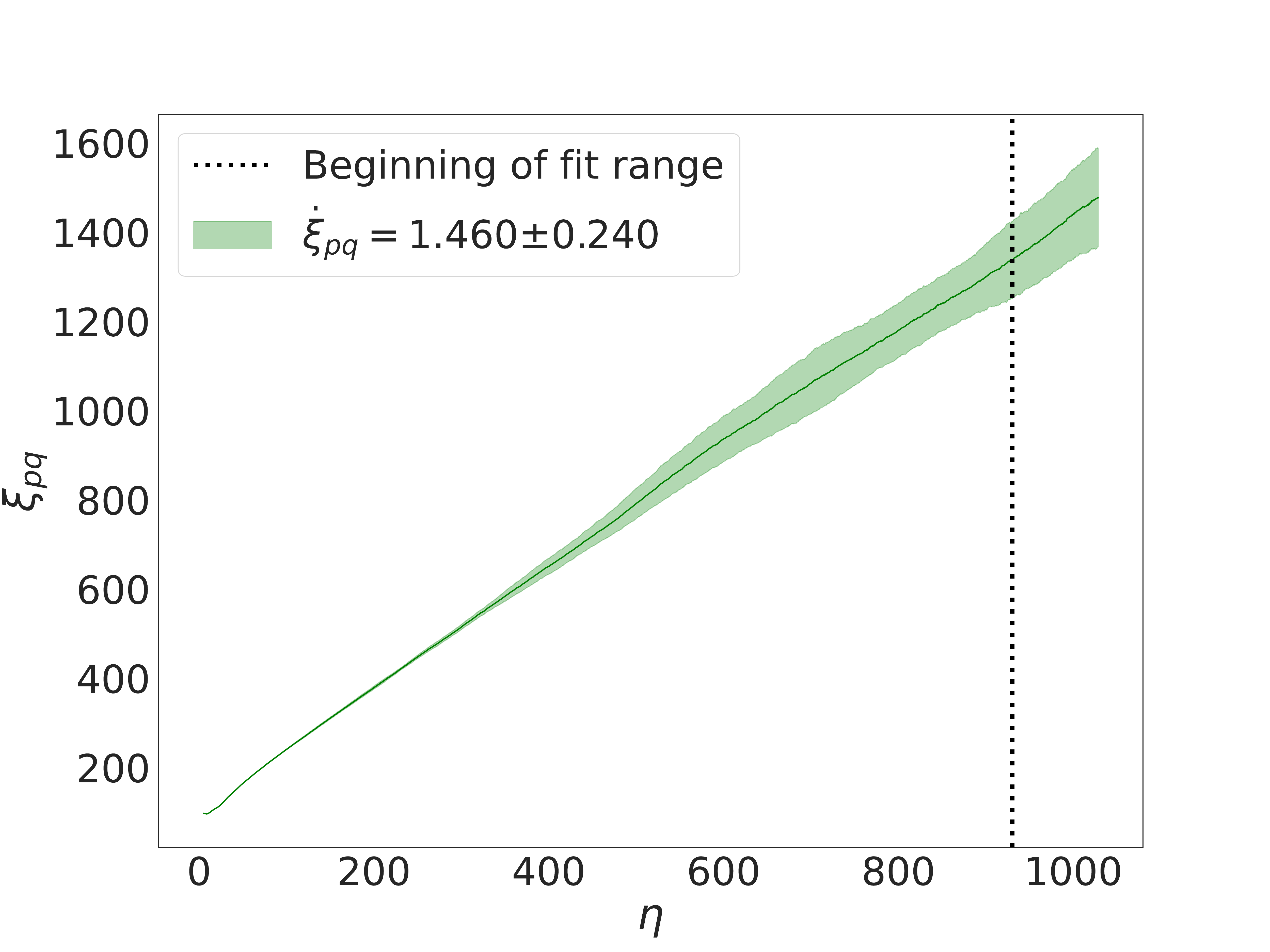}
\includegraphics[width=1.0\columnwidth]{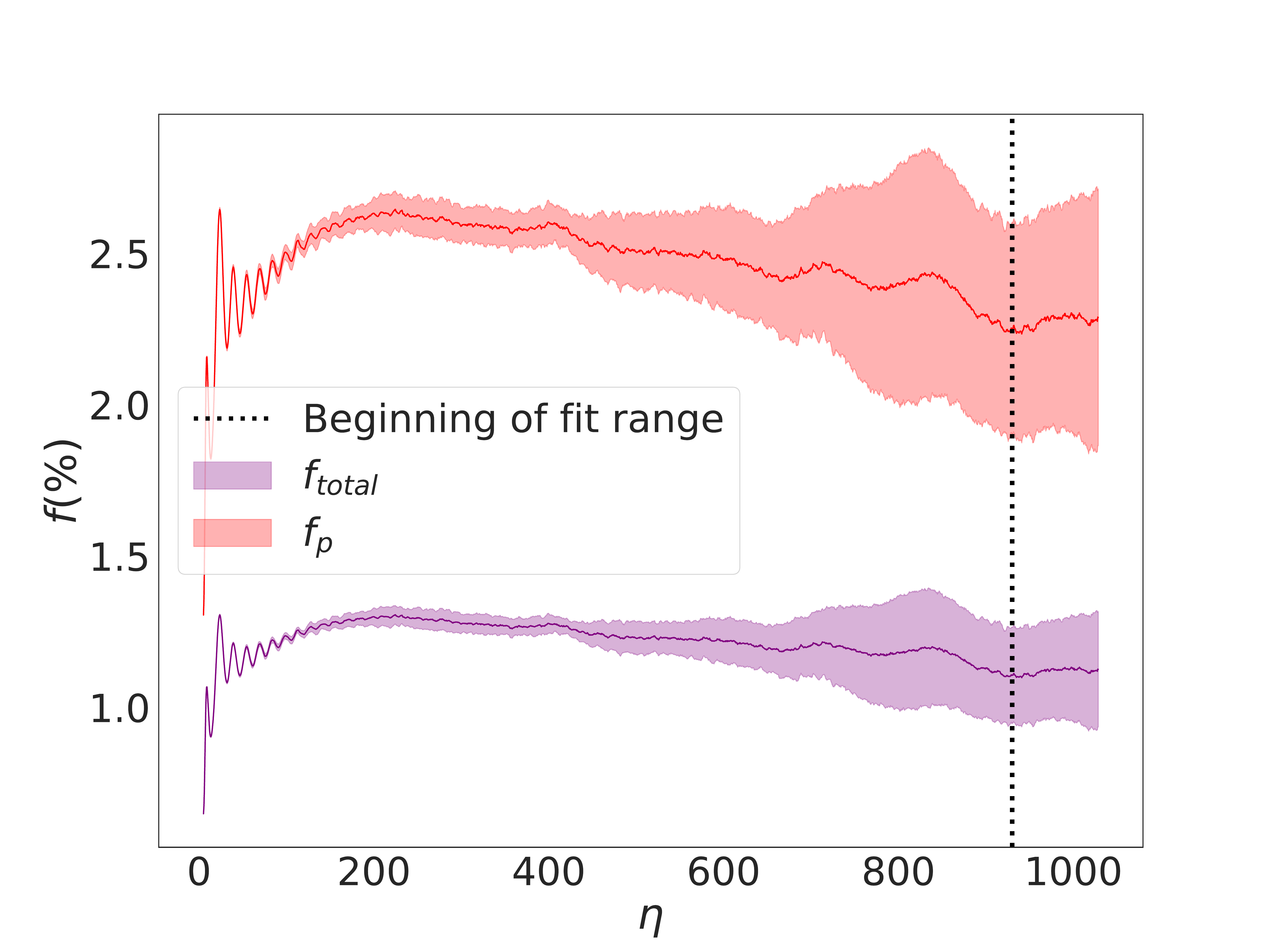}
\includegraphics[width=1.0\columnwidth]{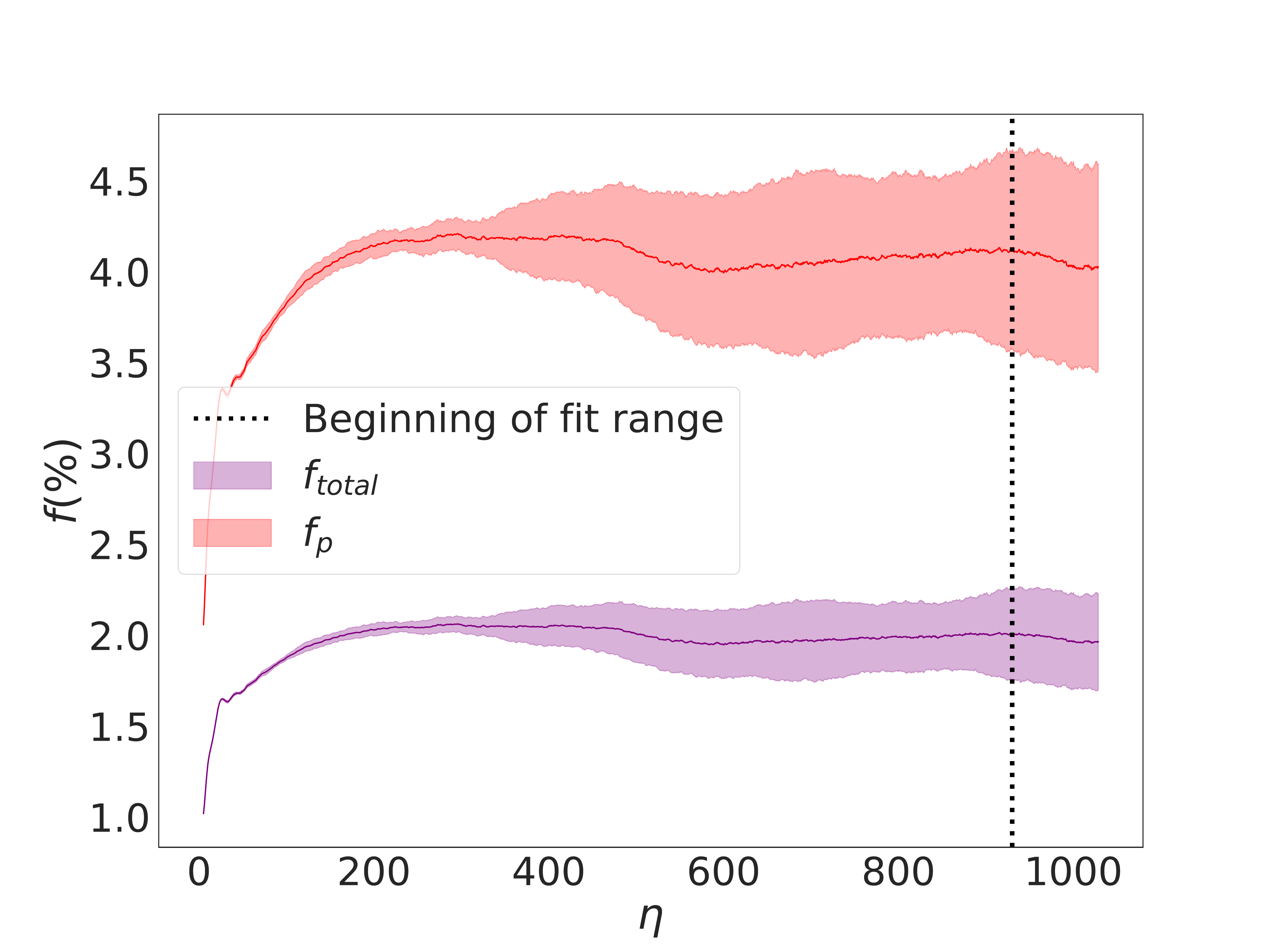}
\caption{The evolution of several quantities derived from the total length of pq-strings present in the box, $L_{pq}$, obtained from the fast computation method, for simulations with constant comoving width. The mean string separations $\xi_{p}$ and $\xi_{q}$ are shown in the top panels (red and blue, respectively); $\xi_{pq}$ in the middle panels (in green), and the relative abundances of pq-strings in the lower panels ($f_{total}$ and $f_p$, in purple and red, respectively). Left and right side panels correspond to the radiation and matter epochs respectively.\label{figure09}}
\end{figure*}

We can now discuss the evolution of the mean string separations of each of the string species in the simulations, $\xi_p$, $\xi_q$ and $\xi_{pq}$, along with their relative abundances. These are depicted in figure \ref{figure08} for core growth followed by physical simulations, while the equivalent PRS ones are shown in figure \ref{figure09}. As before, we report the results for both the radiation and the matter eras. The asymptotic rate of change of $\xi$ can be found in their corresponding panels but is also summarized in table \ref{table4}. The values for $\xi_p$ and $\xi_q$ are unsurprising in the sense that scaling is reached for both types of simulations, and that this scaling is consistently found through the asymptotic values of $\dot{\xi}$ in physical and PRS simulations; again we see that the approach to scaling is somewhat different in the two cases. On the other hand, the values of $\xi_{pq}$ are interesting, clearly indicating that the change from core growth to physical evolution does impact the behavior of bound states. While in core growth the mean $pq$-string separation is increasing linearly, as soon as the transition occurs this quantity decreases, signaling a possible production of bound states, and then inverts its tendency to growing linearly (although with a different slope $\dot{\xi}$).

Thus a change in $pq$-string abundances must occur after the switch to physical evolution, with production of bound states. Since we again observe linear scaling at the very end of the simulation, this means the bound state abundances must have stabilized by then. This is confirmed by the lower panels of figures \ref{figure08} and \ref{figure09}, which show the behavior $f_{total}$ and $f_p$. Both decrease in the core growth phase to much lower abundance than in the PRS case. For instance $f_p$ in the radiation era drops to $0.5\%$, indicating a much more sparse $pq$-string network than in the $\beta=0$ case. However, as soon as the transition happens, $p$ and $q$-strings begin binding more efficiently, producing more bound states. At the end of the simulation, both physical and PRS abundances are in agreement, within their statistical uncertainties.

From this we conclude that changing $\beta$ does have an impact in the formation and destruction of $pq$-strings, and therefore in their abundances. This follows from the fact that different values of $\beta\neq1$ will introduce a time-dependence in the action which spoils energy conservation. Comparing with $\beta=0$ simulations, it is clear the core growth period can result in lower abundances. Overall, the effect of physical evolution still results in a sparse $pq$-string network and no evidence of frustration is found, with all string species scaling. We note that with a smaller dynamic range this behaviour might not be apparent, which might explain the more tentative results of earlier works \cite{Urrestilla:2007yw, Lizarraga:2016hpd}.

\subsection{\label{robustRes}Computation of $L_{pq}$ with the robust method}

The results in the previous section were obtained with the fast method. Presently we turn our attention to computing the total length of $pq$ strings, $L_{pq}$, through the robust method, and its implication for the calculation of $\xi_{pq}$ and average segment length $l_{pq}$. This is again dome with $4096^3$ simulations, exploring the differences between radiation and matter, physical and core growth evolution.

One practical limitation of this method is that it is computationally harder at earlier timesteps, due to string network density being larger; as the network evolves and becomes less dense, the computational cost of the method decreases. A related aspect is that earlier timesteps also possess more regions (clusters of cells) that can be identified as strings, which also end up making earlier timesteps computationally harder. It is therefore convenient to choose a minimum timestep to output the $p$, $q$ and $pq$ string positions necessary for this method. We will also reduce the temporal rate of outputs, in order to reduce our output files footprint, and stay within the allowed quotas of the computational facility we use (Piz Daint at CSCS, Switzerland). These justify our choice to output string positions every 40 timesteps, for the conformal time range of $\eta \in [500,1024]$. Overall, our robust method simulations accounted for more than $80\%$ of the  125 000 node hours (8.5 million core hours) used by the present work.

\begin{figure*}
\includegraphics[width=0.66\columnwidth]{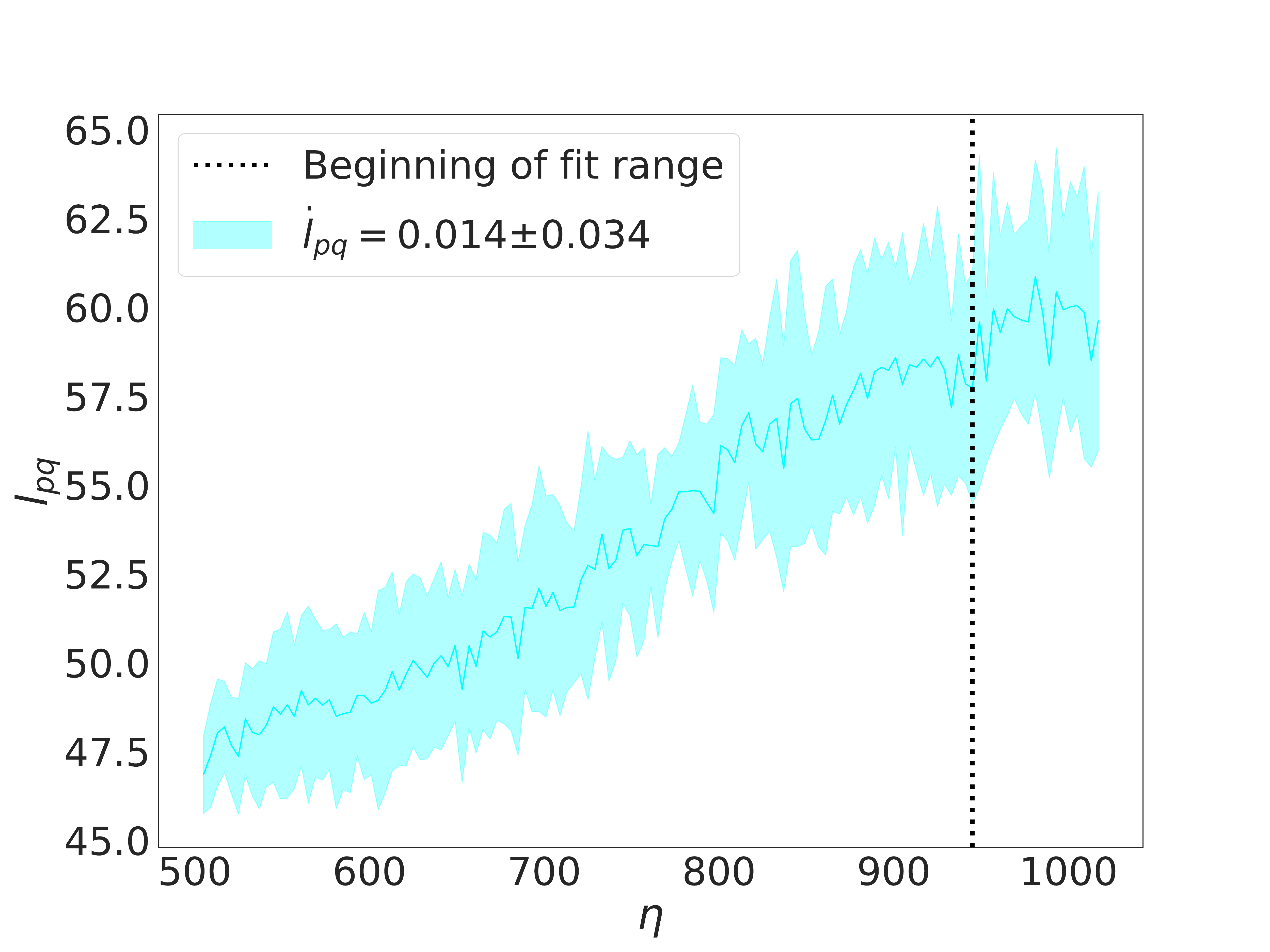}
\includegraphics[width=0.66\columnwidth]{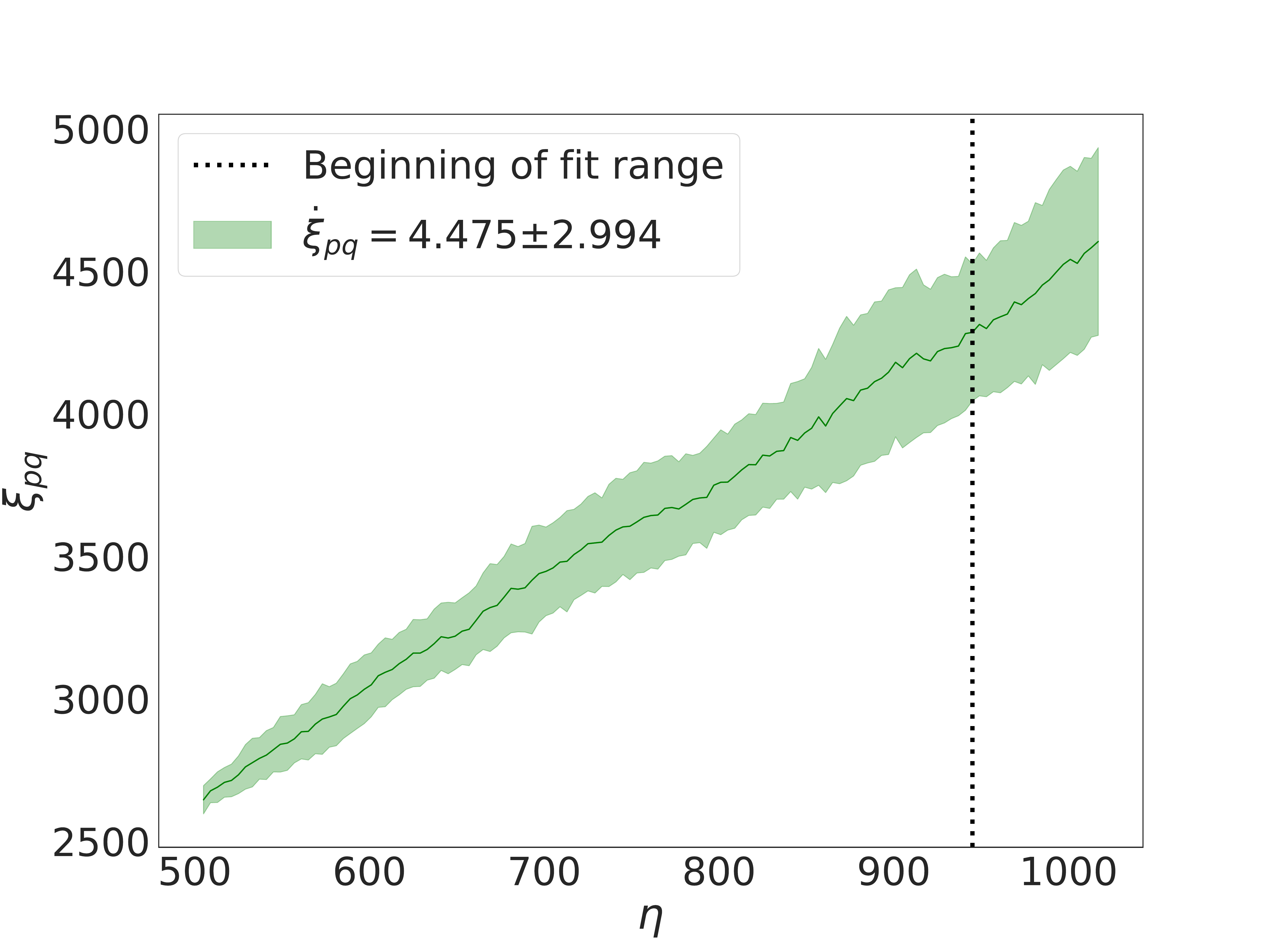}
\includegraphics[width=0.66\columnwidth]{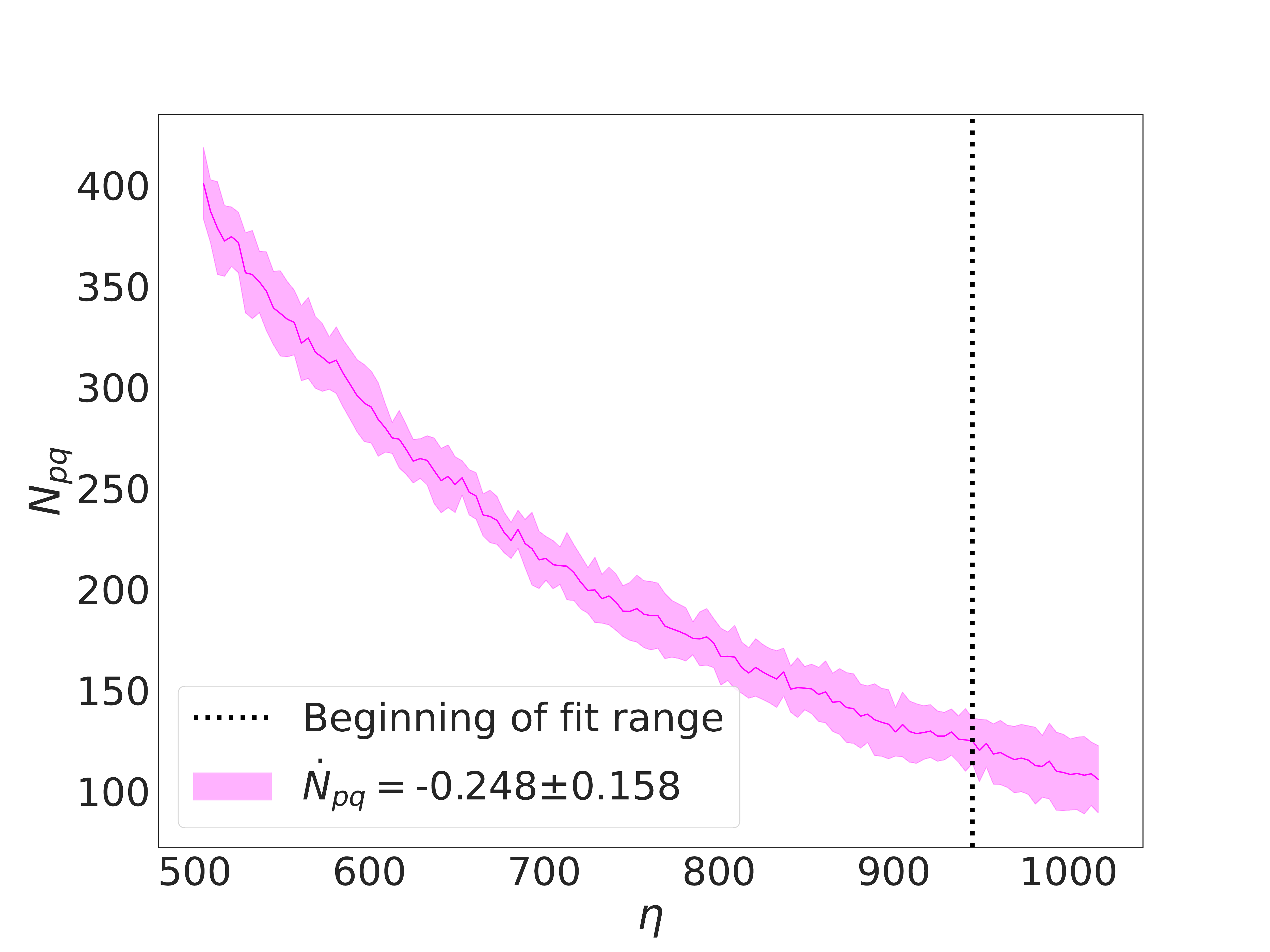}

\includegraphics[width=0.66\columnwidth]{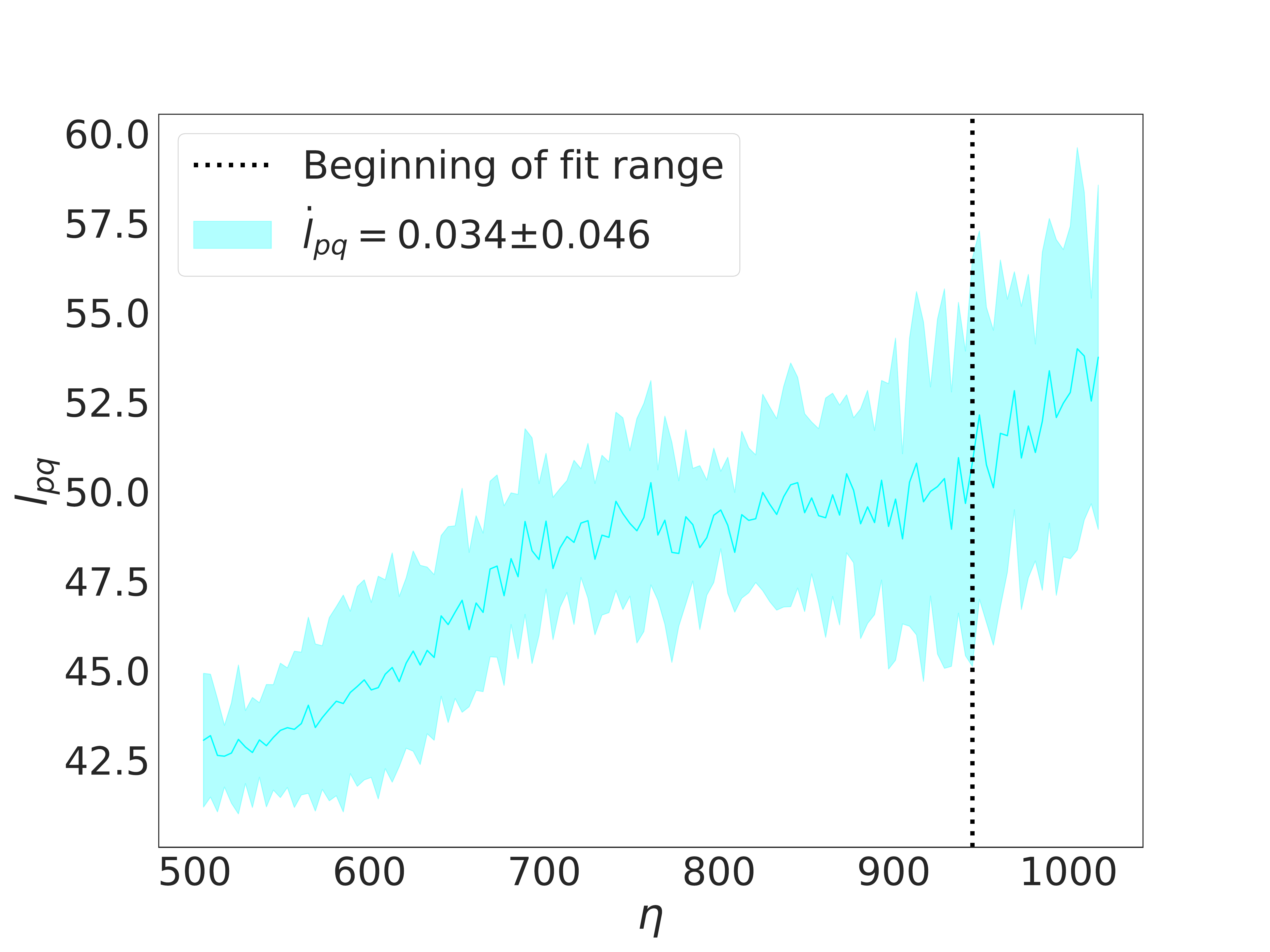}
\includegraphics[width=0.66\columnwidth]{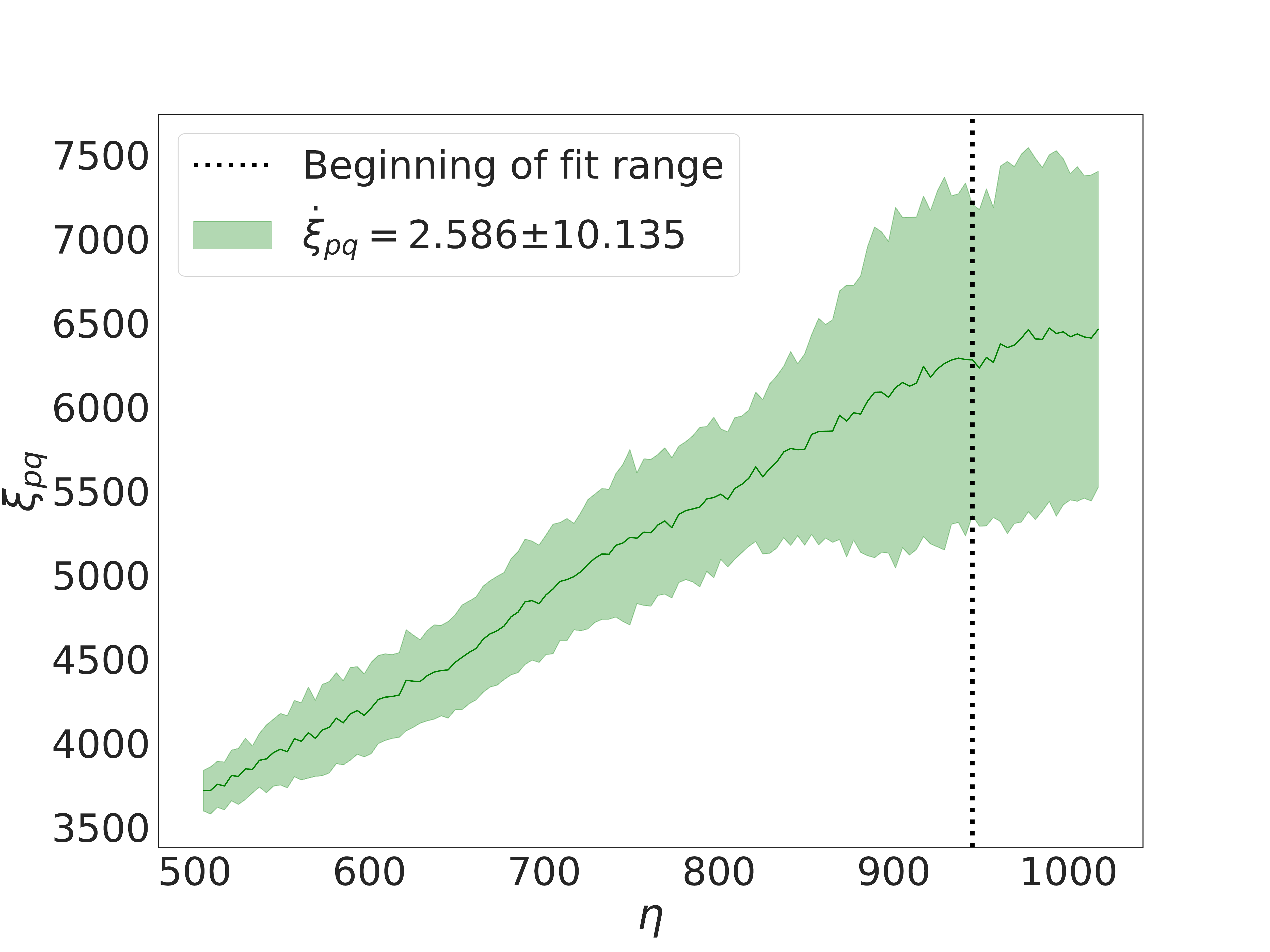}
\includegraphics[width=0.66\columnwidth]{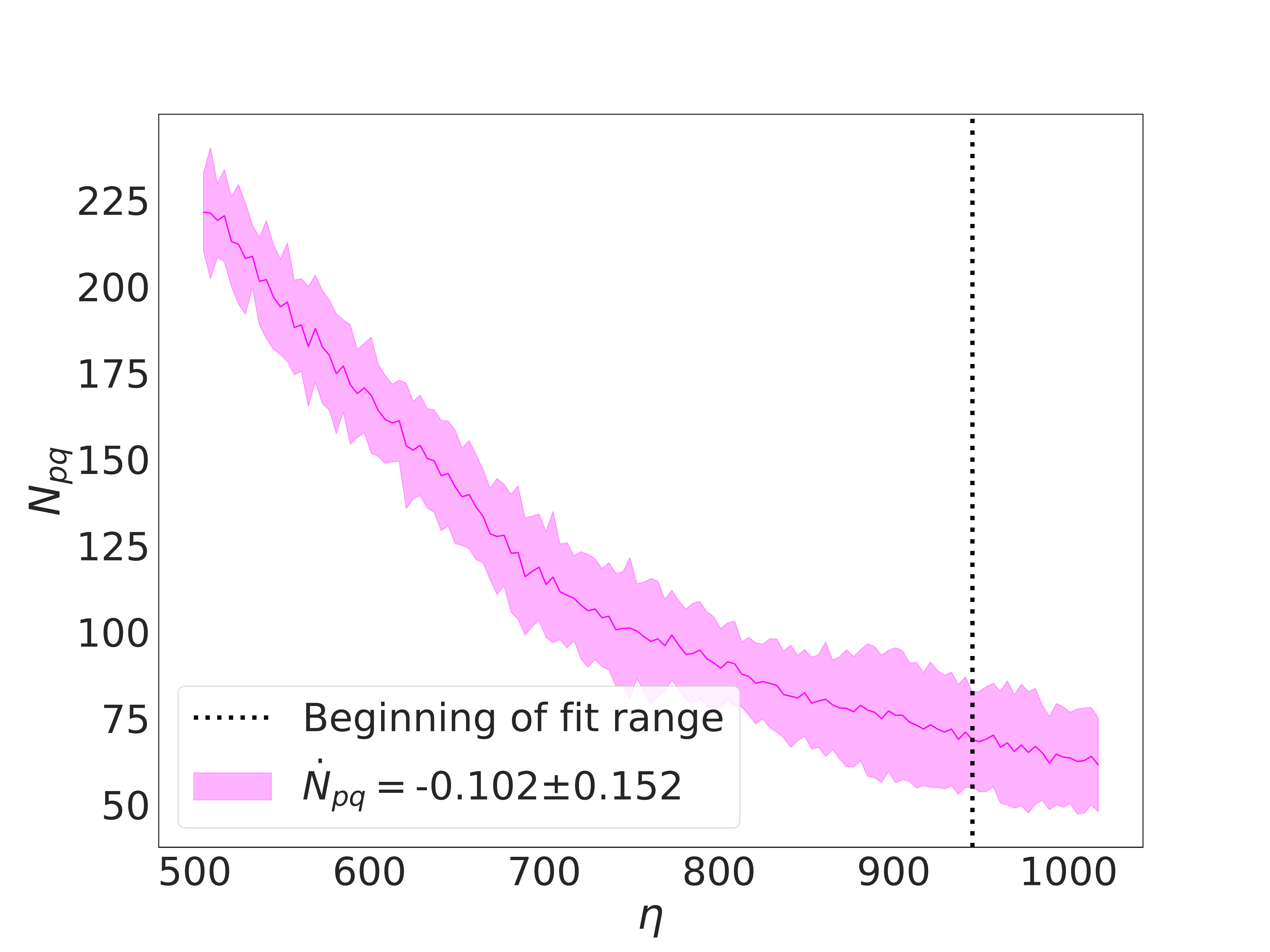}

\includegraphics[width=0.66\columnwidth]{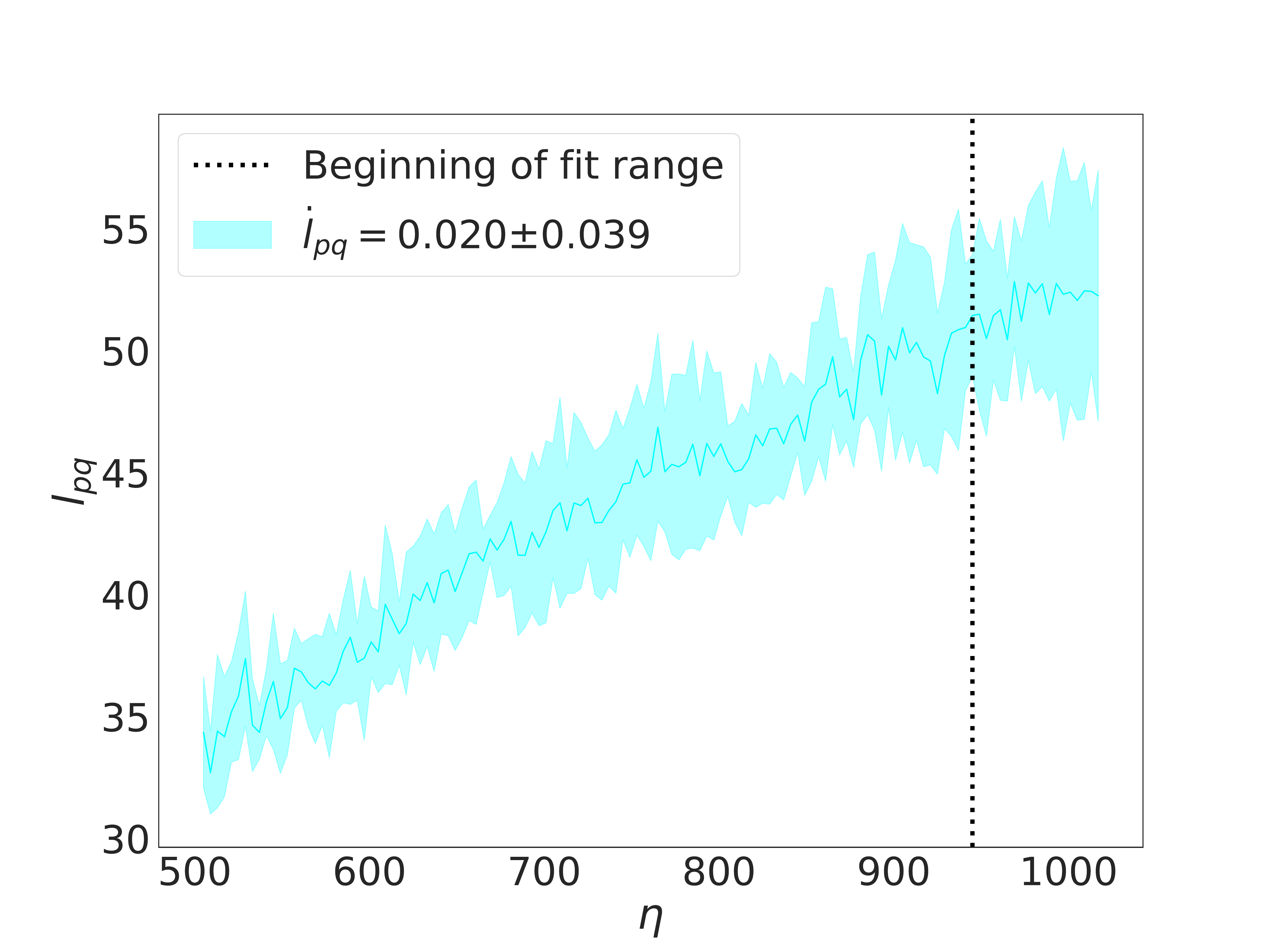}
\includegraphics[width=0.66\columnwidth]{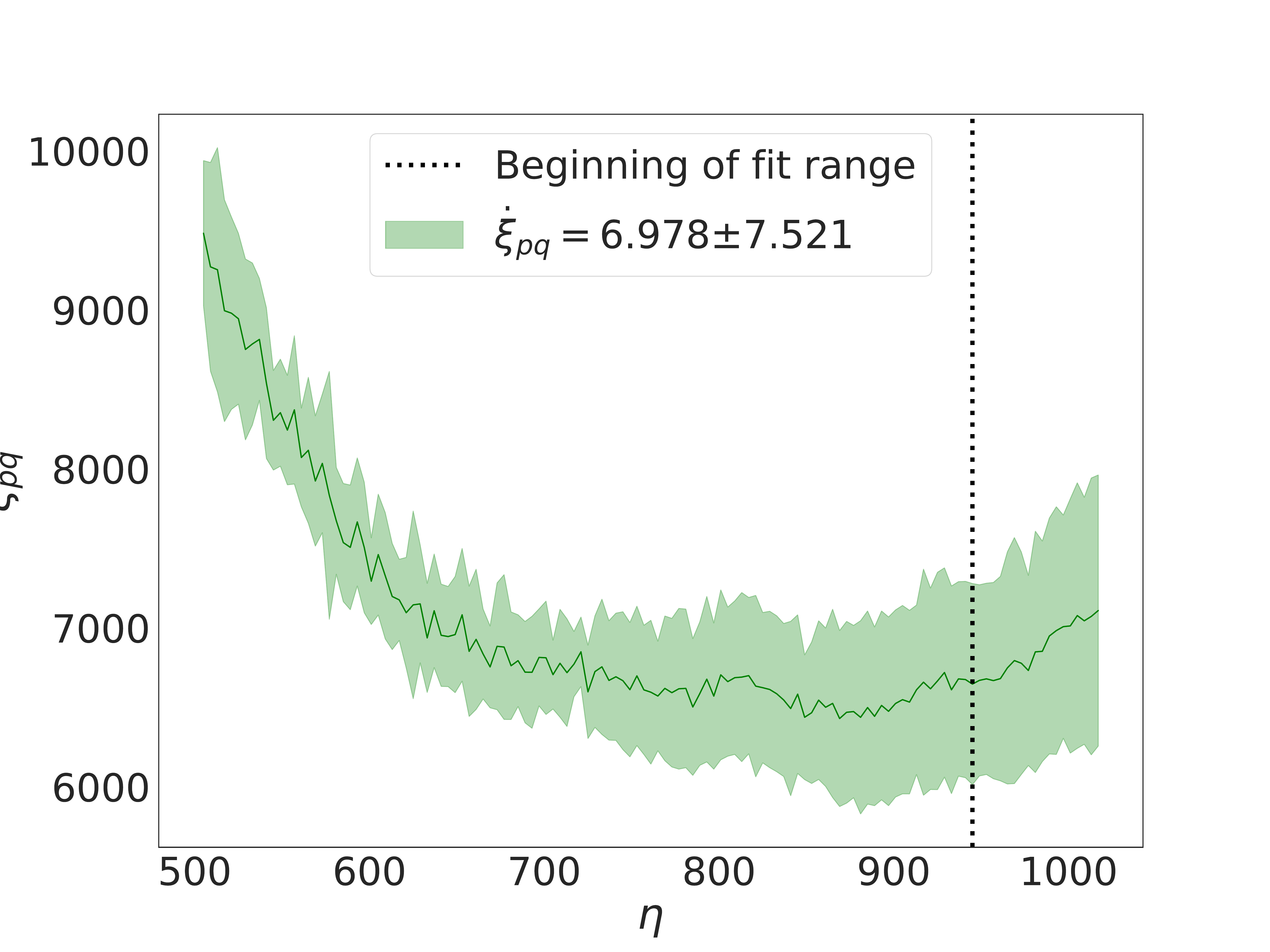}
\includegraphics[width=0.66\columnwidth]{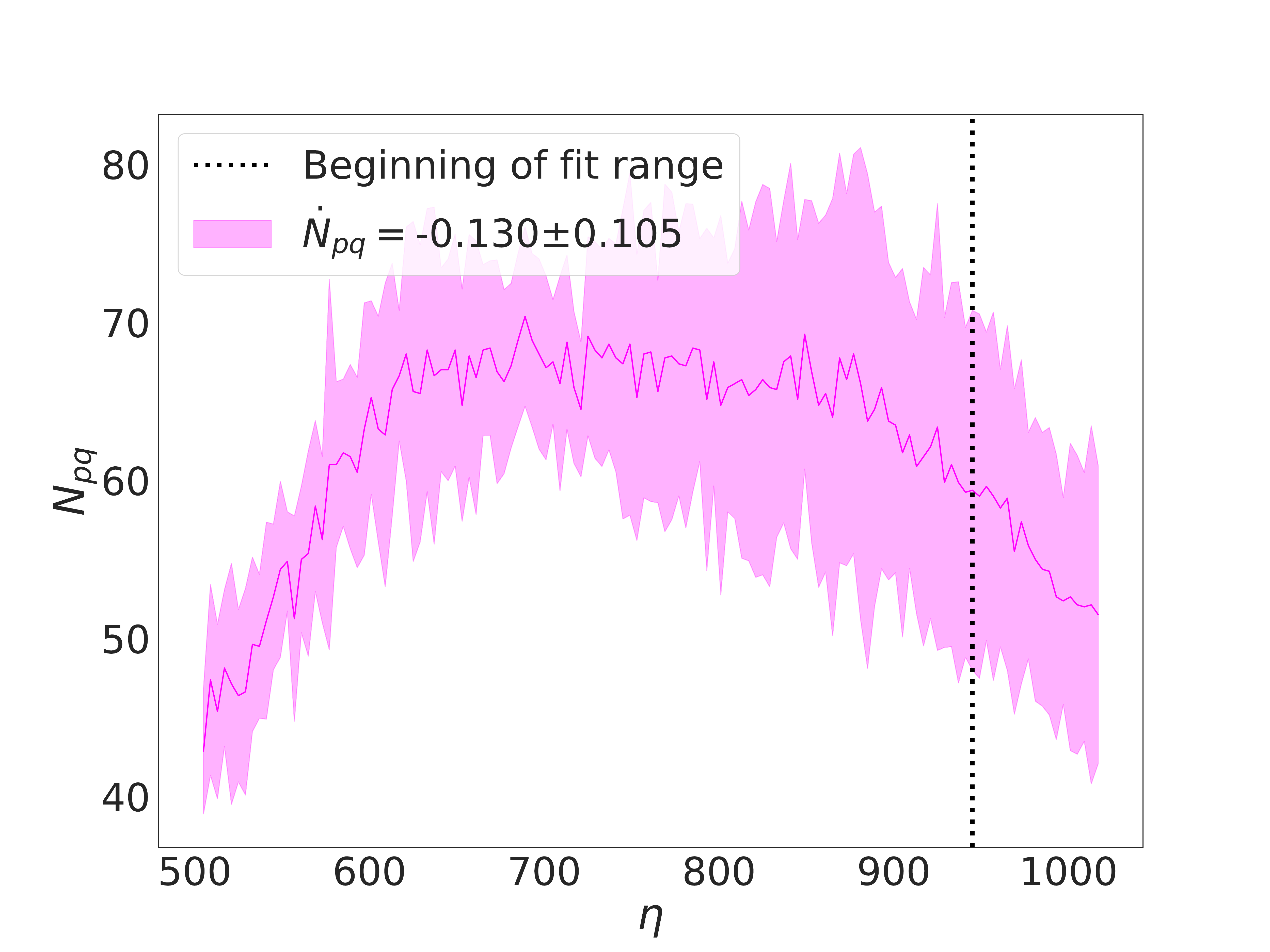}
\caption{The evolution of the mean segment length $l_{pq}$ (left panels), the mean pq-string separation $\xi_{pq}$ (middle panels) and number of pq-segments (right-hand panels) after applying all corrections described in the text. Top and middle panels correspond to constant comoving width runs in the matter and radiation epochs respectively. Bottom panels correspond to radiation epoch runs but with core growth followed by physical evolution. In all panels the rate of change  of plotted quantities is computed using the conformal time range $\eta \in [950, 1024]$. \label{figure10}}
\end{figure*}

It is worth recalling the corrections underlying the robust method: a one-cell tolerance on the overlap of the strings, a fix for the off-by-one error on the number of cells and excluding clusters of less than $41$ cells (with length of $20.0$) which are indicative of accidental crossings. Clearly there is no guaranteed way to evade all systematic error sources, and there is some ambiguity in this methodology, specifically in the choice of the thresholds. A full exploration of the impact of choosing different amounts of overlap and different minimum segments length is beyond the scope of the present work. That said, we have confirmed visually that the exclusion of accidental crossings is set by the appropriate threshold on the segment length and that the one-cell overlap seems sufficient (when comparing with interaction potential isosurfaces) to pinpoint the relevant segment locations.

The results for three different simulation sets, two with constant comoving width in the matter and radiation epochs and one with core growth followed by physical evolution in the radiation era can be found in figure \ref{figure10}. The values of $l_{pq}$ do show linear evolution over time, which is consistent with the earlier tentative results of \cite{Urrestilla:2007yw}. In \cite{Lizarraga:2016hpd} this linear growth is only clearly observed for their 'artificial' initial conditions (rich in $pq$-strings); the situation is far less clear when they choose random initial phases. Both of these works suggest that linear growth would be expected with a larger dynamic range. Our present work does confirm this suspicion, highlighting the need for sufficiently large dynamical range and resolution for accurately describing segment lengths. The values of $\xi_{pq}$ show linear evolution, and showcase how the mean string separation decreases after core growth and begins to linearly increase after about $\eta \approx 800$. Overall the tendency is in qualitative agreement with the fast estimator, even if the values of $\dot{\xi}_{pq}$ (in the range $\eta \in [950,1024]$) are somewhat affected by the systematic themselves.

One final interesting remark which stems from this figure is that $l_{pq}$ is far less impacted by the post core growth evolution than $\xi_{pq}$. While so far we have not studied this in detail (or modelled the dynamics of the bound states), we may speculate that this could be related to the fact that these bound states do not form a Brownian network (and most standard definitions of correlation lengths or other characteristic lengths typically assume, explicitly or implicitly, Brownian networks). A future study of the properties of these bound states is clearly warranted. We can also comment on the transient regime that seems to appear on growth into physical simulations, using again the plots of $l_{pq}$, $\xi_{pq}$ and $N_{pq}$ on figure \ref{figure10}. It seems that after core growth (in other words, once the physical evolution starts) more segments appear, and these segments grow until the pq-network relaxes into a tentative scaling regime. When the scaling regime is reached the number of segments decreases, but the average length keeps increasing.

\subsection{\label{Velpq}Measuring string velocities}

Finally we discuss the evolution of string velocities, which are crucial for an accurate analytic modeling of the evolution of these networks, and also as a consistency test of the occurrence of a scaling solution. This has been shown to be the case for the evolution of the simpler cosmic string networks \cite{Book,Martins:1996jp}, and more recent work indicates that this is even more so for analogous models for strings with additional degrees of freedom on the string worldsheet, e.g. wiggly strings \cite{Almeida} and superconducting strings \cite{Currents1,Currents2}.

\begin{figure*}
\includegraphics[width=1.0\columnwidth]{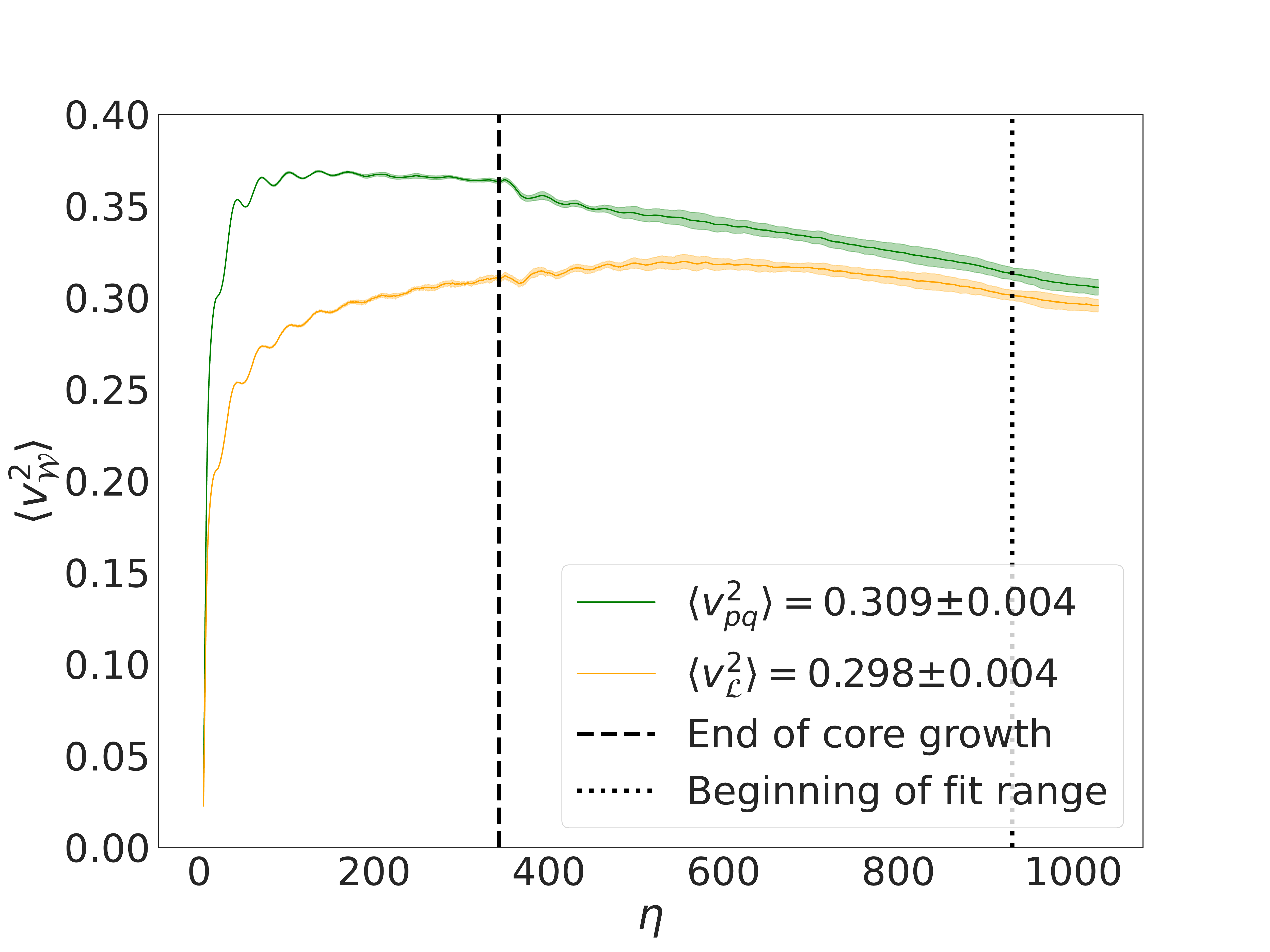}
\includegraphics[width=1.0\columnwidth]{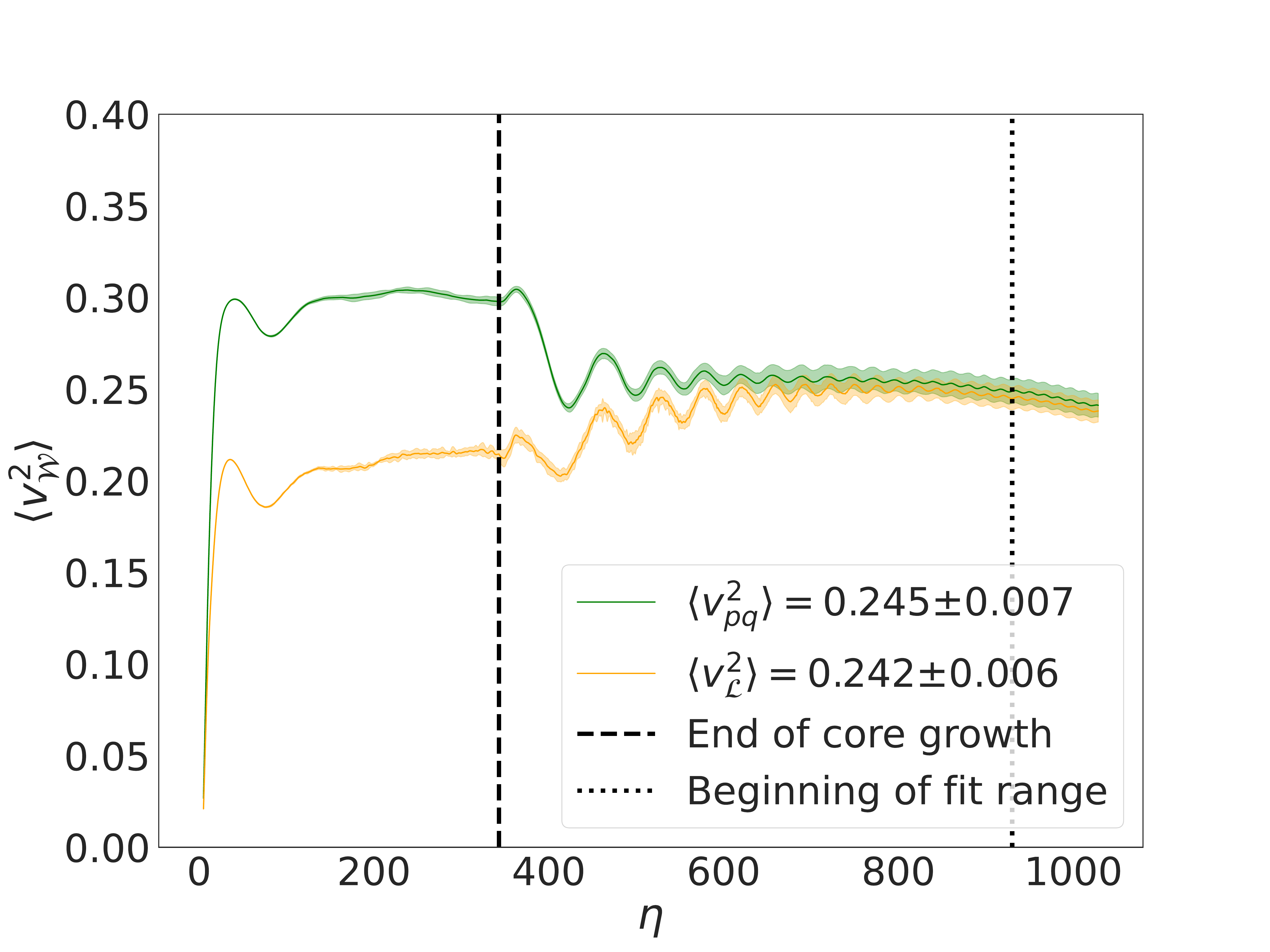}
\includegraphics[width=1.0\columnwidth]{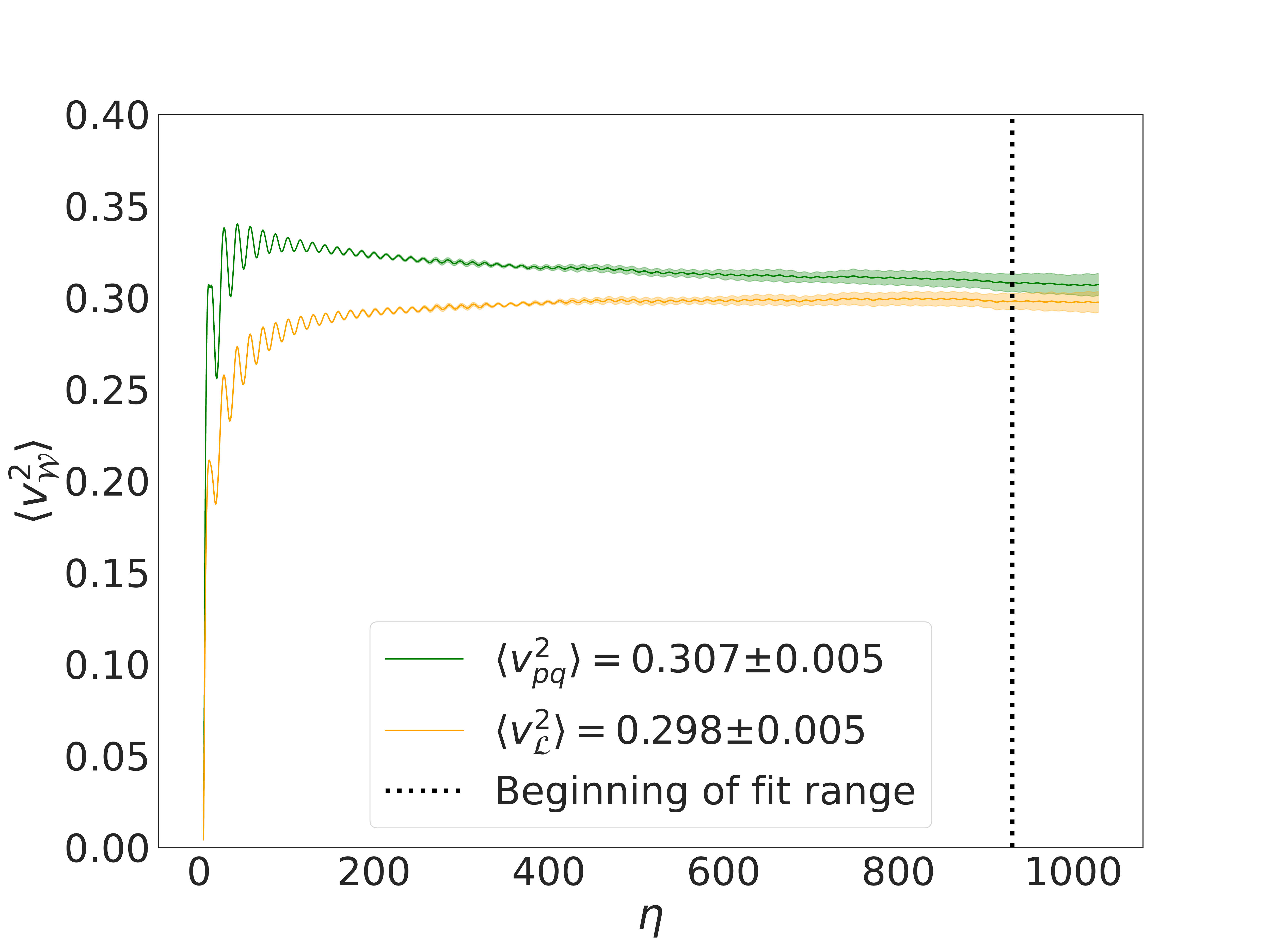}
\includegraphics[width=1.0\columnwidth]{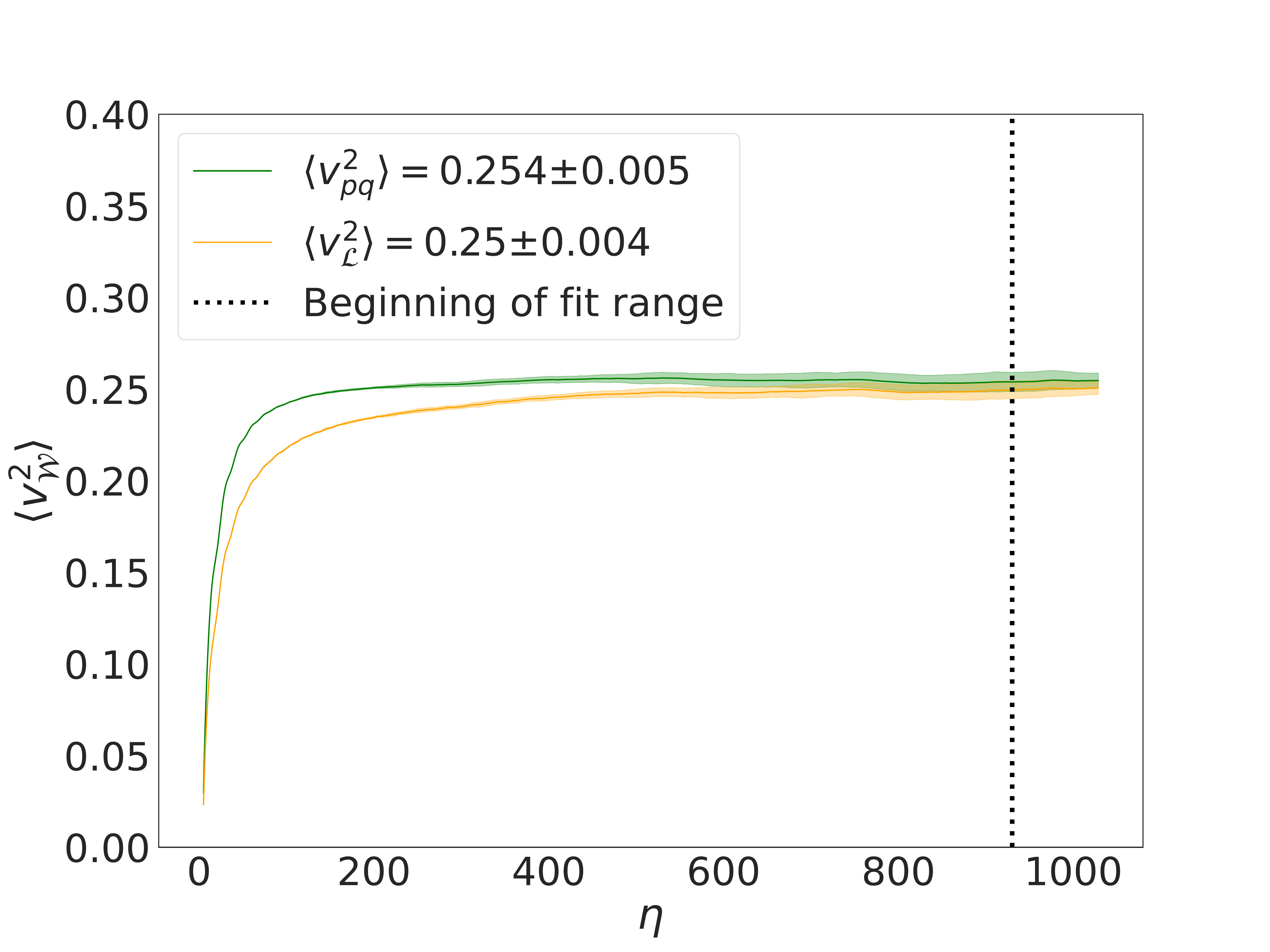}
\includegraphics[width=1.0\columnwidth]{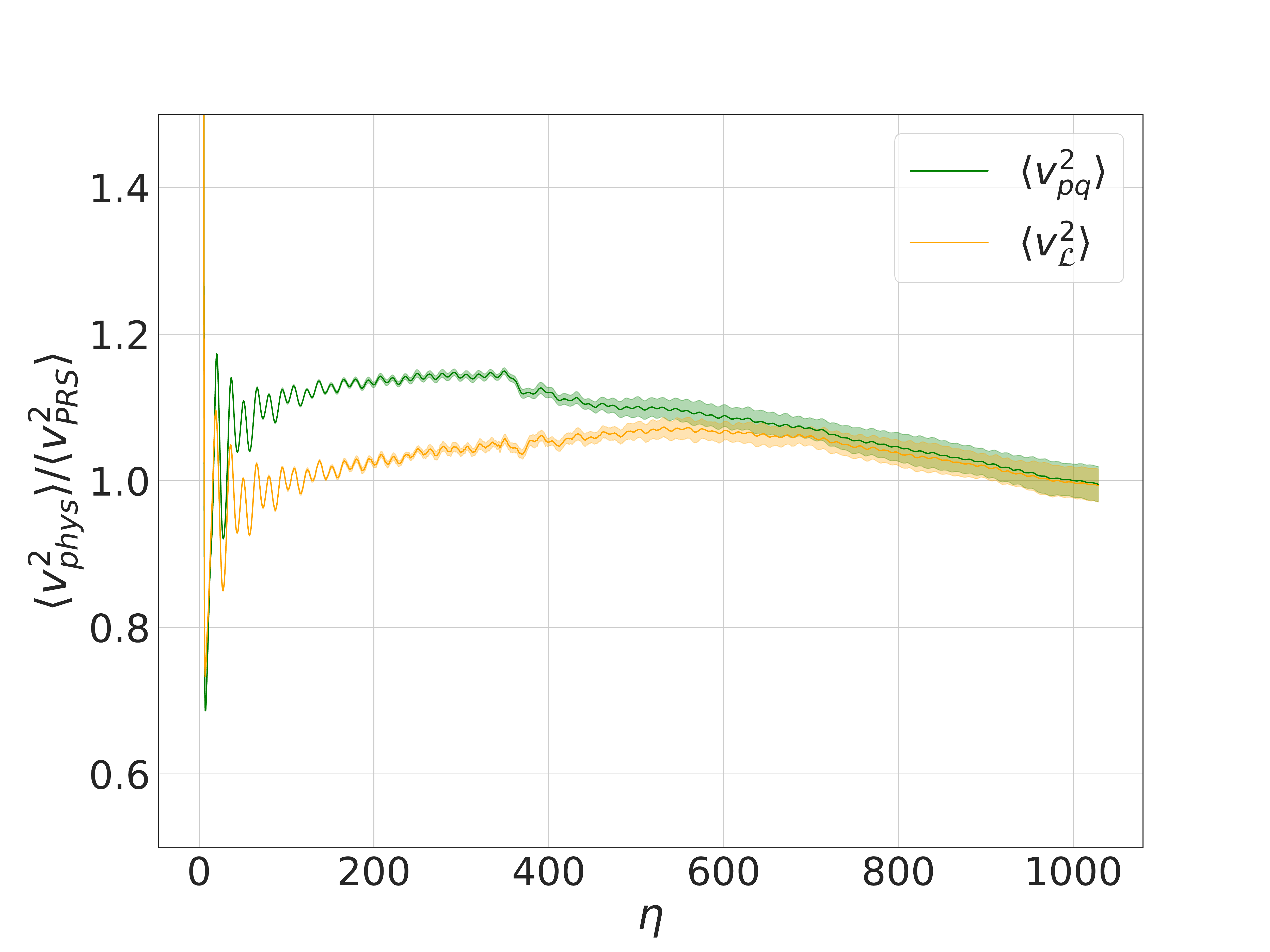}
\includegraphics[width=1.0\columnwidth]{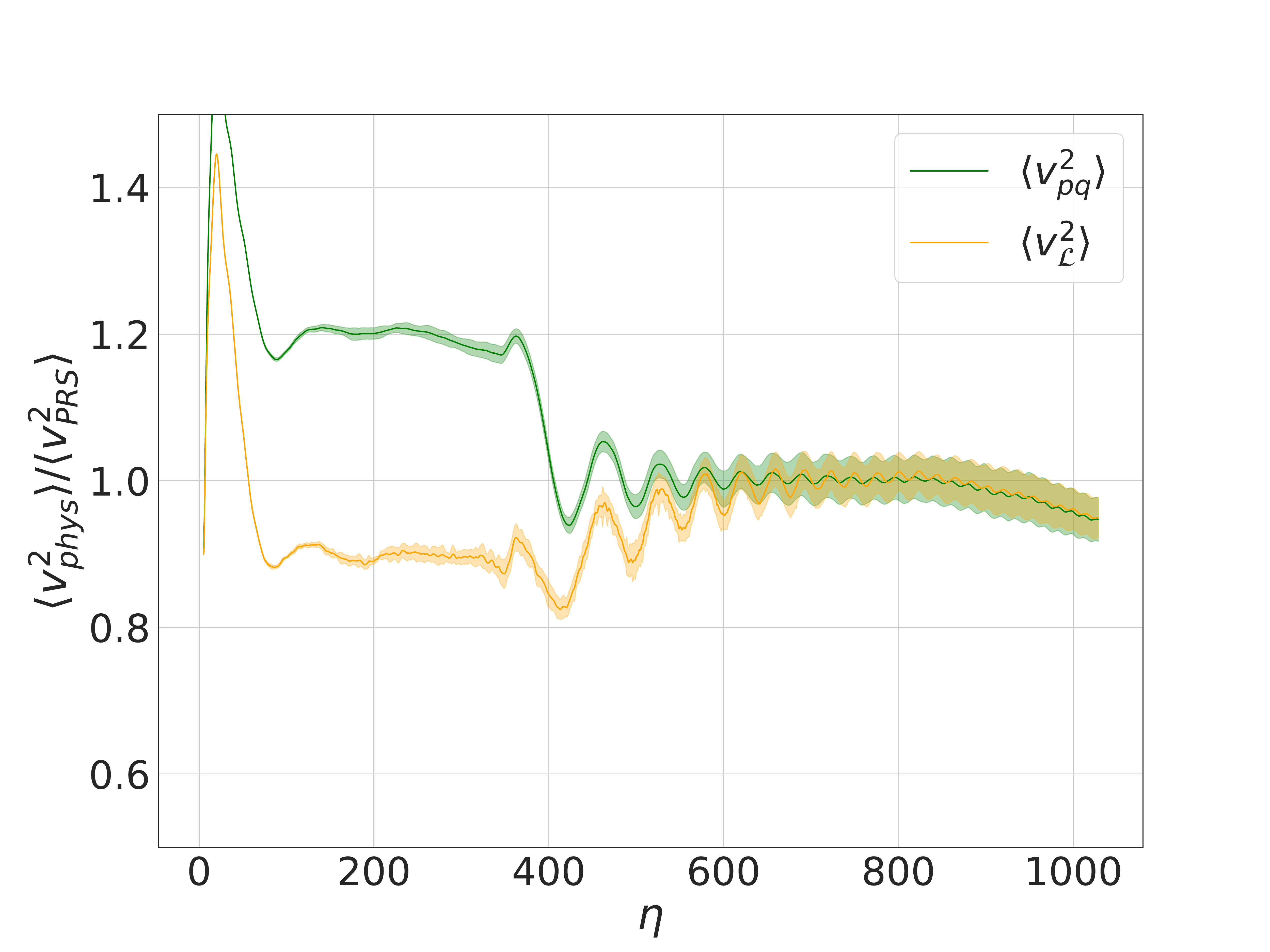}
\caption{The evolution of the mean squared velocity $\langle v^2_{\mathcal{W}} \rangle$ for the full network (by specifying the full Lagrangian as a weight function, in orange) or the $pq$-segments (by using the interaction potential as the weight, in green). Top panels use core growth and subsequent physical evolution, middle panels correspond to constant comoving width simulations, and bottom panels show the ratios of the two. Left and right panels correspond to the radiation and matter eras respectively.\label{figure11}}
\end{figure*}

The velocities can be computed both for the full network (weighted by the Lagrangian) or for the $pq$-segments (weighted by the interaction potential). The evolution of these quantities in both the growth plus physical evolution simulations and the constant comoving width simulations can be found in the top and middle panels of figure \ref{figure11}, respectively. Here, and also in table \ref{table5} we can find the asymptotic values of the velocities, again computed in conformal time range $\eta\in[950,1024]$. Note that in some cases the velocities are not exactly stable and are still slightly decreasing in the range used. This occurs for $\langle v^2_{pq} \rangle$, which decreases in all cases except in the matter era, for constant comoving width.

Again we see agreement in this time range for both physical and PRS simulations, although (as in the case of the mean string separations) the agreement in the asymptotic values is closer than the agreement in the approach to them. On the other hand, there is a clear disagreement between core growth $pq$-string velocities and constant width ones. To quantify the disagreements between core growth and physical evolution we can compute the ratio between the quantities in the top and middle panels; the bottom panels of \ref{figure11} depict the ratios of the two. This makes it clear that the asymptotic values for physical and PRS evolution agree (within their statistical uncertainties), while the core growth ones are clearly discrepant.

\begin{table*}
  \begin{center}
    \begin{tabular}{|c|c|c|c|c|c|c|}
      \hline
$\beta$ & m & Size, $\Delta x$ & $\langle v^2 \rangle_{pq}$ & $\langle v^2 \rangle_\mathcal{L}$ & Reference \\
        \hline
  $0$ & 1/2 & $4096^3, \Delta x=0.5$ & $0.307\pm0.005$ & $0.298\pm0.005$ & This section   \\ 
  $0$ & 1/2 & $1024^3, \Delta x=0.5$ & $0.319\pm0.008$ & $0.293\pm0.006$ & Previous section \\ 
  $0$ & 1/2 & $1024^3, \Delta x=0.5$ & $\sim0.33$      & $0.306\pm0.004$ & \cite{Lizarraga:2016hpd} \\ 
  $1$ & 1/2 & $4096^3, \Delta x=0.5$ & $0.309\pm0.008$ & $0.298\pm0.004$ & This section     \\
      \hline
  $0$ & 2/3  & $4096^3, \Delta x=0.5$ & $0.254\pm0.005$ & $0.250\pm0.004$ & This section    \\ 
  $0$ & 2/3  & $1024^3, \Delta x=0.5$ & $0.247\pm0.006$ & $0.253\pm0.009$ & Previous section \\
  $0$ & 2/3  & $1024^3, \Delta x=0.5$ & $\sim 0.27$     & $0.264\pm0.006$ & \cite{Lizarraga:2016hpd} \\
  $1$ & 2/3  & $4096^3, \Delta x=0.5$ & $0.249\pm0.006$ & $0.246\pm0.006$ & This section \\
      \hline
    \end{tabular}
\caption{The asymptotic values of the mean velocity squared $\langle v^2_\mathcal{W} \rangle$ for the full network (weighted by the Lagrangian) and $pq$-segments (weighted by the interaction potential) obtained from this simulations from this section, Abelian-Higgs simulations from \cite{Correia:2020yqg}, and $pq$-strings simulations from \cite{Lizarraga:2016hpd}.}
    \label{table5}
    \end{center}
\end{table*}

\section{\label{conc}Conclusions}

We have presented a detailed numerical study of the evolution of the $U(1) \times U(1)$ cosmic strings model, previously introduced by Saffin \cite{Saffin:2005cs}. This provides a convenient model to simulate networks with multiple string types and also a simplified proxy for the richer phenomenology expected from cosmic superstrings.

The work built upon our previously reported multi-GPU Abelian-Higgs code \cite{Correia:2020yqg}, which was suitably extended and modified to accommodate the $U(1) \times U(1)$, and subject to several validation tests. We have verified the validity of Gauss' law in both string sectors, complemented by visualizing bound states \cite{Movie}. Two separate methods were developed to compute the mean $pq$-segment separation and the average number of bound state segments in the simulation: a computationally fast one (though limited by some systematic effects) and a robust though more computationally costly one. We have used the largest field theory simulations of this model so far, specifically $4096^3$, $\Delta x = 0.5, \Delta t=0.1$ boxes. This substantially increased the available dynamic range, which for the physical evolution case was further maximized by the comparatively standard numerical method of introducing an early phase of core growth. 

One side benefit of our methodology is a comparison of the evolution under PRS (constant comoving width) and physical (true equations of motion) numerical settings---as well as of the more crude core growth assumption, which for numerical convenience often precedes the physical evolution phase of the simulations. Given the kinematic conditions for bound state formation (see for example \cite{PhysRevD.99.063516} and references therein) one might expect that this choice will have an impact in the abundances measured.  The combination of the two also enabled us to quantify how changing energy conservation in the action itself can impact the formation and destruction of bound states. 

In the case of core growth we find that nearly all bound states unwind into basic constituent strings. On the other hand, as soon as the true physical evolution begins, bound states start re-forming and we even reach slightly higher relative abundances than in the PRS case. For instance, $f_{total}$ reaches around $5\%$ in the case of core growth followed by physical evolution versus about $4\%$ in constant comoving width, for the matter epoch---although given the statistical uncertainties this difference is not statistically significant (it's only about one standard deviation). In any case, the core growth phase results are distinctly different from the PRS and physical cases. This allows us to conclude that changing how the comoving string width behaves (and therefore deviating from energy conservation at the level of the action) does have an impact on the evolution of realistic cosmic string networks, at least at the level of their bound states. With respect to the core growth case, our results are complementary to those of \cite{Klaer,Asier} which for global strings highlight a dramatic overestimation of average velocities (their work focuses on the radiation epoch, while we further show it in the matter epoch). In addition, the bound states disappear in the core growth case. Our results confirm that while core growth can be numerically convenient for the early evolution of defect simulations, it is clearly inadequate for direct cosmological studies.

The PRS algorithm is often used as the \textit{de facto} standard in numerical simulations, even for many non-Abelian-Higgs string models (apart from the model studied in the present work, semilocal strings provide another example) and one assumes that it correctly reproduces the dynamics of a network. There are very few detailed studies of this even in the simplest Abelian-Higgs case, and for the U(1)xU(1) case there was no such previous study that we are aware of. To our knowledge, this assumption has only been challenged for Abelian-Higgs networks at criticality \cite{Hindmarsh}, where bound states are not expected to form (and thus the impact of the PRS on kinematic constraints of bound state formation cannot be seen). This is relevant because, even if one assumes that the PRS algorithm is adequate for the Abelian-Higgs case, it does not necessarily follow that this will also hold for the U(1)xU(1) case, due to the presence of the bound states. Testing this assumption is therefore essential in order to assess the reliability of these simulations. Our results do suggest some differences in the derived quantities, although these differences may not be statistically significant---they are comparable to those between different length estimators. To our knowledge, this is the first time that a quantification of these possible PRS limitations has been done, and it will impact future field theory simulations beyond the simple Abelian-Higgs model.

Admittedly these differences are subtle, but still show themselves in the bound state abundance (see the matter epoch plots of Figures \ref{figure08} and \ref{figure09}) and in velocity estimations (see the lower plots of Figure \ref{figure11}). Moreover, they seem to diminish as the networks evolve: in other words, they affect how the network approaches a scaling solution more than the asymptotic properties of that scaling solution. On the one hand, as already mentioned in the previous paragraph, they might not be very significant statistically, given other uncertainties (e.g., related to the choice of estimators). On the other hand, even small differences, especially in the network velocity, are of importance for the calibration of semi-analytical models of cosmic superstrings, as our previous work with such calibrations for Abelian-Higgs strings has shown \cite{Correia:2019bdl}. To give one specific example, one may anticipate that using values for the network velocity and correlation length from the final timestep of a simulation only or using analogous values from a range of timesteps (in the final part of a simulation) may lead to quantitatively different model calibrations. Note that in order to calculate the rate of change of quantities such as the mean string separation one must necessarily choose a range of timesteps in which to calculate this slope, and the choice of such a range---and it will always be a matter of choice---could impact the results. To put it briefly, the PRS algorithm can be quantitatively reliable for simulations with a sufficiently large dynamic range (implying a large box size, of at least $4096^3$), but is unlikely to be better than qualitatively reliable for smaller simulations.

Turning to the main scientific results, we have presented robust evidence of scaling for the lightest strings, measured through a complete and self-consistent set of diagnostics for the string correlation lengths and velocities. In particular, our velocity estimator improves upon that of \cite{Lizarraga:2016hpd}. We find very good agreement with previous works \cite{Lizarraga:2016hpd,Urrestilla:2007yw}, bearing in mind that the simulations in these previous works had a much shorter dynamic range, and moreover they do not always report statistical uncertainties in their diagnostics. In short, our quantitative analysis confirms their more tentative results. 

The authors of \cite{Lizarraga:2016hpd}, and even \cite{Urrestilla:2007yw} earlier on, hypothesized that with a larger dynamic range than the ones they simulated would lead to a linearly growing average length of the bound state segments, consistent with scaling behaviour for this component. (In previously reported lower-resolution simulations, such behaviour had only been identified with carefully engineered initial conditions, rich in those segments.) The results we report confirm this expectation, without needing to resort to such engineered (but artificial) initial conditions rich in $pq$-strings, such as in \cite{Lizarraga:2016hpd}. Our reported evidence that the average segment length does exhibit linear growth is only possible due to the larger dynamic range of our simulations, which was beyond the computational resources of the earlier works. Whether or not these segments can be analytically modelled as a Brownian network is an interesting open problem which we intend to address in future work. 

Finally, we find low relative abundances of the bound states, in agreement with both earlier works. The earlier analysis of \cite{Urrestilla:2007yw} further hypothesized that a sufficiently high damping could lead to the formation of bound states (at least at an initial stage). We do not see any evidence of a large population of bound states forming at early stages of the network evolution in any of the sets of initial conditions which we have simulated. That said, we do find tentative evidence for an asymptotic constant value of the fraction of bound states. Given the relatively large uncertainties no firm conclusions can be drawn at present, but we also find a tentative indication of this value being different in the radiation and the matter eras (and being larger in the latter), which would be commensurate with the fact that the network has different characteristic lengthscales and velocities in the two epochs. Improving the statistical uncertainties in these bound state diagnostics by running more and/or larger simulations will be necessary for a definitive answer, and both of these are achievable with our code. 

In terms of analytic modelling, the differences in bound state abundance are also not dramatic but still show a 1-2 percent reduction when using PRS. In this case assessing the possible impact of any such differences is more difficult, since the absolute numbers of the bound states are low, our current statistical uncertainties on the overall bound state fraction are comparatively large, and no detailed analytic modelling of the bound states in this networks currently exists in the literature. In any case, these possible differences  between diagnostics extracted from PRS and physical evolution simulations should be kept in mind.

One clear task for future work is the extension to the case of unequal tensions, which, if done at high resolution, would naturally require computing resources comparable to the ones used in the present work. Numerically, it would suffice to set one of the simulation parameters such that the tension of $q$-strings would be, for example, exactly half of that of $p$-strings. The only possible bottleneck would be the issue of whether or not the estimators for the string characteristic lengths and velocities which have been discussed in the present work are still optimal in the unequal tensions case. Both the equal and unequal tension strings cases can be used to compare with semi-analytic modelling of multi-tension strings \cite{Pourtsidou:2010gu,PhysRevD.99.063516} provided relevant assumptions (in terms of inter-commutation probabilities) are assumed. We do note again that the rather small amount of $pq$-strings deserves further study and a more detailed quantification, since it is likely to be the limiting factor for quantitative calibrations of analytic models. Moreover, such analytic models generically depend on kinematic constraint hypotheses on junction formation. Therefore these simulations can enable forthcoming tests of the arguments and hypotheses of \cite{CKS,Zipping,PhysRevD.99.063516}. For now we can say that our simulations support the expectation that all string species seen so far can reach scaling.

In conclusion, the work presented herein demonstrates that our GPU-accelerated field theory code can by successfully extended beyond the simple Abelian-Higgs approximation. Work in progress also shows that an analogous programme of $8192^3$ simulations is viable for these networks. Together, these developments enable a detailed study of realistic string networks, containing additional degrees of freedom such as charges and currents on their worldsheets, together with a quantitative assessment of their observational signatures, leading to reliable constraints on theoretical paradigms in which they form---or possibly specific tests for their presence.

\begin{acknowledgments}
This work was financed by FEDER---Fundo Europeu de Desenvolvimento Regional funds through the COMPETE 2020---Operational Programme for Competitiveness and Internationalisation (POCI), and by Portuguese funds through FCT - Funda\c c\~ao para a Ci\^encia e a Tecnologia in the framework of the project POCI-01-0145-FEDER-028987 and PTDC/FIS-AST/28987/2017. J.R.C. is supported by an FCT fellowship (SFRH/BD/130445/2017). CJM acknowledges FCT and POCH/FSE (EC) support through Investigador FCT Contract 2021.01214.CEECIND/CP1658/CT0001. We gratefully acknowledge the support of NVIDIA Corporation with the donation of the Quadro P5000 GPU used for this research.

We acknowledge PRACE for awarding us access to Piz Daint at CSCS, Switzerland, through Preparatory Access proposal 2010PA4610, Project Access proposal 2019204986 and Project Access proposal 2020225448. Technical support from Jean Favre at CSCS is gratefully acknowledged.
\end{acknowledgments}

\bibliography{artigo}
\end{document}